\documentclass[review,3p,authoryear,11pt]{elsarticle}

\usepackage{amssymb}
\usepackage{natbib}
\usepackage{amscd}
\usepackage{amsmath}
\usepackage{amsfonts}
\usepackage{bm}
\usepackage{bbm}
\usepackage{amssymb}
\usepackage{latexsym}
\usepackage{mathrsfs}
\usepackage{amsthm}
\usepackage{mathrsfs}
\usepackage{booktabs}
\usepackage{caption}
\usepackage{rotating}
\usepackage{multirow}
\usepackage{lscape}
\usepackage{xcolor}
\usepackage{graphicx}
\usepackage{subfig}
\usepackage{cancel}

\biboptions{round}

\newtheorem{proposition}{Proposition}[section]
\newtheorem{remark}{Remark}[section]

\def\bdelta{\mbox{\boldmath $\delta$}}

\def\bomega{\mbox{\boldmath $\omega$}}

\def\bkappa{\mbox{\boldmath $\kappa$}}
\def\bpsi{\mbox{\boldmath $\psi$}}
\def\blambda{\mbox{\boldmath $\lambda$}}

\def\bXi{\mathbf{\Xi}} 

\def\bu{\mathbf{u}} 

\def\by{\mathbf{y}} 
\def\bY{\mathbf{Y}}
\def\0{\mbox{\bf{0}}}
\def\bs{\mathbf{s}}

\def\bA{\mathbf{A}}

\def\bQ{\mathbf{Q}}

\def\bB{\mathbf{B}}

\def\bU{\mathbf{U}}
 
\def\vec{\mbox{vec}}

\newcommand{\xp}{\mathbb{E}}

\def\qmo{``}
\def\qmc{''}
\def\qmcsp{'' }
\newcommand\bbone{\ensuremath{\mathbbm{1}}}
\newcommand{\suchthat}{\;\ifnum\currentgrouptype=16 \middle\fi|\;}

\begin{document}

\begin{frontmatter}



\title{\LARGE Switching--GAS Copula Models\\ with Application to Systemic Risk}



\author[bernardi]{Mauro Bernardi\corref{corresp}}
\author[catania]{Leopoldo Catania}
\cortext[corresp]{Department of Statistical Sciences, University of Padova, Via C. Battisti, 241/243, 35121, Padova, Italy. \texttt{mauro.bernardi@unipd.it}, tel.; +39.049.82.74.165, web page: \texttt{http://homes.stat.unipd.it/maurobernardi/}.}
\address[bernardi]{Department of Statistical Sciences, University of Padova, Padua, Italy.}
\address[catania]{Department of Economics and Finance, University of Rome, Tor Vergata, Rome, Italy.}

\begin{abstract}
Recent financial disasters have emphasised the need to accurately predict extreme financial losses and their consequences for the institutions belonging to a given financial market. The ability of econometric models to predict extreme events strongly relies on their flexibility to account for the highly nonlinear and asymmetric dependence observed in financial returns. We develop a new class of flexible Copula models where the evolution of the dependence parameters follow a Markov--Switching Generalised Autoregressive Score (SGASC) dynamics. Maximum Likelihood estimation is consistently performed using the Inference Functions for Margins (IFM) approach and the Expectation--Maximisation (EM) algorithm specifically tailored to this class of models. The SGASC models are then used to estimate the Conditional Value--at--Risk (CoVaR) and the Conditional Expected Shortfall (CoES), measuring the impact on an institution of extreme events affecting another institution or the market. The empirical investigation, conducted on a panel of European regional portfolios, reveals that the proposed SGASC model is able to explain and predict the evolution of the systemic risk contributions over the period 1999--2015.
\end{abstract}

\end{frontmatter}


\section{Introduction}
\label{sec:intro}
%
\noindent Recent financial disasters have emphasised the need to accurately predict extreme financial losses and their consequences for the institutions' financial health and, more generally, for the safety of the broader economy. Major financial crisis, such as the Global Financial Crisis (GFC) of 2007--2008 and the European Sovereign Debt Crisis (ESDC) of 2010--2011, usually spread over the whole economy leading to sharp economic downturns and recessions. In fact, during huge crisis episodes, the failure of banks and financial institutions is not rare and may trigger other non--financial institutions through the balance sheet and liquidity channels, threatening the stability of real economy, see, e.g., \cite{adrian_brunnermeier.2014}, \cite{adrian_shin.2010}, \cite{brunnermeier_pedersen.2009}, \cite{brunnermeier_etal.2009} and \cite{brunnermeier.2009}. The ability of econometric models to predict such extreme events strongly relies on their flexibility to model the highly nonlinear and asymmetric dependence structure of financial returns, see e.g., \cite{mcneil_etal.2015}. Despite its widespread use, the simple linear correlation fails to capture the important tails behaviour of the joint probability distribution, see, e.g., \cite{mcneil_etal.2015} and \cite{embrechts_etal.1999,embrechts_etal.2002}. Hence, modelling the tail dependence and the asymmetric dependence between pairs of assets have been becoming increasingly more important in a multivariate environment, especially after the recent crisis episodes. The departure from the linear correlation as measure of dependence usually implies to go beyond the multivariate Elliptical assumption for the joint distribution of asset returns. In this respect, the copula methodology allows to model a huge variety of dependence structures, see, e.g., \cite{durante_sempi.2015}. Another interesting feature of the dependence behaviour of stock returns, which displays its relevant role when measuring extreme co--movements, is that it usually evolves smoothly over time as a function of past assets co--movements, see, e.g., \cite{engle.2002} and \cite{tse_tsui.2002}. Due to the exposure of common shocks affecting all the market participants, the conditional correlation between asset returns increases during periods of financial instability, see, e.g., \cite{kotkatvuori-ornberg_etal.2013}, \cite{sandoval_junior_de_paula_franca.2012}, \cite{syllignakis_kouretas.2011} and \cite{kenourgios_etal.2011}. Dynamic copula models are referred to \cite{patton.2006} and \cite{jondeau_rockinger.2006}, although the problem of modelling the joint co--movement of stock returns was already present in \cite{bollerslev.1988}, \cite{bollerslev.1990} and \cite{engle.1990}, among others. Moreover, occasionally, we observe breaks into the dependence structure, which are more evident during crises periods and other infrequent events, as documented, for example, by \cite{bernardi_etal.2013b} and \cite{bernardi_petrella.2015}. As regards dependence breaks, Markov switching (MS) models have been proven to effectively capture non--smooth evolutions of the volatility and correlations dynamics. \cite{chollete_etal.2009} and \cite{rodriguez.2007}, for example, have firstly adopted MS copula with static regime--dependent parameters to analyse financial contagion.\newline
\indent In this paper, we propose to model the regime dependent dynamic of the copula parameters using the score driven framework recently introduced by \cite{creal_etal.2013} and \cite{harvey.2013}. Specifically, we allow the copula dependence parameters to depend on the realisations of a first order Markovian process with a specific Generalised Autoregressive Score (GAS) dynamic in each regime, while retaining an appropriate arbitrary specification for the marginals' conditional distribution dynamics. In this way, we extend the GAS literature by introducing nonlinearity and a stochastic evolution of the dependence structure similarly to \cite{boudt_etal.2012}. We name this new class of models Switching Generalised Autoregressive Score Copula (SGASC) models. During last few years, the conditional score approach has been extensively used to model the time varying behaviour of unobservable parameters in an observation driven environment \citep{cox.1981}. The use of the score as general updating mechanism has been justified in several ways by the literature. \cite{harvey.2013}, for example, represents the score process as a filter of an unobservable component model, while \cite{creal_etal.2013} suggest that the use of the score to update the latent parameter dynamic, can be interpret as a steepest ascent algorithm for improving the model's local fit given the current parameter position, as it usually happens into a Newton--Raphson algorithm. Recently, \cite{blasques_etal.2015} and \cite{blasques_etal.2014b} have demonstrated that the score driven processes are optimal within the class of nonlinear autoregressive dynamic. More precisely, they argue that, only the GAS processes are optimal in the sense of reducing the local Kullback--Leibler divergence between the true and the model implied conditional densities. Furthermore, the Kullback--Leibler optimality property of GAS models holds true under very mild conditions irrespective to the level of the possible model misspecification. Several theoretical results for the maximum likelihood estimate of score process has been developed by \cite{blasques_etal.2014}, \cite{andres.2014}, \cite{blasques_etal.2014e} and \cite{harvey.2013}. Moreover, score driven processes have been proved to be effectively used in many empirical applications. Most applications cover volatility modelling, as for example, \cite{harvey_luati.2014}, \cite{harvey_sucarrat.2014}, \cite{caivano_harvey.2014} and \cite{creal_etal.2011b}. Other empirical applications are in the systemic risk measurement as, for example, \cite{blasques_etal.2014a}, \cite{andre_etal.2014} and \cite{oh_patton.2013}, in the credit risk analysis, \cite{creal_etal.2011}, in macroeconomics, \cite{massacci.2014} and \cite{bazzi_etal.2014}, and in dependence modelling, \cite{harvey_thiele.2014}, \cite{janus_etal.2014}, and \cite{delirasalvatierra_patton.2015}.\newline
\indent The documented superior ability of GAS filters to approximate complicated nonlinear data generating processes in a straightforward and effective way (see, e.g.,  \citealt{koopman_etal.2015}) is particularly useful in the context of dependence modelling through copulas here considered. Specifically, the use of score driven models really helps to deal with cases where it is not clear how to update the parameter dynamics as for the archemedian copulas. Another relevant contribution of this paper is to introduce and estimate the Conditional Value--at--Risk (CoVaR) and the Conditional Expected Shortfall (CoES) risk measures, recently proposed by \cite{adrian_brunnermeier.2011,adrian_brunnermeier.2014} and \cite{girardi_ergun.2013}, for the class of SGASC models. The CoVaR measures co--movements between any two distinct institutions by extending the Value--at--Risk (VaR) to a conditional approach. Following the CoVaR methodology, the risk of an institution is evaluated as its VaR conditional to a relevant extreme event affecting another institution. The appealing characteristic of the CoVaR risk measure is that it inherits the flexibility of the dynamic switching copula framework here developed. The copula approach naturally adapts to environments characterised by different kind of upper and lower tail dependence enabling the CoVaR as an effective measure of the extreme conditional co--movements among financial variables. The literature on co--movement risk measures has proliferated during the last few years, see, e.g., \cite{bernardi_etal.2015}, \cite{bernal_etal.2014}, \cite{castro_ferrari.2014}, \cite{girardi_ergun.2013}, \cite{jager-ambrozewicz.2013}, \cite{sordo_etal.2015}. \cite{bisias_etal.2012} provide an extensive and up to date survey of the systemic risk measures that have been recently proposed.\newline
\indent One of the main appealing characteristics of the copula framework, with respect to standard distributions, relies on its ability to model the marginals' dynamics separately from the joint dependence structure, see, e.g., \cite{nelsen.2007}. Marginals and dependence separability has some additional advantages even from the econometric point of view, since it permits to employ a two--step procedure to estimate the parameters. This two--step procedure is known as Inference Function for margins (IFM) and is usually referred to \cite{godambe_1960} and \cite{mcleish_small.1988}. To estimate the SGASC model parameters, we adapt the IFM two--step procedure of \cite{patton.2006} for conditional copulas to the MS dynamics. More precisely, the marginals' conditional distributions parameters are estimated in the first step, while the copula GAS parameters are considered in the second step by adapting the Expectation--Maximization algorithm of \cite{dempster_etal.1977}.\newline
\indent The model superior ability to address the issue of tracking the potentially non--linear dependence dynamics usually observed in financial markets is assessed in the empirical part of the work. The empirical analysis considers a panel of European regional indexes and focuses on the evaluation of the systemic risk contributions of the considered European countries and their evolution during the recent GFC of 2007--2008 and the ESDB of 2010--2011. Our analysis confirms that the proposed SGASC model is able to explain and predict the systemic risk evolution of the considered countries in an effective way. In particular, we find results similar to those recently obtained by \cite{engle_etal.2015} using a modified version of the Marginal Expected Shortfall risk measure and \cite{lucas_etal.2014} who consider the European Sovereign debt CDS market. We found that the systemic risk contribution of each European country growths during the ESDC of 2010-2011 reaching its highest level in mid--2014. After a short decreasing period in the last part of 2014, the magnitude as well as the relative importance each country had in terms of systemic risk contribution sharply changed. In the very last part of our sample, especially after June 26, 2015 when the Greek government unilaterally broke off negotiations with the Eurogroup, we found that the overall systemic risk reaches again the mid--2014 levels.\newline
\indent The remainder of the paper is organised as follows. Section \ref{sec:model} introduces the SGASC model. We discuss the marginal models and then we detail the Markov--switching framework to model the joint behaviour of the series as well as the dynamic GAS specification. Section \ref{sec:hmm_inference} concerns the estimation methodology and presents the EM algorithm to estimate the GAS parameters. Section \ref{sec:risk_measures}
deals with the systemic risk measurement framework by providing the co--movement risk measures for the SGASC model. Section \ref{sec:empirical_study} presents data and discusses the main empirical results. Section \ref{sec:conclusion} concludes.
%
\section{The Model}
\label{sec:model}
%
\noindent Let $\bY_t=\left(Y_{1,t},\dots,Y_{d,t}\right)\in\mathbb{R}^d$ be a $d$--dimensional stochastic vector and let $\by_{1:t-1}=\left(\by_1,\dots,\by_{t-1}\right)$ be the past history up to time $t-1$ of the weakly stationary stochastic process $\{\bY_s, s>0\}$. Assume also that exists a first order ergodic Markov chain $\{S_s,s>0\}$ defined on the discrete space $\Omega=\{1,2,\dots,L\}$ with transition probability matrix $\bQ=\{q_{l,k}\}$, where $q_{l,k}=\mathbb{P}\left(S_t=k\mid S_{t-1}=l\right)$, $\forall l,k\in\Omega$ is the probability that state $k$ is visited at time $t$ given that at time $t-1$ the chain was in state $l$, and initial probabilities vector $\bdelta=\left(\delta_1,\dots,\delta_L\right)^\prime$, $\delta_l=\mathbb{P}\left(S_1=l\right)$, i.e., the probability of being in state $l=\left\{1,2,\dots,L\right\}$ at time 1. The SGASC model assumes
\begin{align}
\label{eq:joint_copula_model_density}
\mathbf{Y}_t\mid\left(\bY_{1:t-1}=\by_{1:t-1},S_{t}=l\right)\sim\mathcal{C}\left(\bU_{t};\bkappa_t^l,\bpsi^l\right),
\end{align}
where $\bU_t=\left(U_{1,t},\dots, U_{d,t}\right)$ and $U_{i,t}=F_{i}\left(Y_{i,t};\boldsymbol{\vartheta}_{i}\right)$ denotes the Probability Integral Transformation (PIT) of $Y_{i,t}$ according to its marginal conditional distribution function $F_{i}\left(Y_{i,t};\boldsymbol{\vartheta}_{i}\right)$ and $\mathcal{C}\left(\bU_{t} ;\bkappa_t^l,\bpsi^l\right)$ is the copula distribution conditional to the regime $S_t=l$. Hereafter, we assume that $\boldsymbol{\vartheta}_{i}$ generically denotes the parameters of the marginal distribution of $Y_i$, for $i=1,2,\dots,d$, and the copula distribution may depend on both static $\bpsi^l\in\mathcal{D}_{\psi}\subseteq\mathbb{R}^{m}$ and dynamic $\boldsymbol{\kappa}_t^l\in\mathcal{D}_{\kappa}\subseteq\mathbb{R}^{n}$ state dependent parameters. The SGASC model can be though of as a more general case of the static finite mixture of dynamic copulas models of \cite{creal_etal.2013} or an improvement of the model of \cite{jondeau_rockinger.2006}. The SGASC specification however differs from those classes of models since it introduces a GAS--type dynamic into a finite mixture of copulas, while retaining a Markovian structure on the dependence parameters that switches among two or more endogenous states, as in \cite{chollete_etal.2009}, \cite{pelletier.2006} and \cite{rodriguez.2007}.\newline
\indent In this section, we first introduce the general framework to model the univariate marginal conditional distributions and, then, we present the new SGASC specification. It is worth noting that, the marginal specification we are going to define should be intended for empirical purposes, and it is not strictly related to our new SGASC specification. Indeed, even if the selected marginals' specification is highly flexible, and it usually fits well equity returns distribution, alternative marginals specifications can be chosen.
%
\subsection{Marignal models}
\label{sec:model_marginal}
%
\noindent We assume each marginal stochastic process $\{Y_{i,s},s>0\}$, $i=1,2,\dots,d$ to follows an autoregressive process of order 1, AR$\left(1\right)$, with Skew Student--t innovations \citep{fernandez_steel.1998} and conditional time--varying variance driven by the GJR--GARCH$\left(1,1\right)$ specification of \cite{glosten_etal.1993}. Specifically, the AR(1)--GJR--GARCH(1,1)--$s\mathcal{ST}$ marginal model can be represented as
\begin{align}
\label{eq:marginal_model}
\varepsilon_{i,t} &= \frac{Y_{i,t}-\mu_{i,t}}{\sigma_{i,t}}\sim s\mathcal{ST}\left(0,1,\upsilon_i,\eta_i\right),\\
\mu_{i,t}&=\phi_{0,i}+\phi_{1,i} y_{i,t-1},\\
\label{eq:cond_volat_dyn}
\sigma_{i,t}^2&=\varpi_i+\vartheta_{1,i}\varepsilon_{i,t-1}^2+
\vartheta_{2,i}\bbone_{\left(-\infty,0\right]}\left(\varepsilon_{i,t-1}\right)\varepsilon_{i,t-1}^2+\vartheta_{3,i}\sigma_{i,t-1}^2,
\end{align}
where $\left(\upsilon_i,\eta_i\right)\in\left(0,+\infty\right)\times\left(0,+\infty\right)$ represent the degrees--of--freedom and the skewness parameters, respectively. The standardised skew Student--t $s\mathcal{ST}\left(0,1,\upsilon_i,\eta_i\right)$ density for $\varepsilon_{i,t}$ is
\begin{align}
\tilde t_{\upsilon_i}\left(\varepsilon_{i,t},0,1,\eta_i\right)&=\frac{2}{\eta_i+\frac{1}{\eta_i}}t_{\upsilon_i}\left(\frac{\varepsilon_{i,t}}{\eta_i},0,1\right)\bbone_{\left[0,+\infty\right)}\left(\varepsilon_{i,t}\right)
+\frac{2}{\eta_i+\frac{1}{\eta_i}}t_{\upsilon_i}\left(\eta_i\varepsilon_{i,t},0,1\right)\bbone_{\left(-\infty,0\right)}\left(\varepsilon_{i,t}\right),
\label{eq:skew_student_fernandez}
\end{align}
where $t_{\upsilon_i}\left(\cdot,0,1\right)$, denotes the standardised Student--t density with $\upsilon_i$ degrees of freedom. To preserve the stationarity and to ensure the positiveness of the conditional volatility process $\sigma_{i,t}^2$ for any $t=1,2,\dots,T$, the following conditions are imposed to the GJR--GARCH$\left(1,1\right)$ dynamics in equation \eqref{eq:cond_volat_dyn}: $\varpi_i>0$, $\vartheta_{1,i}\geq0$, $\vert\vartheta_{2,i}\vert<1$, $0\leq\vartheta_{3,i}<1$, with $\vartheta_{1,i}+\vartheta_{2,i}\varsigma_i + \vartheta_{3,i}<1$, for $i=1,2,\dots,d$, where $\varsigma_i=\int_{-\infty}^{0} \tilde t_{\upsilon_i}\left(\varepsilon,0,1,\eta_i\right)d\varepsilon=\frac{\eta_i}{1+\eta_i^2}$.
The skewing mechanism of \cite{fernandez_steel.1998} provides a flexible way to build a skewed distribution starting from any arbitrary univariate unimodal symmetric density function, by including an additional parameter $\eta_i\in\mathbb{R}^+$ that controls for the skewness. The choice of modelling the conditional mean as a first order autoregressive process in equation \eqref{eq:marginal_model} is motivated by the need to account for the serial dependence sometimes displayed by the financial returns, as described by \cite{embrechts_etal.2003}.
For easy of readability, we group all the marginal model parameters in the vectors $\boldsymbol{\vartheta}_i=\left(\phi_{0,i},\phi_{1,i},\varpi_{i},\vartheta_{1,i}, \vartheta_{2,i}, \vartheta_{3,i},\eta_i,\upsilon_i\right)$, for $i=1,2,\dots,d$.
%
\subsection{Switching GAS copula model}
\label{sec:biv_model}
%
\noindent Conditionally on the state of the Markov chain $S_t=l$, for $l=1,2,\dots,L$, we assume that the copula dependence parameter $\bkappa_t^l$ follows the GAS dynamic updating mechanism specified as
\begin{align}
\bkappa_t^l&=\blambda\left(\tilde\bkappa_t^l\right)\label{eq:sgasc_mapping}\\
\tilde\bkappa_t^l&=\bomega^l+\mathbf{A}^l\tilde{\bs}_{t-1}^l+\mathbf{B}^l \tilde\bkappa_{t-1}^l\label{eq:sgasc_dynamic}\\
\tilde{\bs}_t^l&=\mathbf{\tilde S}_t\left(\tilde\bkappa_t^l,\bpsi^l\right)\tilde{\nabla}_t\left(\bu_t;\tilde{\bkappa}_t^l,\bpsi^l\right),\label{eq:sgasc_scaled_score}
\end{align}
where $\bu_t=\left(u_{1,t},\dots,u_{d,t}\right)$, $\bomega^l\in\mathbb{R}^n$ and $\bA^l$, $\bB^l$ are matrices of parameters of dimension $\left(n\times n\right)$ such that the eigenvalues of $\bB^l$ are in modulus strictly less than one to preserve the stationarity of the GAS dynamic, and $\tilde{\bs}_t^l$ is the scaled--score of observation $\bu_t$. The parameter $\bomega^l$ controls for the state specific level of the GAS dynamic, while the matrices $\mathbf{A}^l$ and $\mathbf{B}^l$ control for the updating step and the persistence of the process, respectively.
Moreover, $\blambda:\mathbb{R}^n\to\mathcal{D}_\kappa$ in equation \eqref{eq:sgasc_mapping} is an absolutely continuous deterministic invertible function that maps $\mathbb{R}^n$ into the natural parameter space $\mathcal{D}_\kappa$. When $n\ge2$ it is convenient to specify $\blambda$ as a vector valued function being able to map each component of $\bm{\kappa}_t$ into the proper space. In general, for the $i$--th component of $\blambda$, we consider the modified logistic function defined by
\begin{equation}
\lambda_{i}^{\left(\underline{b},\bar b \right)}\left(x\right)=\underline{b} + \frac{\left(\bar b-\underline{b}\right)}{1+e^{-x}},
\label{eq:mapping}
\end{equation}
which maps $\mathbb{R}$ into the interval $\left(\underline{b},\bar b\right)$. The scaled--score function $\tilde{\bs}_t^l$ in equation \eqref{eq:sgasc_scaled_score} can usually be obtained as the product of the observation score with respect to the natural parameter $\bkappa_t$ and a proper transformation of the jacobian of the mapping function, in the following way
\begin{equation}
\mathbf{\tilde S}_t\left(\tilde\bkappa_t^l,\bpsi^l\right)\tilde{\nabla}_t\left(\tilde{\bkappa}_t^l,\bpsi^l\right) =
z\left(\dot\blambda_t^l\right)\mathbf{S}_t\left(\bu_t;\bkappa_t^l,\bpsi^l\right){\nabla}_t\left(\bu_t;{\bkappa}_t^l,\bpsi^l\right),
\end{equation}
where
\begin{align}
\nabla_t\left(\bu_t;{\bkappa}_t^l,\bpsi^l\right)=\frac{\partial\ln c\left(\bu_t;\bm{\kappa}^l_t,\bpsi^l\right)}{\partial\bm{\kappa}^l_t},\qquad\qquad\quad
\dot{\blambda}_t^l=\frac{\mathrm{d}\blambda\left(\bm{\tilde\kappa}^l_t\right)}{\mathrm{d}\bm{\tilde\kappa}^l_t},\nonumber
\end{align}
and $z\left(\cdot\right)$ depends on the choice of scaling mechanism, see \cite{creal_etal.2013}. Hence, $\nabla_t\left(\bu_t;\bkappa_t^l,\bpsi^l\right)$ denotes the conditional score of the copula probability density function $c\left(\cdot\right)$, evaluated at $\bm{\kappa}_t$ and $\mathbf{S}_t\left(\bm{\kappa}_t,\bpsi\right)$ is a positive definite, possible parameter--dependent, scaling matrix. Convenient choices for the scaling matrix are usually given by
\begin{equation}
\mathbf{S}_t\left(\bkappa^l_t,\bpsi^l\right)=\left[\mathcal{I}\left(\bkappa^l_t,\bpsi^l\right)\right]^{-\zeta},\label{eq:sgasc_scale}
\end{equation}
where $\mathcal{I}\left(\tilde{\bkappa}^l_t,\bpsi^l\right)$ is the Fisher information matrix that, for well behaved densities, could be written as
\begin{align}
\mathcal{I}\left(\bkappa^l_t,\bpsi^l\right)&=-\mathbb{E}_{t-1}\left[\frac{\partial^2\ln c\left(\bu_t;\bkappa^l_t,\bpsi^l\right)}{\partial \bm{\kappa}_t\partial \bm{\kappa}_t^\prime}\right]\nonumber\\
&=\mathbb{E}_{t-1}\left[\nabla_t\left(\bu_t;\bkappa^l_t,\bpsi^l\right)\times\nabla_t\left(\bu_t;\bkappa^l_t,\bpsi^l\right)^\prime\right],
\label{eq:informationMatrix}
\end{align}
and $\zeta$ is usually set equal to $\{0,\frac{1}{2},1\}$. \cite{creal_etal.2013} suggest to use the inverse Fisher information matrix, which corresponds to set $\zeta=1$ in equation \eqref{eq:sgasc_scale}, or its pseudo--inverse square root, which corresponds to $\zeta=\frac{1}{2}$, in order to scale the conditional score for a quantity that accounts for its variance. In our empirical tests, we find this latter scaling mechanism much more efficient than using an identity scaling matrix ($\zeta=0$). However, sometimes the Fisher information matrix is not available in closed form, and we need to resort to simulation or numerical evaluations methods, which should be traded--off with approximation degree and code efficiency. In the supplementary material accompanying this paper we report the analytical formulae to compute the score and the Fisher information matrix for several bivariate copula specifications: Gaussian, Student--t, Gumbel and Clayton. The Frank and Plackett Fisher information matrices are instead evaluated using the grid approach proposed by \cite{creal_etal.2013}, as detailed in the accompanying supplementary material.
%
\section{Estimation and inference}
\label{sec:hmm_inference}
%
\noindent As mentioned in the Introduction, model parameters are estimated using a two step procedure that consists of a first step where the parameters involved in the marginals are estimated, followed by a second step where the dependence parameters are jointly estimated along with the latent Markovian states. The resulting Inference Functions for Margins (IFM) two step procedure has been proved to be asymptotically consistent for conditional copulas by \cite{patton.2006} to which we refer for further details. The IFM procedure can be applied in this context as well as in general dynamic copula models since the parameters of the copula distribution in equation \eqref{eq:joint_copula_model_density} are separable from those of the marginals. In the specific case of the dynamic MS model here considered we also take advantage from the fact that the Markovian dynamic is imposed only on the dependence parameters of the copula function and not on the marginals. To estimate the dependence parameters subject to the Markovian structure as well as to the GAS dynamic specified in the previous section we adapt the Expectation--Maximization algorithm of \cite{dempster_etal.1977}. In this section we present the EM algorithm, assuming that, the parameters of the marginal distribution $\boldsymbol{\vartheta}_i$ have been previously consistently estimated by maximum likelihood, see, e.g., \cite{francq_zakoian.2010}. Hereafter, the pseudo--observations $\bu_t$ are replaced by $\hat \bu_{t}=\left({F}_1\left(y_{1,t};\hat{\boldsymbol{\vartheta}}_1\right),\dots, {F}_d\left(y_{d,t};\hat{\boldsymbol{\vartheta}}_d\right)\right)^\prime$, where $\hat{\boldsymbol{\vartheta}}_i$ is the ML estimate of $\boldsymbol{\vartheta}_i$ for all $i=1,2,\dots,d$.\newline
\indent The EM algorithm is a powerful and easy programmable tool for ML estimation on data having missing structures, such as finite mixtures and Markov--Switching models, and it releases a non--decreasing sequence of the log--likelihood function converging to the maximum. For a general and up--to--date reference on the EM algorithm see the book of \cite{mclachlan_krishnan.2007}. In what follows, we present the EM algorithm for estimating the parameters of the SGASC model described in Section \ref{sec:model}.\newline
\indent For the purpose of application of the EM algorithm the vector of observations $\by_{1:T}$, where $T$ is the sample size, is regarded as being incomplete. Following the implementation described in \cite{mclachlan_peel.2000} the following missing data are consequently introduced $z_{t}=\left(z_{t,1},z_{t,2},\dots,z_{t,L}\right)^\prime$ and $zz_{t}=\left(zz_{t,1,1},zz_{t,1,2},\dots,zz_{t,l,k},\dots,zz_{t,L,L}\right)^\prime$ being defined as
\begin{align}
z_{t,l}&=\left\{\begin{array}{cc}1 &  \text{if}\quad S_t=l,\\ 0 & \text{otherwise}\end{array}\right.\nonumber\\
zz_{t,l,k}&=\left\{\begin{array}{cc}1 & \text{if}\quad S_{t-1}=l, S_{t}=k,\\ 0 & \text{otherwise.}\end{array}\right.\nonumber
\end{align}
Similarly to the latent class approaches, the class membership is unknown and conveniently treated as the value taken by a latent Multinomial variable with one trial and $L$ classes, where the temporal evolution of class membership is driven by the hidden Markov chain $S_t$ for $t=1,2,\dots,T$.
Augmenting the observations $\by_{1:T}$ with the latent variables $\left\{z_t,zz_t,t=1,2,\dots,T\right\}$ allows for replacing the log--likelihood function with the complete--data log--likelihood, which becomes
\begin{align}
\label{eq:copula_complete_loglike}
\log\mathcal{L}_{\sc}\left(\bXi\right)
&=\sum_{l=1}^Lz_{1,l} \log\left(\delta_l\right)
+\sum_{l=1}^L\sum_{k=1}^L\sum_{t=1}^Tzz_{t,l,k}\log\left(q_{l,k}\right) \nonumber\\
&\qquad+\sum_{l=1}^L\sum_{t=1}^Tz_{t,l}\log\mathit{c}\left(\hat\bu_t;\bXi^l\right)
+\sum_{t=1}^T\sum_{i=1}^d\log f_i\left(y_{i,t};\hat{\boldsymbol{\vartheta}}_{i}\right),\nonumber
\end{align}
where $\bXi=\{\bXi^l\}_{l=1}^L$, with $\bXi^l=\left(\bomega^l,\vec{\left(\bA^l\right)},\vec{\left(\bB^l\right)},\bpsi^l\right)$ is a vector containing the parameters of the GAS dynamics for the copula dependence parameters $\boldsymbol{\kappa}_t^l$ and $\bpsi^l$, for $l=1,2,\dots,L$.
The EM algorithm consists of two major steps, one for expectation (E--step) and one for maximization (M--step), see, e.g., \cite{mclachlan_krishnan.2007}. On the $\left(m+1\right)$--th iteration the EM algorithm proceeds as follows:
\begin{itemize}
\item[E--step:] computes the conditional expectation of the complete data log--likelihood given the observed data and the current parameters estimate
\begin{equation}
\mathcal{Q}\left(\bXi,\bXi^{\left(m\right)}\right)=\xp^{p\left(z_t\mid\by_{1:T},\bXi^{\left(m\right)}\right)}
\left[\log\mathcal{L}_{\sc}\left(\bXi\right);\by_{1:T}\right].
\label{eq:al_hmm_estep}
\end{equation}
\item[M--step:] choose $\bXi^{\left(m+1\right)}$ by maximizing the preceding expected values with respect to $\bXi$
\begin{equation}
\Xi^{\left(m+1\right)}=\arg\max_{\bXi}\mathcal{Q}\left(\bXi,\bXi^{\left(m\right)}\right).
\label{eq:al_hmm_mstep}
\end{equation}
\end{itemize}
\indent The E--step in equation \eqref{eq:al_hmm_estep} requires the computation of the so--called $\mathcal{Q}$--function, which calculates the conditional expectation of the complete--data log--likelihood given the observations and the current estimate of the parameter vector $\bXi^{\left(m\right)}$
$\forall t=1,2,\dots,T$ and $l=1,2,\dots,L$. Exploiting the previous factorisation we obtain the following representation of the function $\mathcal{Q}$
\begin{align}
\label{eq:MALD_Q_function}
\mathcal{Q}\left(\bXi,\bXi^{\left(m\right)}\right)
&=\xp^{p\left(z_{t}\mid\by_{1:T},\bXi^{\left(m\right)}\right)}
\left\{\log\mathcal{L}_{\sc}\left(\bXi\right);\by_{1:T}\right\}\nonumber\\
&\propto\sum_{l=1}^L\hat{z}_{1,l} \log\left(\delta_l\right)
+\sum_{l=1}^L\sum_{k=1}^L\sum_{t=1}^T\widehat{zz}_{t,l,k}\log\left(q_{l,k}\right)
+\sum_{l=1}^L\sum_{t=1}^T\hat{z}_{t,l}\log\mathit{c}\left(\hat\bu_t;\bXi^l\right),\nonumber
\end{align}
where
\begin{equation}
\hat{z}_{t,l}=\mathbb{P}\left(S_t=l\mid\by_{1:T},\bXi^{\left(m\right)}\right),\qquad\quad
\widehat{zz}_{t,l,k}=\mathbb{P}\left(S_{t-1}=l,S_t=k\mid\by_{1:T},\bXi^{\left(m\right)}\right),\nonumber
\end{equation}
for $l,k=1,2,\dots,L$, and $\forall t=1,2,\dots,T$, denote the current smoothed probabilities of the states evaluated using the well--known Forward--Filtering Backward--Smoothing (FFBS) algorithm detailed in \cite{fruhwirth_schnatter.2006}, and Capp\'e \textit{et al.} \citeyearpar{cappe_etal.2005}.\newline
\indent The M--step in equation \eqref{eq:al_hmm_mstep} maximizes the function $\mathcal{Q}\left(\bXi,\bXi^{\left(m\right)}\right)$ with respect to $\bXi$ to determine the next set of parameters $\bXi^{\left(m+1\right)}$. The updated estimates of the HMM parameters, i.e., the vector of initial probabilities $\bdelta$ and the transition probability matrix of the hidden Markov chain $\mathbf{Q}$ are:
\begin{align}
\hat{\delta}_l^{\left(m+1\right)}=\hat{z}_{1,l}\qquad\quad
\hat{q}_{l,k}^{\left(m+1\right)}=\frac{\sum_{t=2}^T\widehat{zz}_{t,l,k}}{\sum_{k=1}^L\sum_{t=2}^T\widehat{zz}_{t,l,k}},\nonumber
\end{align}
for $l,k=1,2,\dots,L$, while the parameters $\bXi$ can be obtained as the solution of the following optimisation problem
\begin{eqnarray}
\label{eq:rho_param_opt}
\bXi^{(m+1)}
=\arg\max_{\bXi}\sum_{l=1}^L\sum_{t=1}^T\widehat{z}_{t,l}\log\mathit{c}\left(\hat\bu_t;\bXi^l\right).\nonumber
\end{eqnarray}
\noindent Convergence of the algorithm to the ML estimates is guaranteed since the last optimisation step delivers a parameter update that increases the log--likelihood function. Standard errors for the SGASC dependence and the marginals parameters can be evaluated using the procedure detailed in \cite{patton.2006}, where the Information Matrix of $\bXi$ is evaluated numerically after one run of the direct Maximum Likelihood estimator for HMM detailed by \cite{zucchini_macdonald.2009}.
%
\section{Systemic risk measures}
\label{sec:risk_measures}
%
\noindent In this section, we first introduce the two systemic risk measures we consider throughout the paper, namely the Conditional Value--at--Risk (CoVaR) and the Conditional Expected Shortfall (CoES) and then, we describe how CoVaR and CoES can be calculated assuming the joint returns follow the SGASC model. CoVaR and CoES have been introduced in the systemic risk literature by \cite{adrian_brunnermeier.2011,adrian_brunnermeier.2014}, and subsequently extended to a parametric dynamic framework by \cite{girardi_ergun.2013}. The CoVaR measures the spillover effects between institutions by providing information on the Value--at--Risk of an institution or market, conditional on another institution's distress event. In their seminal paper, \cite{adrian_brunnermeier.2011} propose to estimate the CoVaR measure using a system of two quantile equations extending the traditional approach of direct Value--at--Risk estimate. \cite{bernardi_etal.2013a,bernardi_etal.2015} instead propose a Bayesian dynamic quantile model where both the VaR and CoVaR equations are function of individual and macroeconomic observed risk factors, as in the original CoVaR approach, as well as of unobserved components having their own stochastic dynamics. The copula approach to the CoVaR has been recently proposed by \cite{reboredo_ugolini.2015} to evaluate the systemic risk in European sovereign debt markets. Here, the CoVaR and CoES are further extended to account for both the dynamic evolution of the copula dependence parameters as well as the presence of distinct Markovian regimes. The appealing characteristic of the model based CoVaR risk measurement framework is that it inherits the flexibility and easy computability of the dynamic switching copula. Hereafter, given the bivariate nature of the CoVaR and the CoES systemic risk measures, we consider a bivariate SGASC specification for the stochastic vector $\left(Y_{\mathrm{M},t},Y_{j,t}\right)$, $j=1,2,\dots,N$, where \qmo$\mathrm{M}$\qmcsp denotes the financial System index and \qmo$j$\qmcsp denotes the $j$--th systemically relevant institution.\newline
\indent Formally, let $\left(\tau_1,\tau_2\right)\in\left[0,1\right]^2$ be predetermined confidence levels, then the CoVaR of the financial system at time $t$, denoted by ${\rm CoVaR}_{\mathrm{M}\vert j,t}^{\tau_1,\tau_2}$, satisfies the following equation
\begin{equation}
\mathbb{P}\left(Y_{\mathrm{M},t}\leq {\rm CoVaR}_{\mathrm{M}\vert j,t}^{\tau_1,\tau_2}\mid Y_{j,t}\leq {\rm VaR}_{j,t}^{\tau_2}\right)=\tau_1,
\label{eq:covar_def}
\end{equation}
where ${\rm VaR}_{j,t}^{\tau_2}$ denotes the marginal Value--at--Risk (VaR) of institution $j$ such that $\mathbb{P}\left(Y_{j,t}\leq{\rm VaR}_{j,t}^{\tau_2}\right)=\tau_2$. Roughly speaking, the Conditional Value--at--Risk of the financial system, is the quantile of the distribution of $Y_{M,t}$ conditional on an extreme event affecting institution $j$'s returns $Y_{j,t}$. As in \cite{girardi_ergun.2013}, we define such an extreme event as $Y_{j,t}$ being below its VaR at confidence level $\tau_2$.
\begin{remark}
The definition of Conditional Value--at--Risk in equation \eqref{eq:covar_def} is substantially different from that originally presented in \cite{adrian_brunnermeier.2011} and coincides with that proposed in \cite{girardi_ergun.2013}. As discussed in \cite{girardi_ergun.2013} and \cite{mainik_schaanning.2014} this definition essentially preserves the stochastic ordering introduced by the joint distribution.
\end{remark}
\noindent Given the dynamic context introduced in the previous Sections, the random variable we refer to for the calculation of the forward looking systemic risk measure ${\rm CoVaR}_{\mathrm{M}\vert j,t+1}^{\tau_1,\tau_2}$ is $\left(Y_{\mathrm{M},t+1},Y_{j,t+1}\mid\by_{1:t},S_t\right)$. The ${\rm CoVaR}_{\mathrm{M}\vert j,t+1}^{\tau_1,\tau_2}$ and ${\rm VaR}_{j,t+1}^{\tau_2}$ characterise the conditional and marginal quantiles of the predictive distribution of $\left(Y_{\mathrm{M},t+1},Y_{j,t+1}\mid\by_{1:t},S_t\right)$. The next Proposition provides the natural link between the dynamic model and the approach to systemic risk assessment by characterising the predictive distribution.
\begin{proposition}
Let $\bY_{t}=\left(Y_{\mathrm{M},t+1},Y_{j,t+1}\right)$ follows a bivariate SGASC process, then the one--step ahead predictive cumulative distribution function of $\bY_{t}$ at time $t+1$, given information up to time $t$ is a mixture of component specific predictive cumulative distributions
\begin{eqnarray}
{\rm H}\left(\bY_{t+1}\mid\by_{1:t},S_t\right)
=\sum_{l=1}^L\pi_{t+1\vert t}^{(l)}\mathcal{C}\left(\bU_{t+1}\mid S_{t+1}=l,\by_{1:t}\right),
\label{eq:hmm_predictive_general}
\end{eqnarray}
with mixing weights
\begin{eqnarray}
\pi_{t+1\vert t}^{(l)}=\sum_{m=1}^L q_{m,l}\mathbb{P}\left(S_{t}=m\mid\by_{1:t}\right),\qquad l=1,2,\dots,L,
\end{eqnarray}
where $\bU_{t+1}=\left(F_{\mathrm{M}}\left(Y_{\mathrm{M},t+1},\boldsymbol{\vartheta}_\mathrm{M}\right),F_{j}\left(Y_{j,t+1},\boldsymbol{\vartheta}_j\right)\right)$, $j=1,\dots,N$ and $q_{m,l}$ is the $\left(m,l\right)$--th entry of the Markovian transition matrix $\mathbf{Q}$, see, e.g.,  \cite{zucchini_macdonald.2009}.
\end{proposition}
\noindent It is worth adding that equation $\eqref{eq:hmm_predictive_general}$ follows immediately from the fact that the cumulative predictive distribution of $\bY_{t+1}$ given the past history of the process is a finite mixture of copulas and that mixtures of copulas are copulas themselves, see, e.g., \cite{durante_sempi.2015}. We now provide expressions for the one--step--ahead \qmo predictive\qmcsp systemic risk measures.\newline
\indent In principle, the calculation of CoVaR requires the prior evaluation of institution's $j$ marginal VaR that can be performed by inverting the marginal cdf of $Y_j$, i.e., ${\rm VaR}_{j,t+1}^{\tau_2}=F_{j}^{-1}\left(\tau_2;\boldsymbol{\vartheta}_j\right)$. From a computational point of view, the copula approach is more tractable because it does not require the evaluation of the marginal VaR which coincides with the PIT at the chosen quantile confidence level $\tau_2$. Conditional on ${\rm VaR}_{j,t+1}^{\tau_2}$, the ${\rm CoVaR}_{\mathrm{M}\vert j,t+1}^{\tau_1,\tau_2}$ is calculated as the value of $y_{t+1}^*$ such that
\begin{equation}
\mathbb{P}\left(Y_{\mathrm{M},t+1}\leq y_{t+1}^*,Y_{j,t+1}\leq {\rm VaR}_{j,t+1}^{\tau_2}\right)=\tau_1\tau_2,
\end{equation}
or in terms of the bivariate SGASC predictive distribution we consider here
\begin{align}
&\sum_{l=1}^L\pi_{t+1\vert t}^{(l)}\mathcal{C}\left(F_{\mathrm{M}}\left(y_{t+1}^*,\boldsymbol{\vartheta}_\mathrm{M}\right),F_{j}\left({\rm VaR}_{j,t+1}^{\tau_2},\hat{\boldsymbol{\vartheta}}_j\right);\bkappa_{t}^l,\bpsi^l\right)\nonumber\\
&\qquad\qquad\qquad\qquad\qquad=\sum_{l=1}^L\pi_{t+1\vert t}^{(l)}\mathcal{C}\left(F_{\mathrm{M}}\left(y_{t+1}^*,\boldsymbol{\vartheta}_\mathrm{M}\right),\tau_2;\bkappa_{t}^l,\bpsi^l\right)=\tau_1\tau_2.
\end{align}
\cite{adrian_brunnermeier.2014} also propose to extend the expected shortfall (ES) risk measure to a systemic framework by evaluating the marginal ES of each institution at the CoVaR level. The CoES is defined as the expected shortfall of $Y_{\mathrm{M},t+1}$ below its ${\rm CoVaR}_{\mathrm{M}\vert j,t+1}^{\tau_1,\tau_2}$ level, conditional to $Y_{j,t+1}$ being below its $\mathrm{VaR}_{j,t+1}^{\tau_2}$ level, i.e.,
\begin{align}
{\rm CoES}_{\mathrm{M}\vert j,t+1}^{\tau_1,\tau_2}&=\mathrm{ES}\left(Y_{\mathrm{M},t+1}\le {\rm CoVaR}_{\mathrm{M}\vert j,t+1}^{\tau_1,\tau_2}\mid Y_{j,t+1}\le {\rm VaR}_{j,t+1}^{\tau_2}\right)
\label{eq:CoES}\\
&=\xp\left(Y_{\mathrm{M},t+1}\mid Y_{\mathrm{M},t+1}\le {\rm CoVaR}_{\mathrm{M}\vert j,t+1}^{\tau_1,\tau_2},Y_{j,t+1}\le {\rm VaR}_{j,t+1}^{\tau_2}\right).
\label{eq:CoES_2}
\end{align}
Our definition of CoES coincides with the one given by \cite{bernardi_etal.2013b}, \cite{bernardi_petrella.2015} and \cite{mainik_schaanning.2014} and substantially differs from the one of \cite{adrian_brunnermeier.2011}. Within the SGASC framework, the forward looking ${\rm CoES}_{\mathrm{M}\vert j,t+1}^{\tau_1,\tau_2}$ can be evaluated by numerical integration of the ${\rm CoVaR}_{\mathrm{M}\vert j,t+1}^{\tau_1,\tau_2}$, as follows:
\begin{equation}
{\rm CoES}_{\mathrm{M}\vert j,t+1}^{\tau_1,\tau_2} = \frac{1}{\tau_2}\int_{0}^{\tau_2} {\rm CoVaR}_{\mathrm{M}\vert j,t+1}^{\tau_1,\gamma}\,d\gamma.
\end{equation}
%
As discussed by \cite{bernardi_etal.2013b} and \cite{mainik_schaanning.2014}, the CoES risk measures inherits the same properties of the Expected Shortfall (ES) such as the sub--additivity with respect to linear combinations, see, e.g., \cite{artzner_etal.1999}. As a direct consequence of the sub--additivity property, the CoES can be effectively used in order to measure the total systemic risk contribution of different assets to the overall financial system. In a different context, \cite{engle_etal.2015} suggest to employ a linear combination of individual Long Run Marginal Expected Shortfall (LRMES) in order to obtain an aggregate measure of the total market systemic risk for the European region. Using same arguments, \cite{brownlees_engle.2015} also rely on the sub--additivity property of the LRMES to get an aggregate version of the Systemic Risk (SRISK) indicator. From our definition of CoES in equations \eqref{eq:CoES}--\eqref{eq:CoES_2}, it is easy to see that the LRMES can be obtained as a special case of the CoES by simply letting ${\rm CoVaR}_{\mathrm{M}\vert j,t+1}^{\tau_1,\tau_2}\to\infty$, i.e., by imposing $\tau_1=1$. To aggregate the individual systemic risk levels to get an overall indicator of total systemic risk for the whole economy, we can define the total market forward looking CoES as
\begin{equation}
{\rm CoES}_{\mathrm{M},t+1}^{\tau_1,\tau_2}=\sum_{j=1}^N w_j {\rm CoES}_{\mathrm{M}\vert j,t+1}^{\tau_1,\tau_2},
\end{equation}
where the scalars $w_j$, $j=1,2,\dots,N$, denote the weights associated to each one of the $N$ systemically relevant financial institutions belonging to the market. The definition of ${\rm CoES}_{\mathrm{M},t+1}^{\tau_1,\tau_2}$ is useful to estimate the total loss the overall market is going to face as a consequence of a crisis affecting a market participant, which is transmitted to the market when we observe a realisation of $Y_\mathrm{M}$ below the CoVaR level.\newline
\indent As discussed in \cite{adrian_brunnermeier.2011,adrian_brunnermeier.2014}, \cite{girardi_ergun.2013} and  \cite{mainik_schaanning.2014}, it is also useful to consider the difference between the CoVaR and the CoES from their \qmo median\qmc value. Here, we consider the $\Delta \mathrm{CoVaR}_{\mathrm{M}\vert j,t+1}^{\tau_1,\tau_2}$ and the $\Delta \mathrm{CoES}_{\mathrm{M}\vert j,t+1}^{\tau_1,\tau_2}$ quantities defined as
\begin{equation}
\Delta\mathrm{CoVaR}_{\mathrm{M}\vert j,t+1}^{\tau_1,\tau_2} = 100 \times\frac{\mathrm{CoVaR}_{\mathrm{M}\vert j,t+1}^{\tau_1,\tau_2} - \mathrm{CoVaR}_{\mathrm{M}\vert j,t+1}^{\tau_1,b^j}}{ \mathrm{CoVaR}_{\mathrm{M}\vert j,t+1}^{\tau_1,b^j}},
\label{eq:deltaCoVaR}
\end{equation}
and
\begin{equation}
\Delta\mathrm{CoES}_{\mathrm{M}\vert j,t+1}^{\tau_1,\tau_2} = 100 \times\frac{\mathrm{CoES}_{\mathrm{M}\vert j,t+1}^{\tau_1,\tau_2} - \mathrm{CoES}_{\mathrm{M}\vert j,t+1}^{\tau_1,b^j}}{ \mathrm{CoES}_{\mathrm{M}\vert j,t+1}^{\tau_1,b^j}},
\label{eq:deltaCoES}
\end{equation}
where $b^j$ represents the benchmark state that we define as $\mathbb{P}\left(Y_{j,t+1}\leq {\rm VaR}_{j,t+1}^{0.5}\right)=0.5$, i.e., the CoVaR of the system when the country specific indexes are below their median value.
The $\Delta\mathrm{CoVaR}$ and the $\Delta\mathrm{CoES}$ measure the percentage increase of the systemic risk conditional on a pre--specified distress event. It follows that the $\Delta\mathrm{CoVaR}$ and the $\Delta\mathrm{CoES}$ can be effectively used to measure how the CoVaR and the CoES change when a particular institution becomes financially distressed. In other words, the $\Delta\mathrm{CoVaR}$ and the $\Delta\mathrm{CoES}$ estimate the dynamic evolution of the specific institution $j$'s contribution to the overall systemic risk. Furthermore, these two quantities can be employed for policy rules and for risk management. \cite{adrian_brunnermeier.2011,adrian_brunnermeier.2014} found strong evidence for the existence of a relation between the $\Delta\mathrm{CoVaR}$ and several macroeconomic indicators. In the empirical part of the paper, we will employ the $\Delta\mathrm{CoVaR}$ and the $\Delta\mathrm{CoES}$ risk measures to investigate the systemic risk contributions of specific countries to the overall risk of the European economic system.
%
\section{Empirical study}
\label{sec:empirical_study}
%
\noindent In what follows, we apply the econometric framework and the methodology described in the previous sections to examine the evolution of the systemic risk in Europe over the past decade. As discussed in \cite{bernardi_etal.2013b}, \cite{bernardi_petrella.2015} and \cite{billio_etal.2012} among others, the high level of interdependence and interconnection among financial institutions has been recognised as the main ingredient that facilitates adverse shocks affecting individual institutions or countries to spread over the overall financial system. Systemic events are particularly relevant since they involve extreme losses for all the market participants threatening the stability of the entire economic and financial system. The evaluation of systemic risk in the European financial system has been recently considered by \cite{reboredo_ugolini.2015} in order to assess the impact of the recent ESDC of 2010--2011. \cite{billio_etal.2012} analyse stock price data on hedge funds, banks, brokers and insurers for both the US and the Euro market. More recently, \cite{engle_etal.2015} and \cite{lucas_etal.2014} analyse the evolution of the European systemic risk using a new systemic risk indicator (LRMES) and the multivariate Generalised--Hyperbolic GAS model, respectively.
%
\subsection{Data}
\label{sec:data}
%
\noindent To investigate the systemic risk contribution of several European countries to the overall European system we consider eleven equally weighed portfolios composed by the fifteen most capitalised companies domiciled in each country. By construction, the considered country specific indexes are representative of the equity market of the country they belong to. The selected countries are Austria (AU), Belgium (BEL), Denmark (DEN), France (FRA), Germany (GER), Hungary (HUN), Italy (IT), Netherland (NET), Spain (SPA), Sweden (SWE) and United Kingdom (UK). To evaluate the systemic risk we also need to identify a proxy for the total Eurozone equity market (MKT) which consists of an equally weighed portfolio across all the companies belonging to our dataset. Details about the composition of the regional indexes are reported in Table \ref{tab:IndexList_list}.\newline
\indent We analyse equity indexes' log--returns from July 8, 1999 to October 16, 2015, covering the recent Global Financial Crisis of 2007/2008 as well as the European sovereign debt crisis of 2010. For each index, the last $H=2000$ observations, covering the period from November 12, 2007 to the end of the sample, are considered to perform the out--of--sample forecasting exercise, while the first part is used to estimate models' parameters and to assess models' performances with respect to nested alternatives. Descriptive statistics for all the considered series are reported in Table \ref{tab:Index_data_summary_stat}. In line with most important stylised facts, frequently detected in financial time series, the returns appear to be negatively skewed and leptokurtic, indicating that their empirical distributions strongly deviate from the Gaussian one. The departure from normality is confirmed by the Jarque--Bera (JB) statistic which always rejects the null hypothesis at the 1\% level of significance. The presence of large volatility clusters followed by periods of low volatility is also well documented by the data. These facts are coherent with the presence of different regimes of \qmo bull\qmcsp and \qmo bear\qmcsp market conditions, which are usually associated with \qmo high\qmcsp and \qmo low\qmcsp dependence, as discussed in \cite{pelletier.2006} among others.
%
\subsection{Marginal specifications}
\label{sec:marginal_spec}
\begin{sidewaysfigure}[!h]
\centering
\subfloat[Austria]{\label{fig:Austria_Uniform}\includegraphics[width=0.2\textwidth]{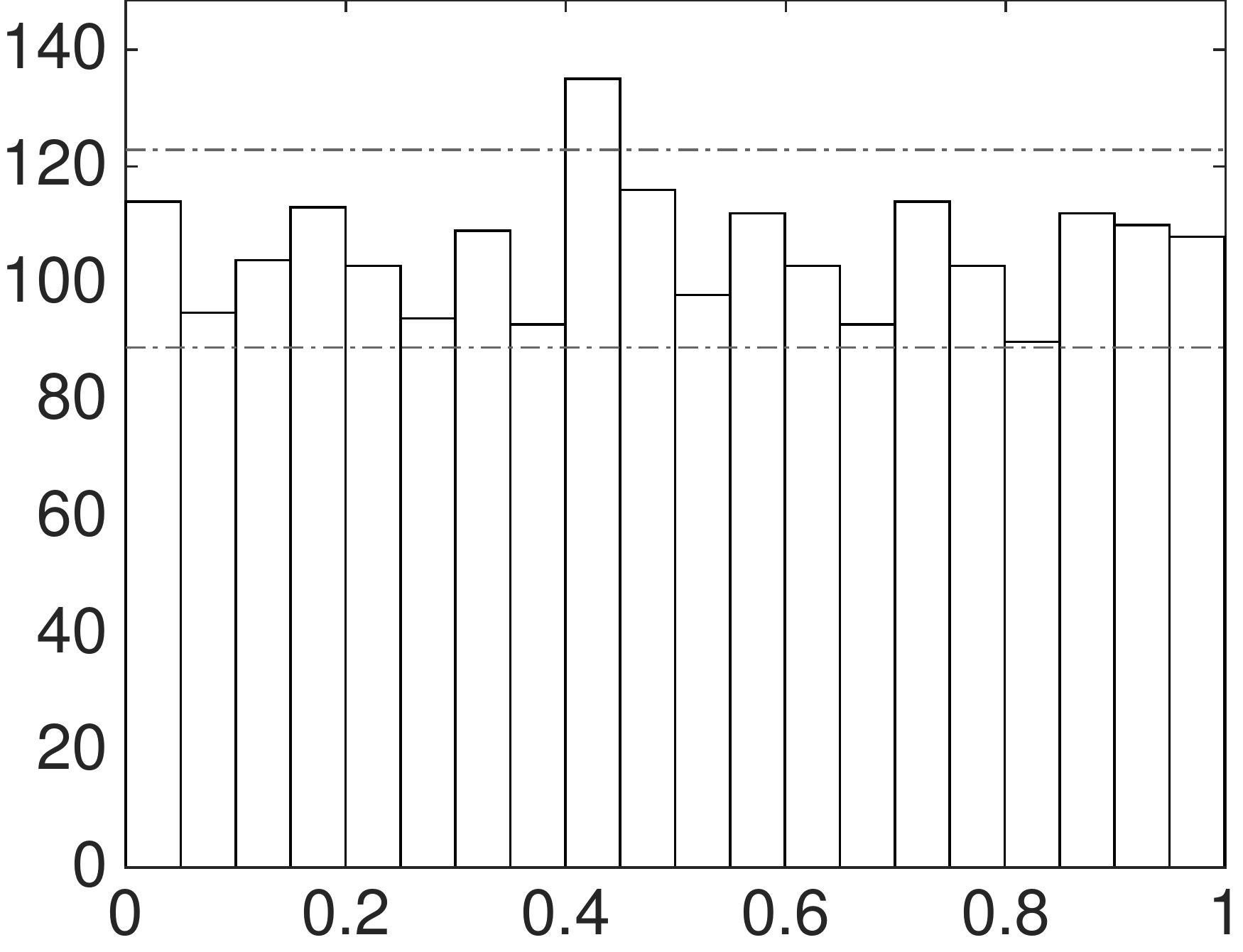}}
\qquad
\subfloat[Belgium]{\label{fig:Belgium_Uniform}\includegraphics[width=0.2\textwidth]{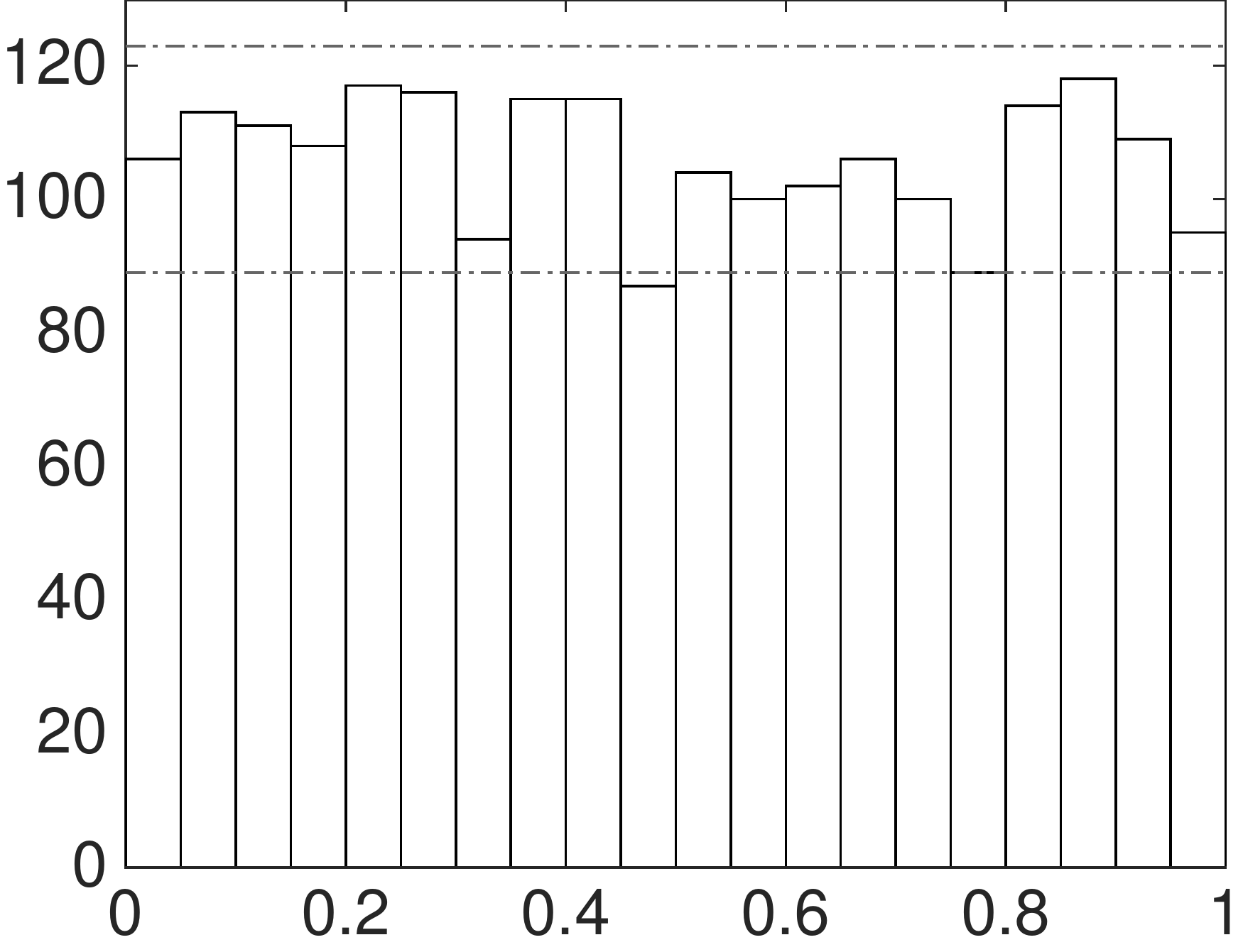}}
\qquad
\subfloat[Denmark]{\label{fig:Denmark_Uniform}\includegraphics[width=0.2\textwidth]{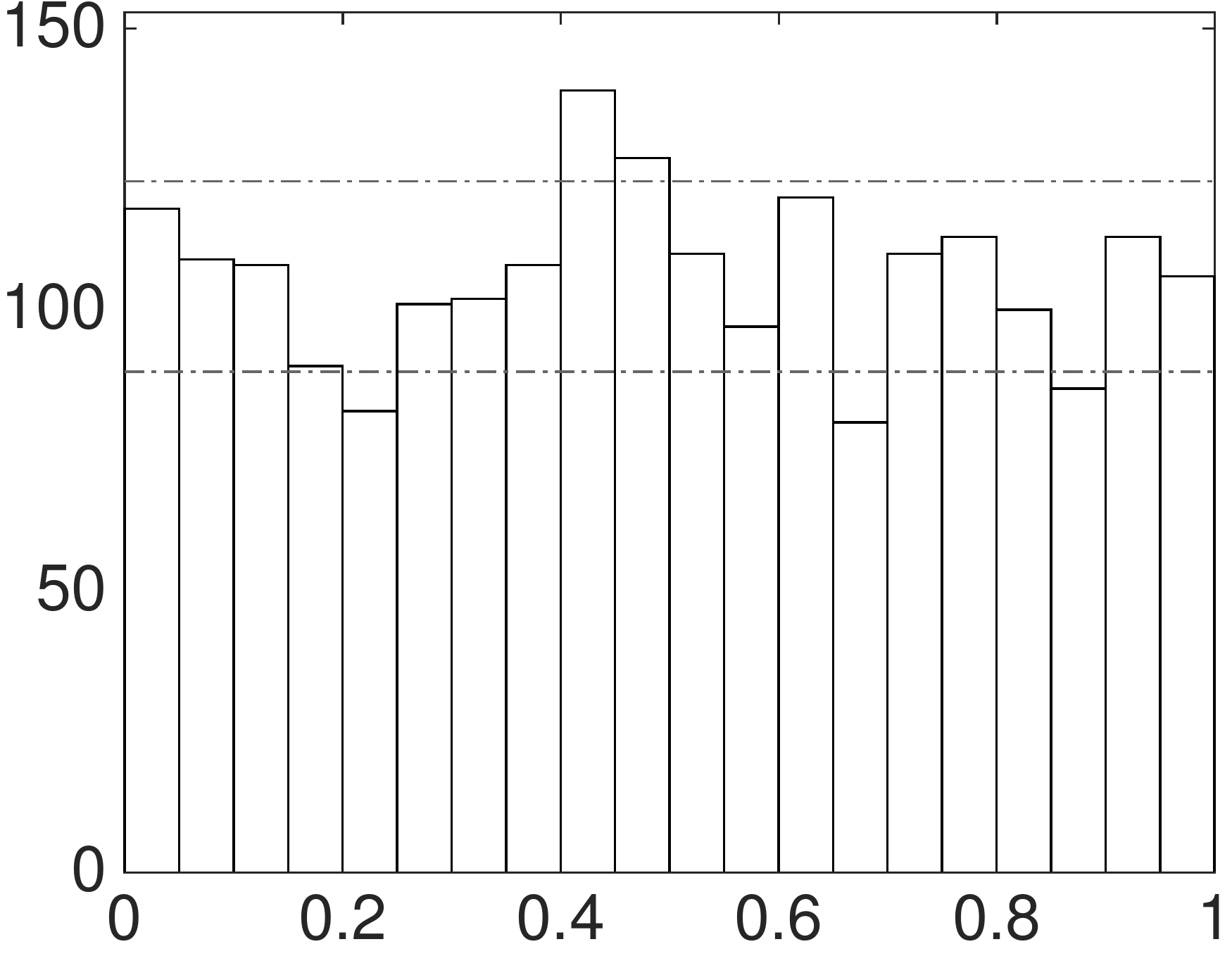}}
\qquad
\subfloat[France]{\label{fig:France_Uniform}\includegraphics[width=0.2\textwidth]{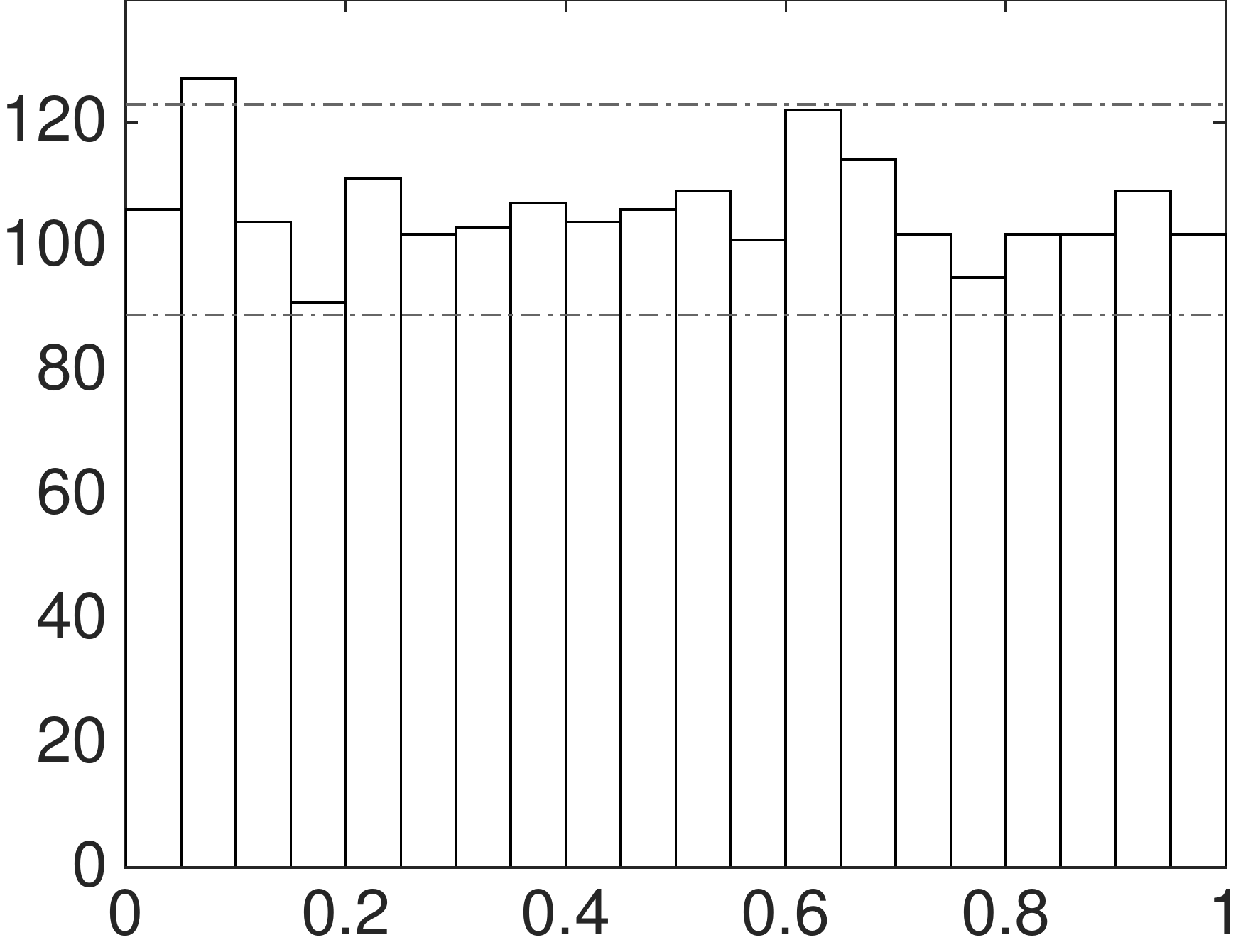}}\\
\subfloat[Germany]{\label{fig:Germany_Uniform}\includegraphics[width=0.2\textwidth]{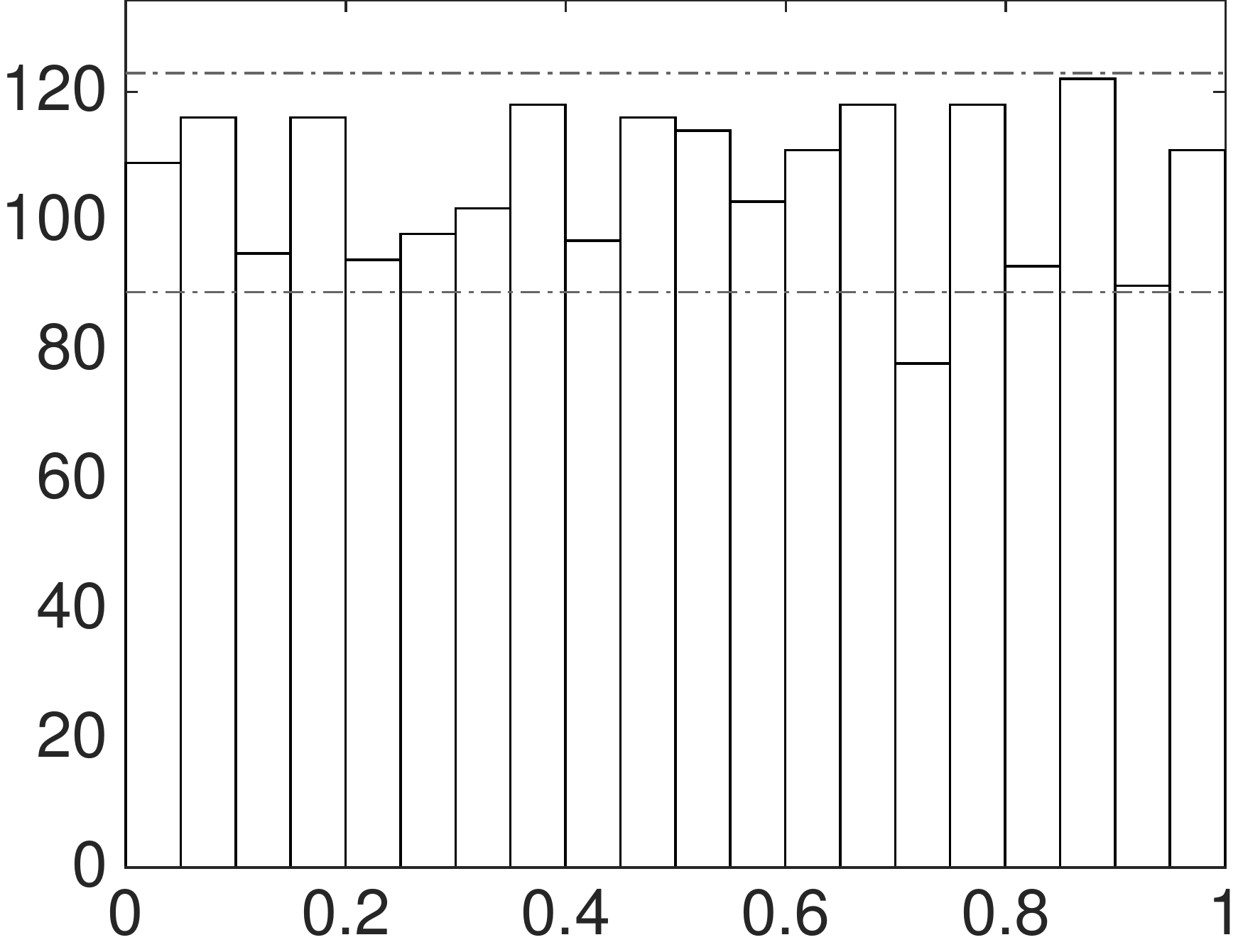}}
\qquad
\subfloat[Hungary]{\label{fig:Hungary_Uniform}\includegraphics[width=0.2\textwidth]{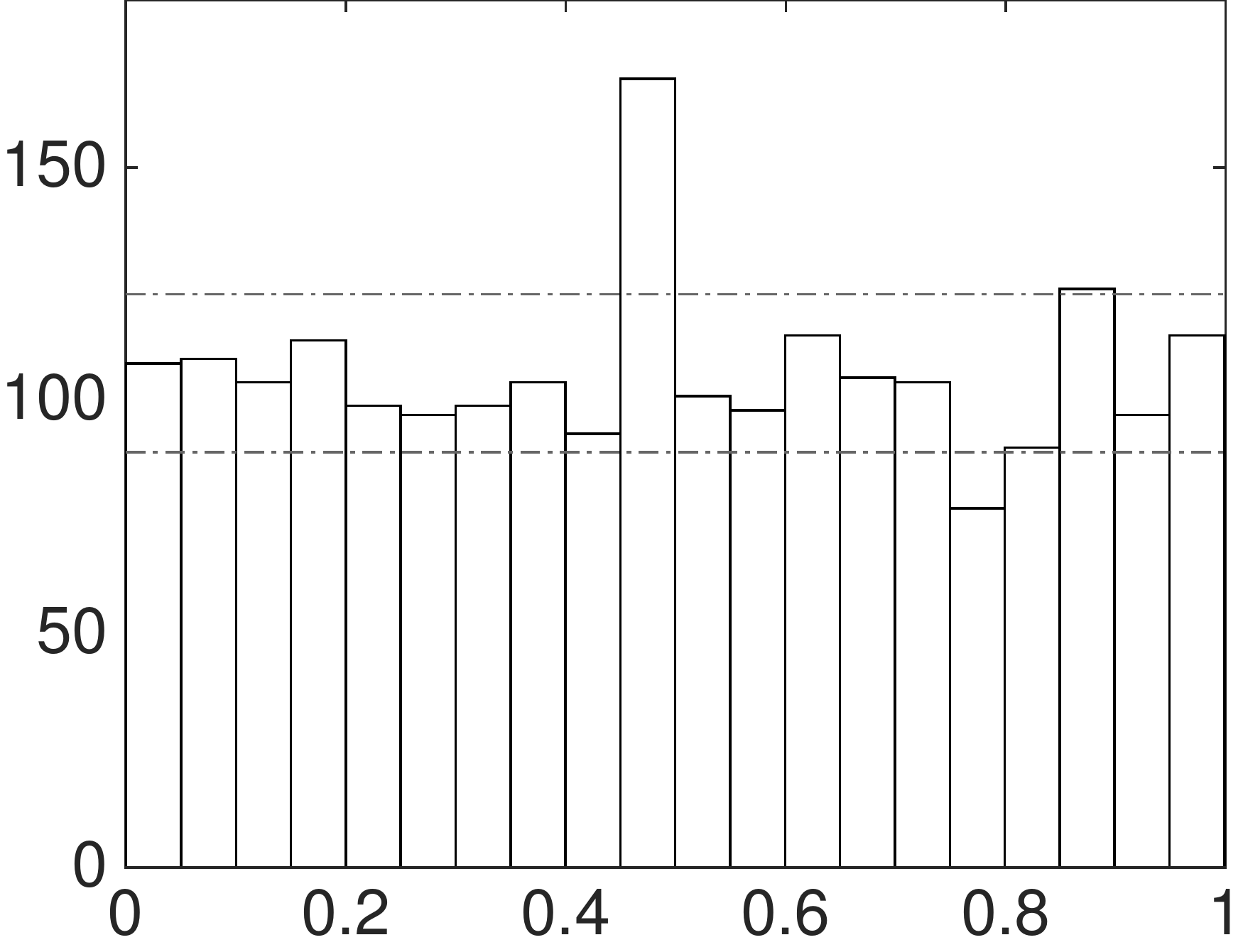}}
\qquad
\subfloat[Italy]{\label{fig:Italy_Uniform}\includegraphics[width=0.2\textwidth]{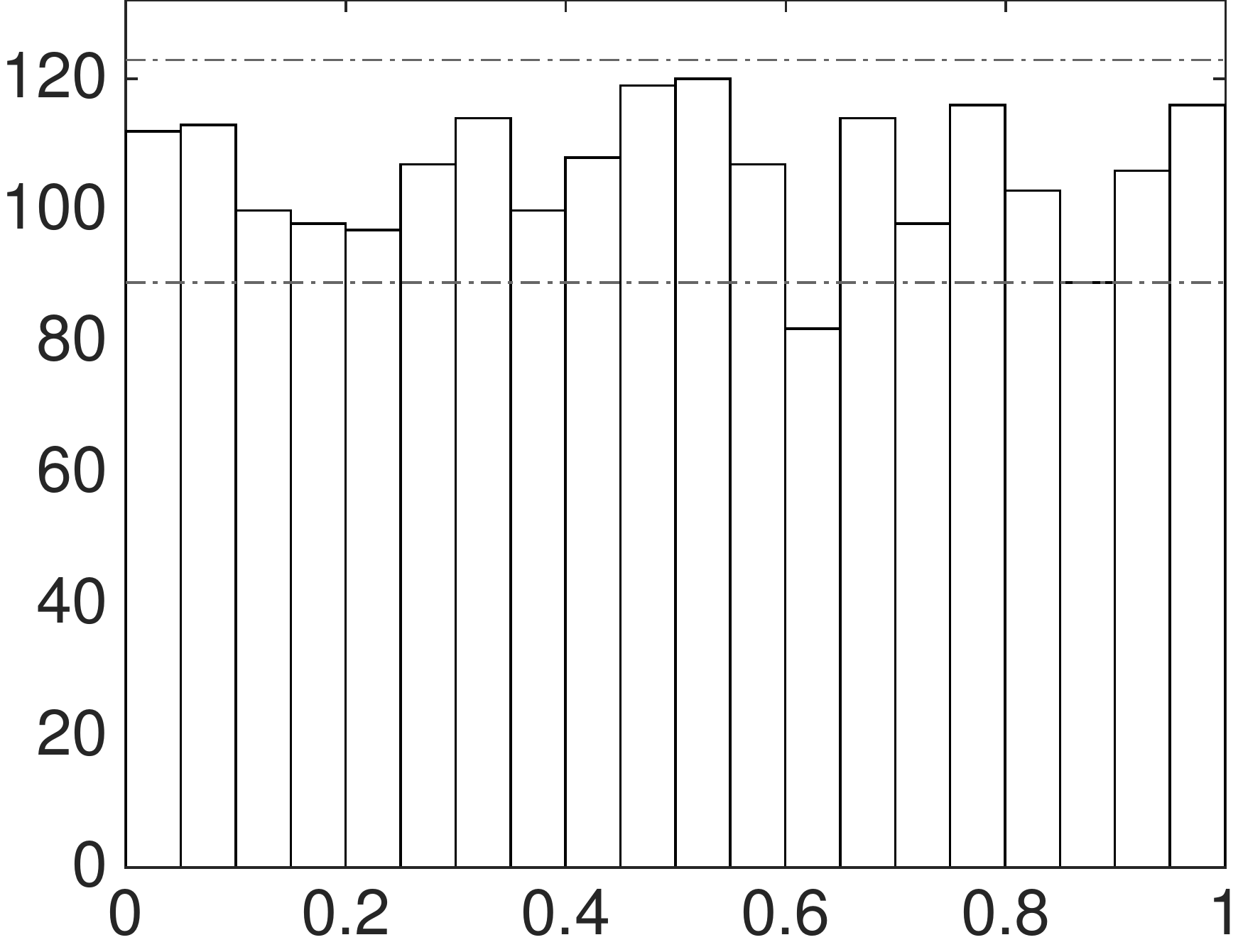}}
\qquad
\subfloat[Netherlands]{\label{fig:Netherlands_Uniform}\includegraphics[width=0.2\textwidth]{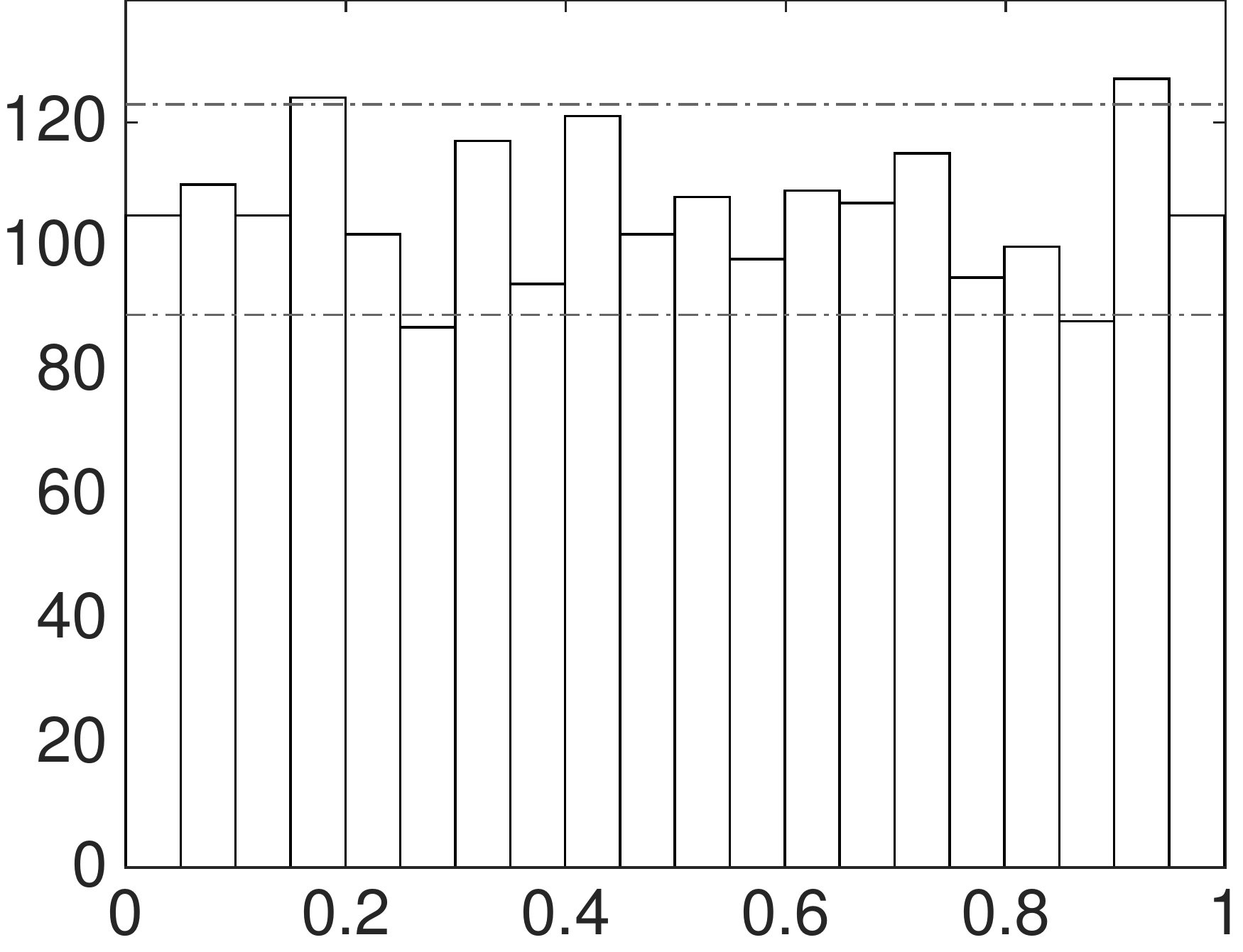}}\\
\subfloat[Spain]{\label{fig:Spain_Uniform}\includegraphics[width=0.2\textwidth]{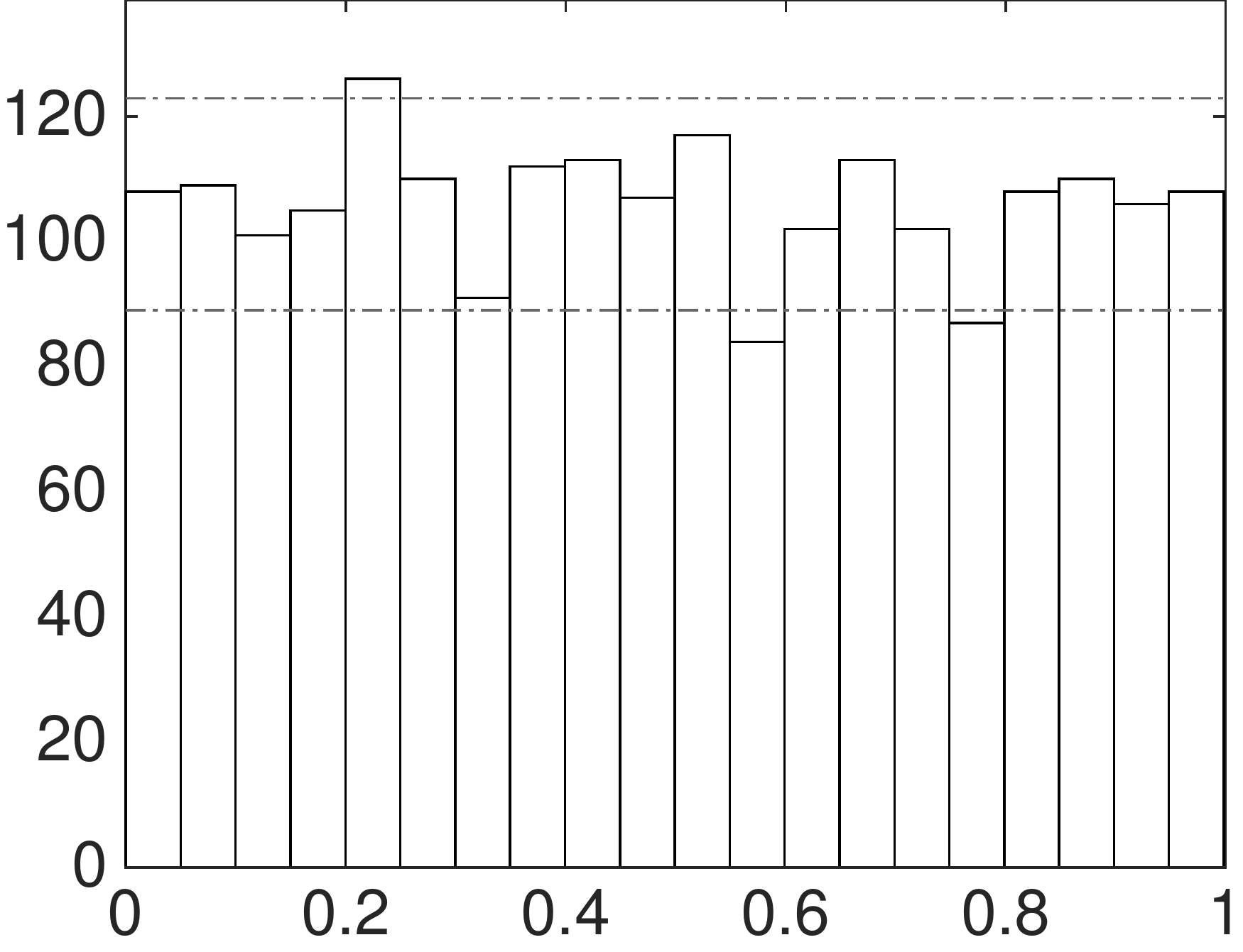}}
\qquad
\subfloat[Sweden]{\label{fig:Sweden_Uniform}\includegraphics[width=0.2\textwidth]{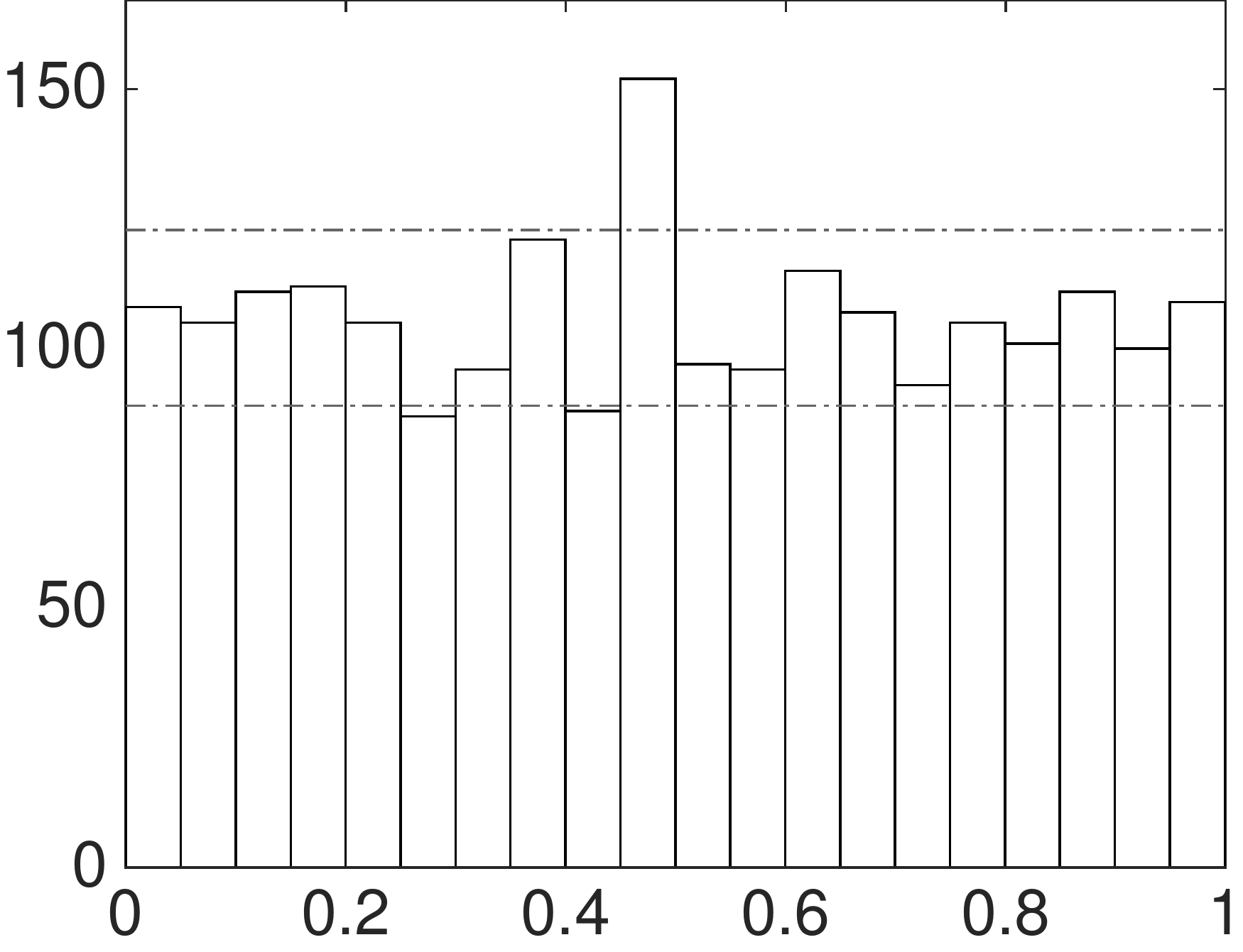}}
\qquad
\subfloat[United Kingdom]{\label{fig:United Kingdom_Uniform}\includegraphics[width=0.2\textwidth]{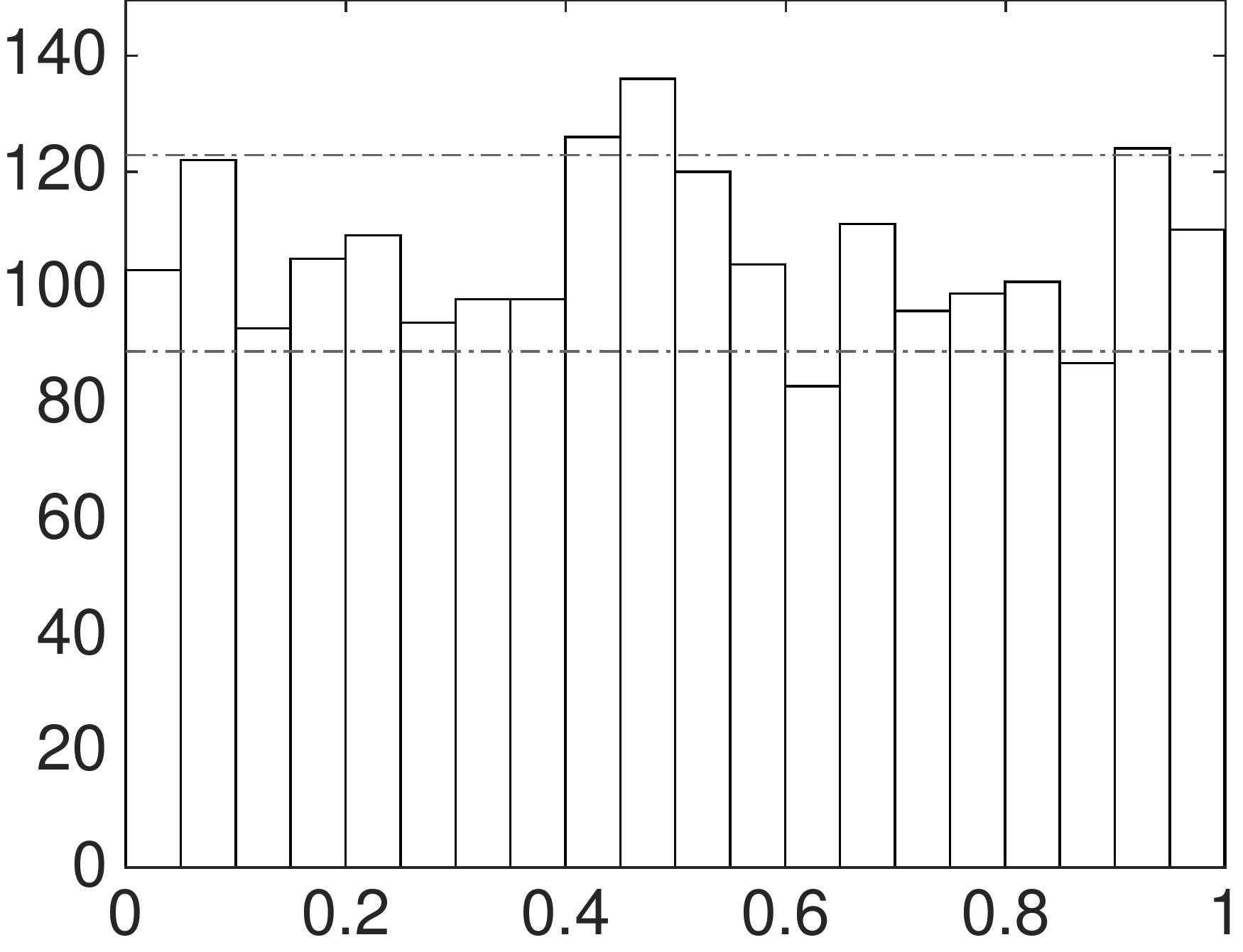}}
\qquad
\subfloat[Eurozone]{\label{fig:Eurozone_Uniform}\includegraphics[width=0.2\textwidth]{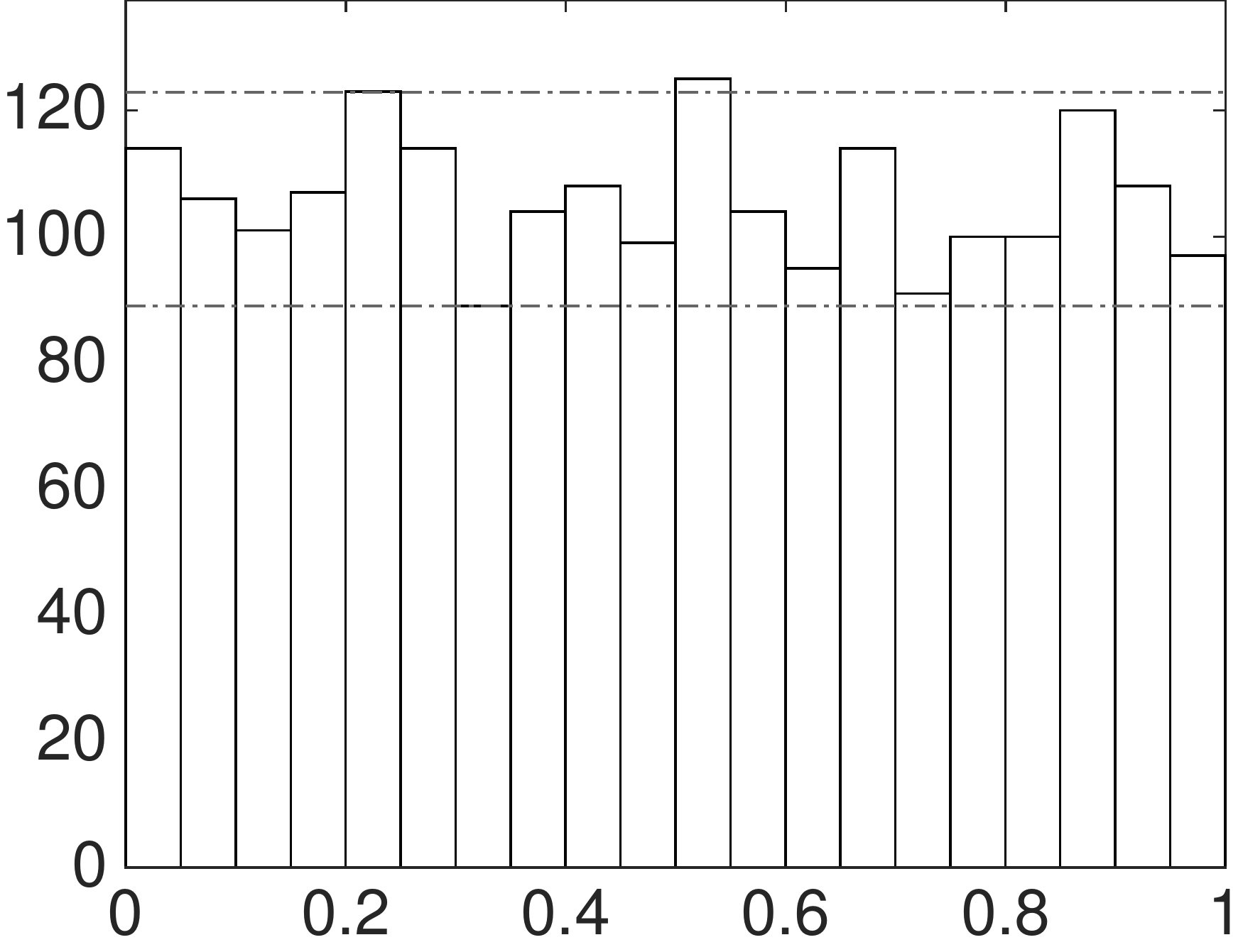}}
\caption{\footnotesize{Marginal empirical PIT distributions. The histogram is divided into 20 bins, red lines represent $5\%$ approximated confidence intervals. For more informations see \cite{jondeau_rockinger.2006} and \cite{tay_etal.1998}.}}
\label{fig:PIT_istogram}
\end{sidewaysfigure}

%
\noindent In order to filter out all the stylised facts affecting the univariate marginal conditional distributions we estimate the AR(1)--GJR--GARCH(1,1)--$s\mathcal{ST}$ detailed in Section \ref{sec:model_marginal}. Estimated coefficients of each marginal model are reported in Table \ref{tab:Marginal_estimates}. Our results seems to be in line with those usually found in the financial econometric literature, such as the strong persistence and the positive reaction of the conditional volatility to past negative innovations. The skewness parameters $\hat{\eta}_i$ of the Skew Student--t distribution, are always significantly smaller then one, justifying our choice of skew innovations. The estimated degree of freedom parameters $\hat{\upsilon}_i$ strongly confirms the excess of kurtosis and the departure from normality of each of the considered series. To check the goodness of fit of the estimated marginal distributions, we test if the PITs implied by the estimated conditional densities are independently and identically distributed uniformly in the unit interval $\left(0,1\right)$. To this end, we implement the same testing procedure employed by \cite{vlaar_palm.1993}, \cite{jondeau_rockinger.2006} and \cite{tay_etal.1998}. Specifically, the iid uniform test is made of two main parts. The first part checks the independence assumption by testing if all the conditional moments of the data up to the fourth one have been captured by the model, while the second part aims to verify whether the Skew Student--t assumption is reliable by applying a Uniform $\left(0,1\right)$ test on the PITs. The first test consists to examine the serial correlation of the quantities $\left(\hat{u}_{i,t} - \bar{\hat{u}}_i\right)^k$ for $k=1,2,\dots,4$, where $\bar{\hat{u}}_i = T^{-1}\sum_{t=1}^T\hat{u}_{i,t}$, by performing a simple linear regression of $\left(\hat{u}_{i,t}-\bar{\hat{u}}_i\right)^k$ on their 20 lags. The null hypothesis of absence of serial correlation is tested using a Lagrange multiplier--type test defined by the statistics $\left(T-20\right){\rm R}_k^2$ where ${\rm R}_k^2$, with $k=1,2,\dots,4$ is the coefficient of determination of the regressions. The corresponding test statistics DGT--AR$\left(k\right)$ for $k=1,2,3,4$, are reported in the first three rows of Table \ref{tab:Uniform_test}. We note that, for almost all the considered series, the test is in favour of the null of absence of serial dependence for the first four conditional moments of the estimated PITs. Concerning the uniform distribution test, following \cite{jondeau_rockinger.2006}, the statistic tests have been evaluated by splitting the empirical distributions in $G=20$ bins. For more information about the test implementation, see, e.g., \cite{tay_etal.1998} and \cite{jondeau_rockinger.2006}. The estimated statistics for the uniform test, DGT--H$\left(20\right)$, are reported in the last row of Table \ref{tab:Uniform_test}. Except for DEN, HUN, SWE and UK, for the remaining country indexes the test indicates that the estimated PITs are uniformly distributed over the interval $(0,1)$ at a confidence level lower then $1\%$. Concerning HUN and SWE, the rejection of the uniform assumption for the PITs is principally related to the large number of zeros those series exhibits, especially in the very first part of our sample. This empirical evidence is also displayed in Figure \ref{fig:PIT_istogram} where the empirical distribution of the PIT series along with the $5\%$ approximated confidence levels are reported. However, since this finding is quite common and only affects the centre of the marginal distribution, we decide to not pre--filter the series and to continue with our empirical investigation.
%
\subsection{Dynamic copula specifications}
\label{sec:dyn_copula_spec}
%
\noindent The SGASC model nests several alternative copula specifications: it reduces to the GAS copula (GAScop) model of \cite{creal_etal.2013} and \cite{harvey.2013} when $L=1$, while for $\alpha^l= \beta^l=0$, $\forall l=1,2,\dots,L$, it reduces to the Markov Switching copula model of \cite{jondeau_rockinger.2006} and \cite{chollete_etal.2009} (MScop), and to the static copula (STATcop), for $\alpha^l=\beta^l=0$ and $L=1$. Concerning the choice of the copula distribution, in an unreported analysis, we find that the Student--t copula seems to be the best choice in order to describe the dependence patterns across financial indexes. This evidence is quite common in the financial econometric literature, see, e.g., \cite{jondeau_rockinger.2006}, \cite{rodriguez.2007} and \cite{demarta_mcneil.2005}.
To select the best model, we compare the GAScop, the STATcop as well as the MScop and the SGASC models with different number of regimes, namely $L=2,3,4$. For each pairs of country and market indexes, Table \ref{tab:copulaAIC} reports the Akaike Information Criteria (AIC) for all the Student--t copula specifications we consider. The information criteria select the SGASC--L2 specification for 6 countries, while the simpler GAScop model is preferred in the remaining five cases. This evidence suggests that countries like France, Italy, Netherlands do not display abrupt changes in their dependence structure with the overall European equity index while Markovian patterns are detected for countries such as, for example, Belgium, Denmark and Germany. Table \ref{tab:copulaAIC} reveals also that the standard static copula specification STATcop is clearly suboptimal when compared with more flexible models.\newline
\indent Before turning to the out--of--sample forecast exercise, two additional points should be addressed. The first concerns the goodness of fit analysis of the copula specification, while the latter refers to the adequacy of the Markov switching dynamics. As regards the goodness--of--fit analysis, we aim to tests the null hypothesis of correct copula specification
\begin{equation}
{\rm H}_0:\widehat{\mathcal{C}}\left(U_{1,t},U_{2,t}\right)=\mathcal{C}_0\left(U_{1,t},U_{2,t}\right),\qquad\forall t=1,2,\dots,T,
\label{eq:copulatest}
\end{equation}
with respect to the true unknown copula distribution $\mathcal{C}_0\left(U_{1t},U_{2t}\right)$. Table \ref{tab:ADTest} reports the p--values of the Anderson--Darling test statistics performed using $1000$ parametric bootstrap resamples as indicated in \cite{manner_reznikova.2012} and \cite{patton.2012}, to which we refer for further details. Not surprisingly, except for Belgium, the goodness--of--fit test results are in line with those obtained by the model selection criterium AIC. This evidence supports our choice of performing the systemic risk analysis using those models selected by the AIC. The latter analysis investigates which SGASC parameter drives the Markov--Switching behaviour of the dependence structure. Specifically, we test the hypothesis of regime independence for the parameters $\{\omega,\alpha,\beta,\nu\}$ using a standard likelihood ratio test. LR p--values are reported in Table \ref{tab:LR}. Once again the LR results strengthen the evidence provided by both the AIC and the goodness--of--fit test. At both extremes, there is no evidence of state dependence for Hungary, while United Kingdom displays well identified regime dynamics. In between those cases, we find indexes with low or moderate Markovian structure. For example, Belgium and Germany display significant switches in both the unconditional mean and persistence of the dependence parameter, while Denmark is characterised by different tail regimes since only the degrees--of--freedom parameter changes according to the regime. Finally, Table \ref{tab:copulaCoef} reports the in--sample estimated coefficients for the selected models. Concerning the SGASC specifications, we observe that almost all coefficients are strongly significant with high persistence in each state. The positive and significantly different from zero impact of the scaled scores to the copula parameters, suggests that the GAS dynamic effectively moves the dependence parameters to the proper direction. Those findings hold also for the simpler GAScop specification.
%
\subsection{Systemic risk contributions}
\label{sec:systemic_risk_contrib}
\begin{sidewaysfigure}[t]
\centering
\subfloat[Austria]{\label{fig:Austria_CoVaR}\includegraphics[width=0.22\textwidth]{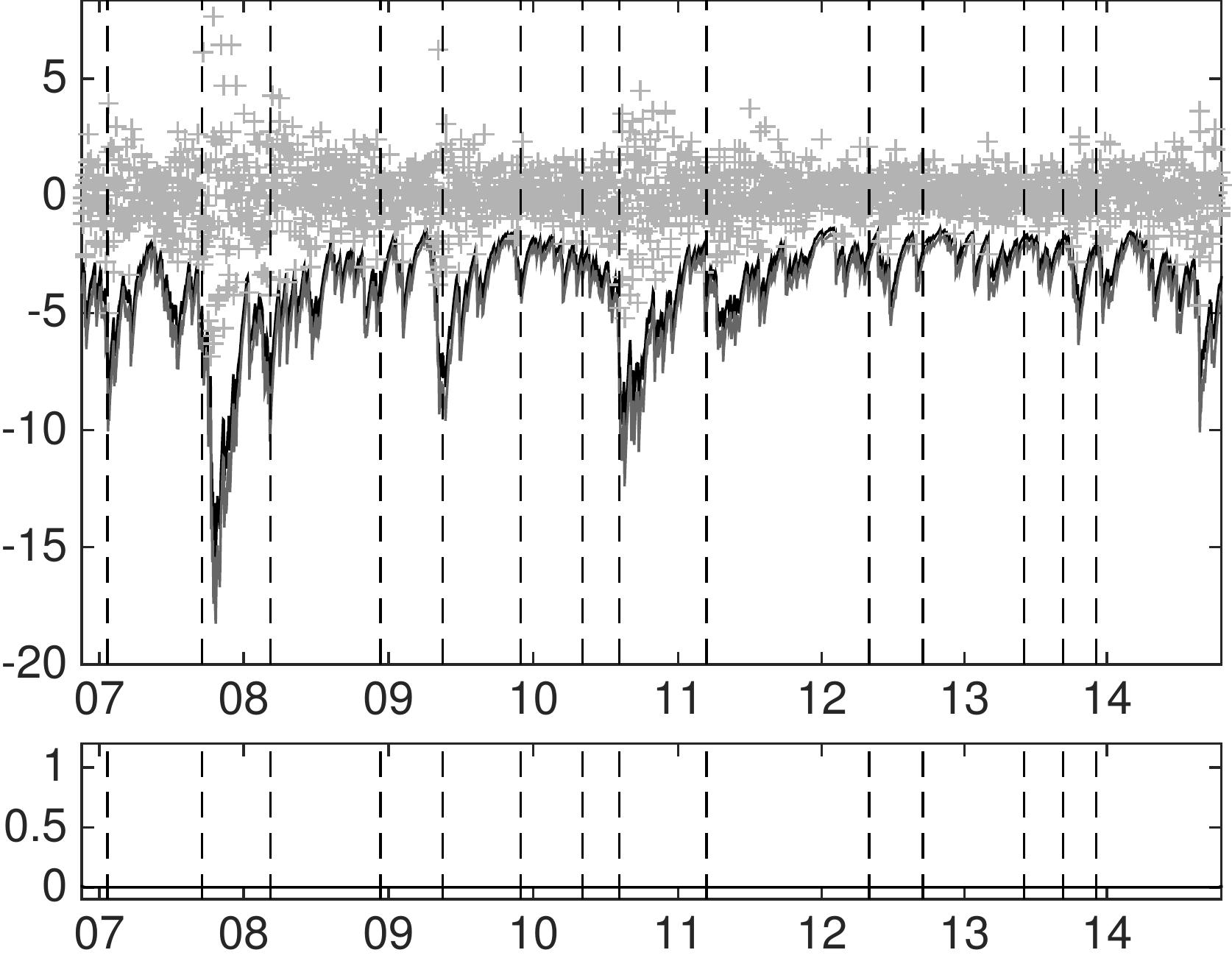}}\quad
\subfloat[Belgium]{\label{fig:Belgium_CoVaR}\includegraphics[width=0.22\textwidth]{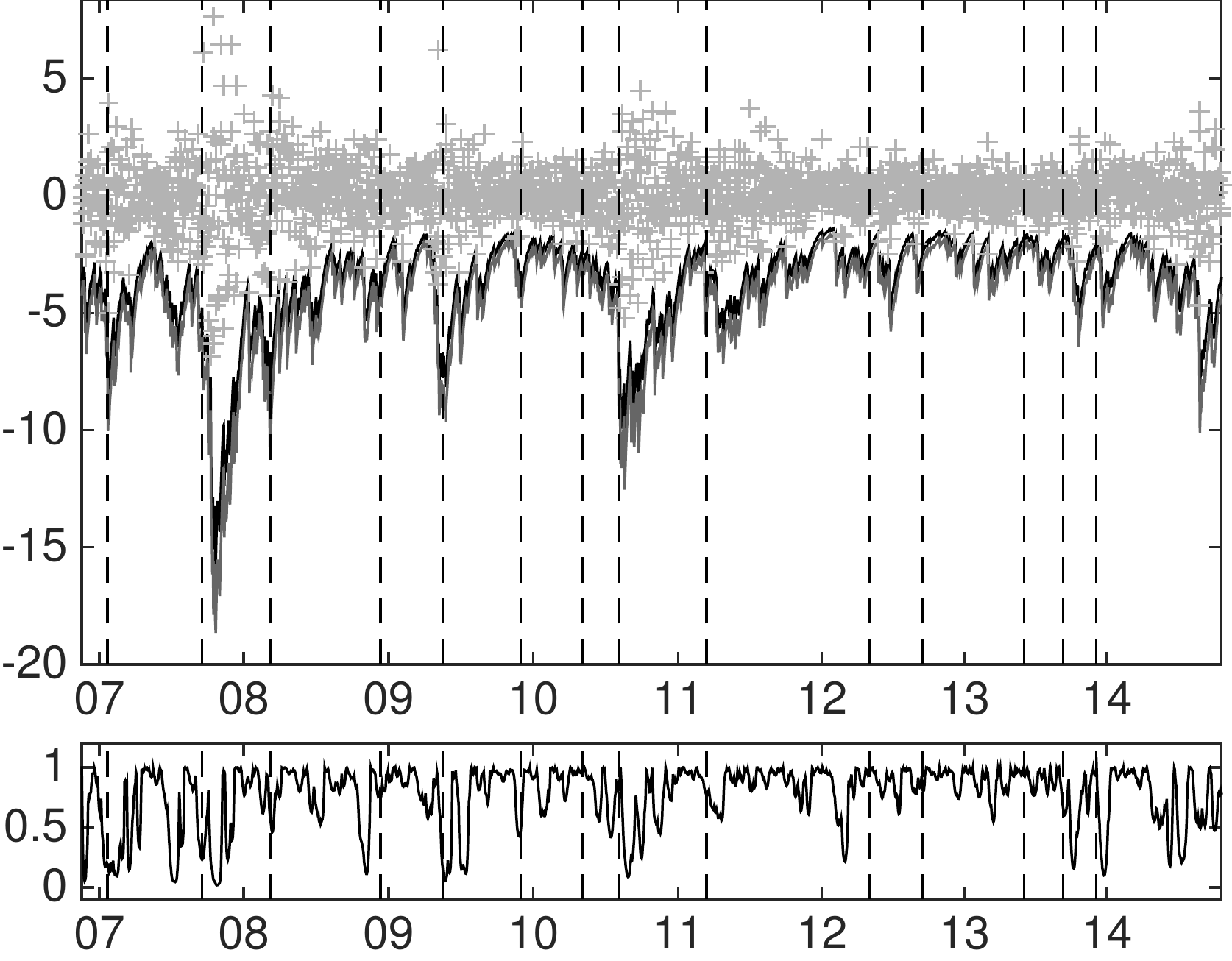}}\quad
\subfloat[Denmark]{\label{fig:Denmark_CoVaR}\includegraphics[width=0.22\textwidth]{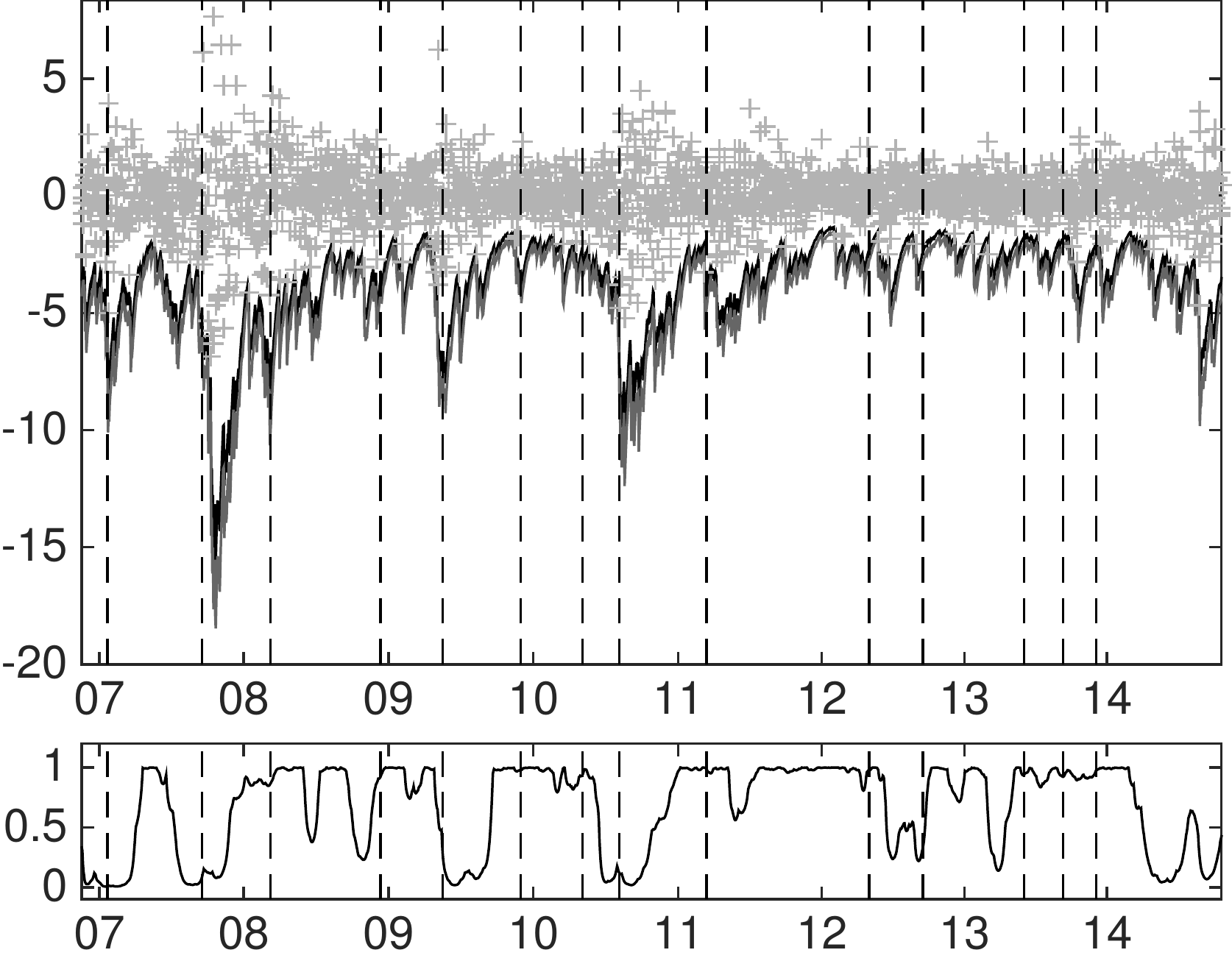}}\quad
\subfloat[France]{\label{fig:France_CoVaR}\includegraphics[width=0.22\textwidth]{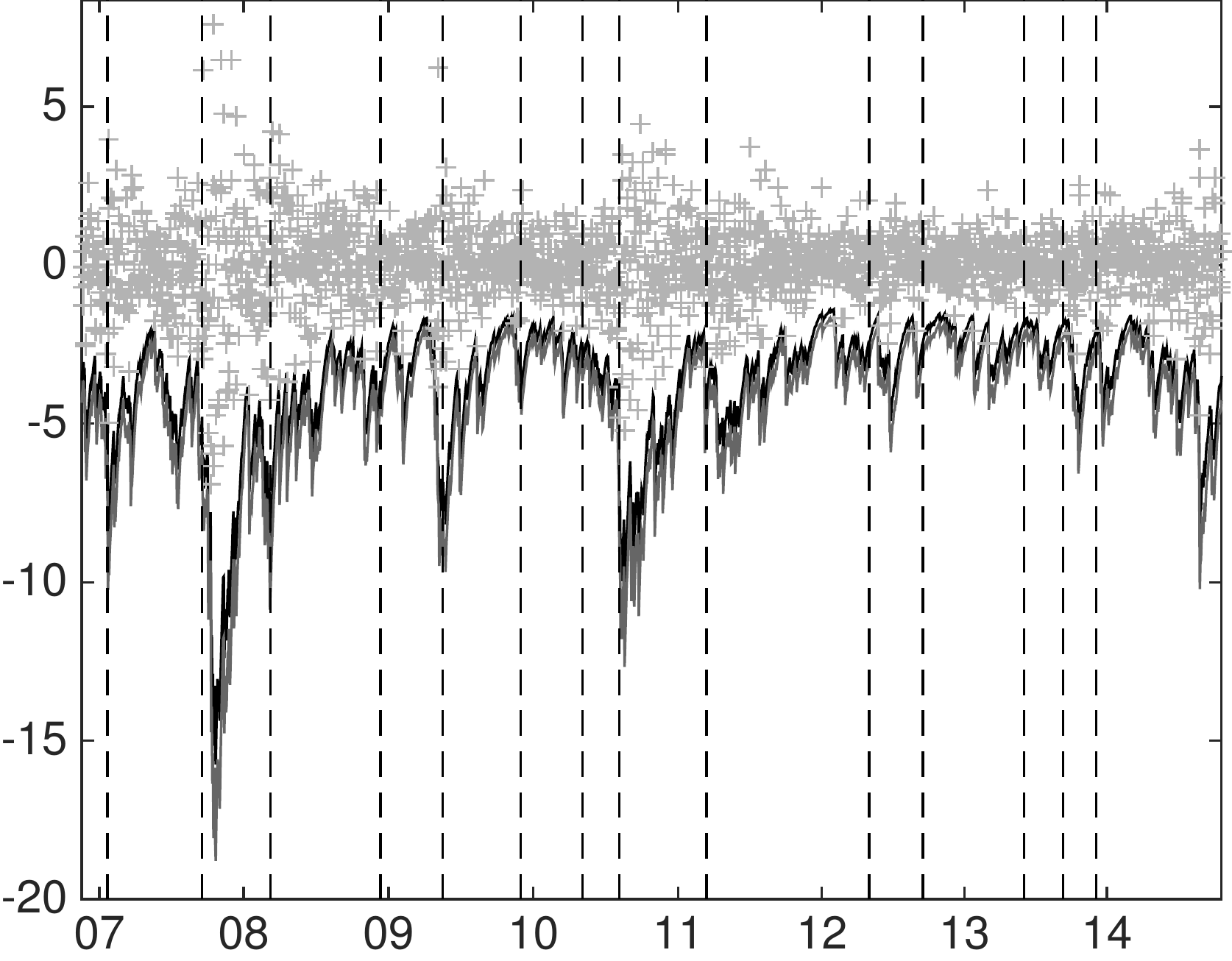}}\\
\subfloat[Germany]{\label{fig:Germany_CoVaR}\includegraphics[width=0.22\textwidth]{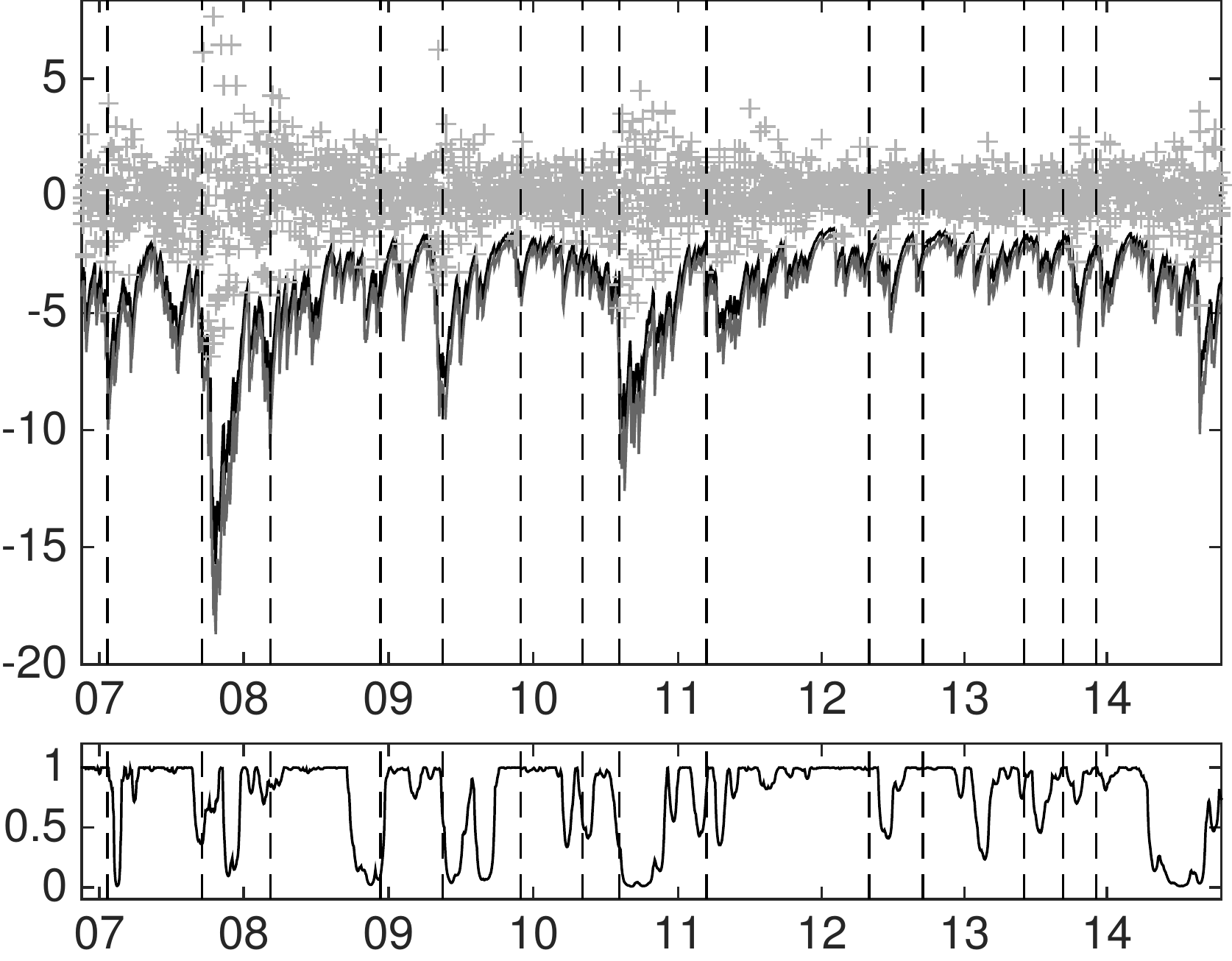}}\quad
\subfloat[Hungary]{\label{fig:Hungary_CoVaR}\includegraphics[width=0.22\textwidth]{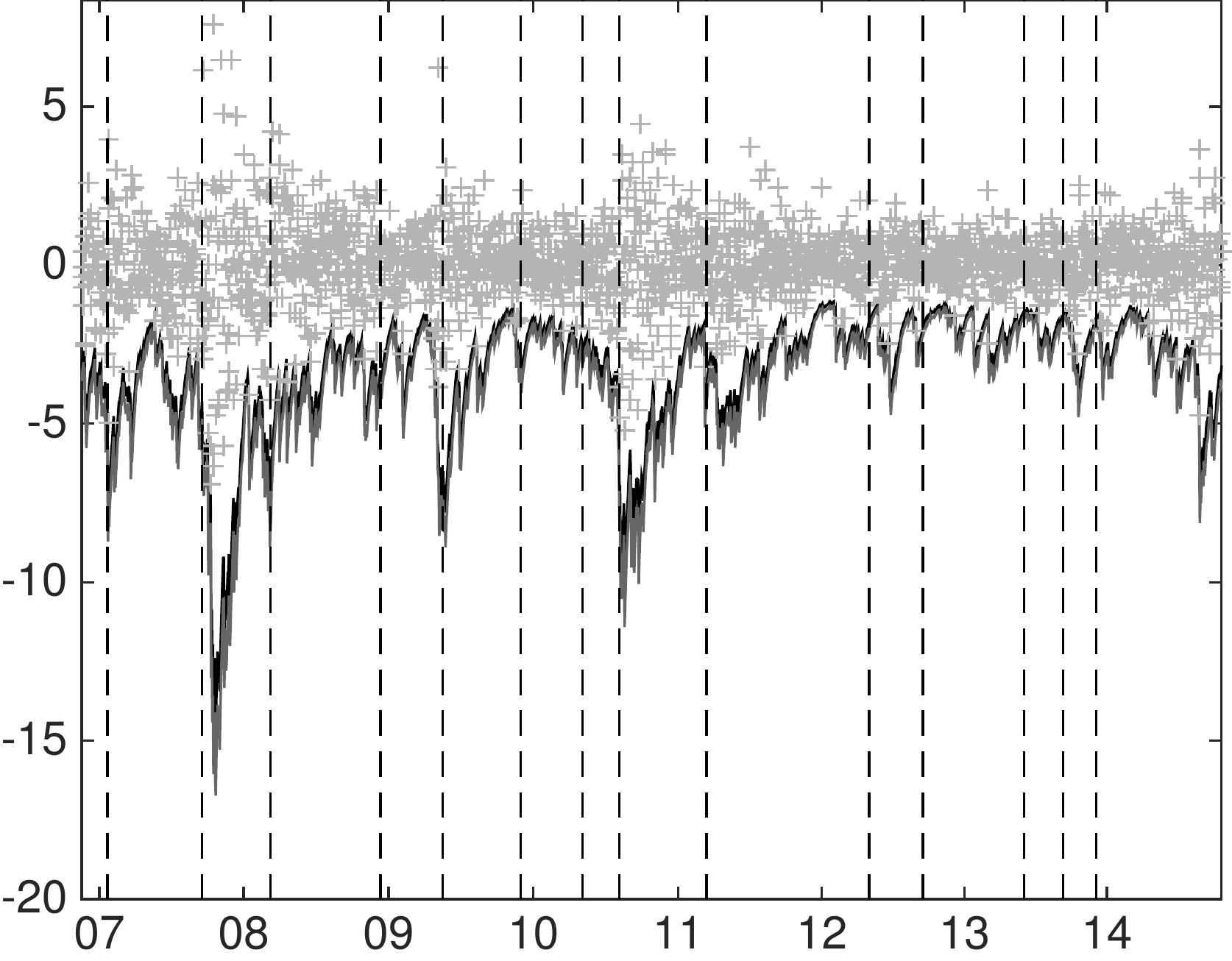}}\quad
\subfloat[Italy]{\label{fig:Italy_CoVaR}\includegraphics[width=0.22\textwidth]{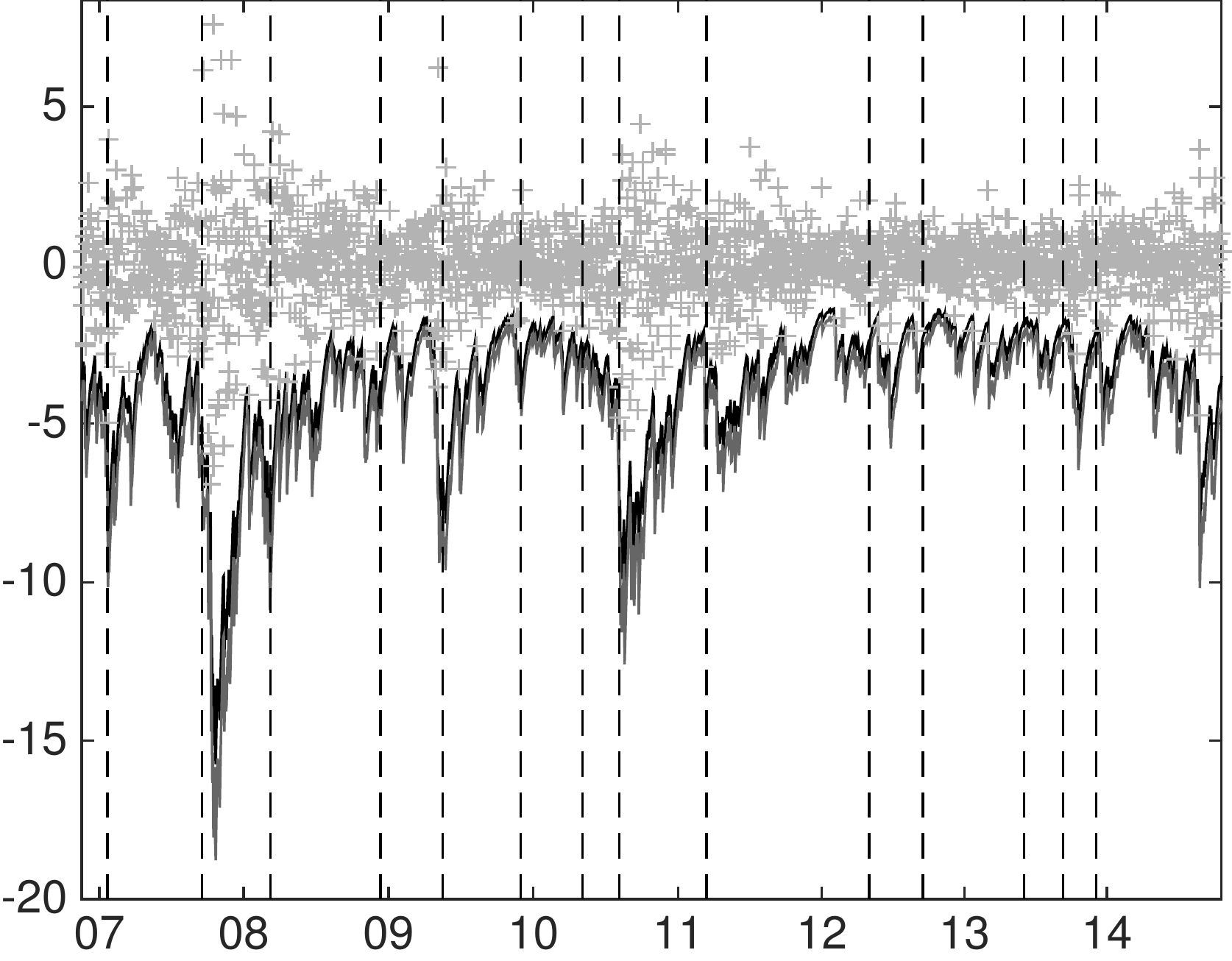}}\quad
\subfloat[Netherlands]{\label{fig:Netherlands_CoVaR}\includegraphics[width=0.22\textwidth]{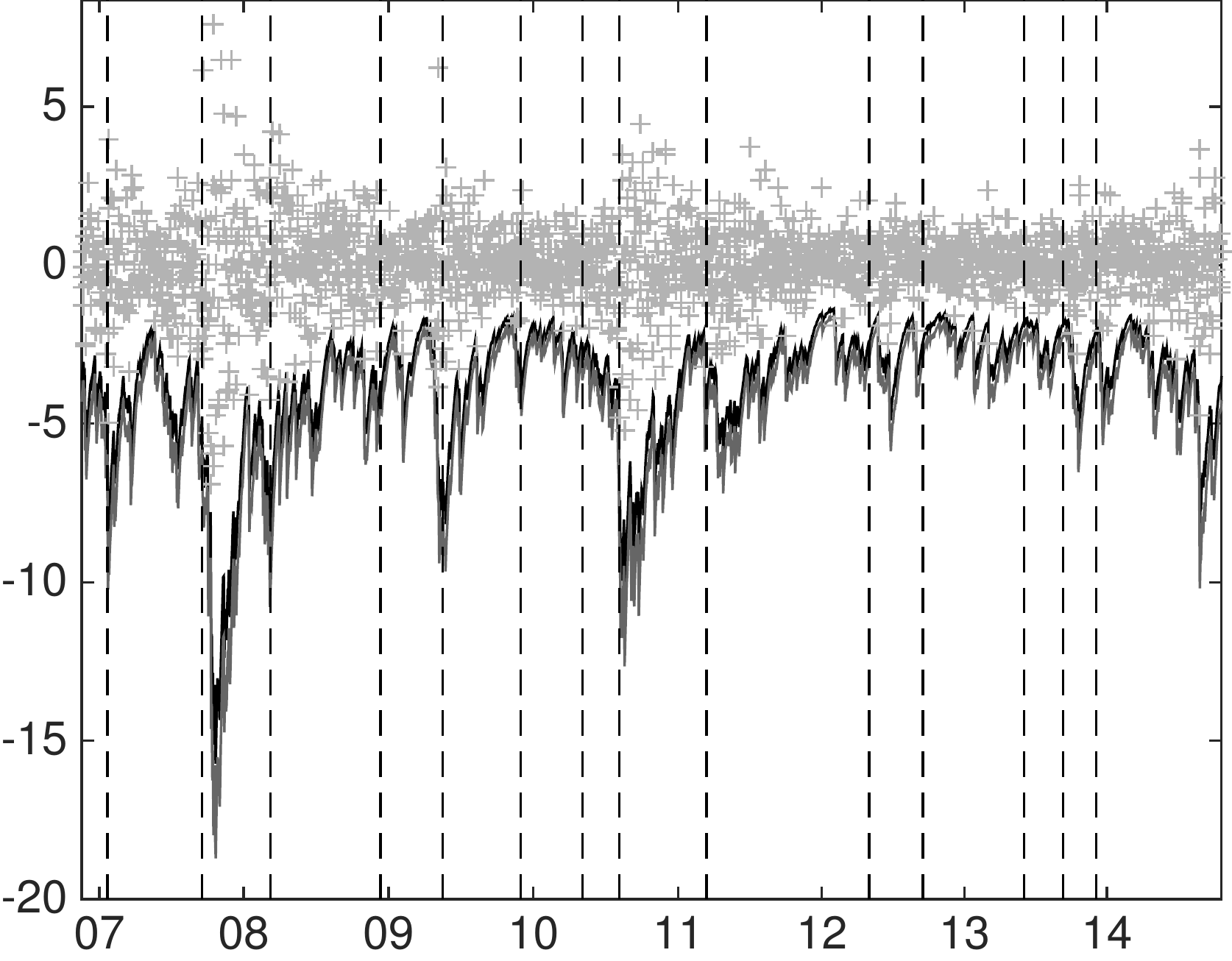}}\\
\subfloat[Spain]{\label{fig:Spain_CoVaR}\includegraphics[width=0.22\textwidth]{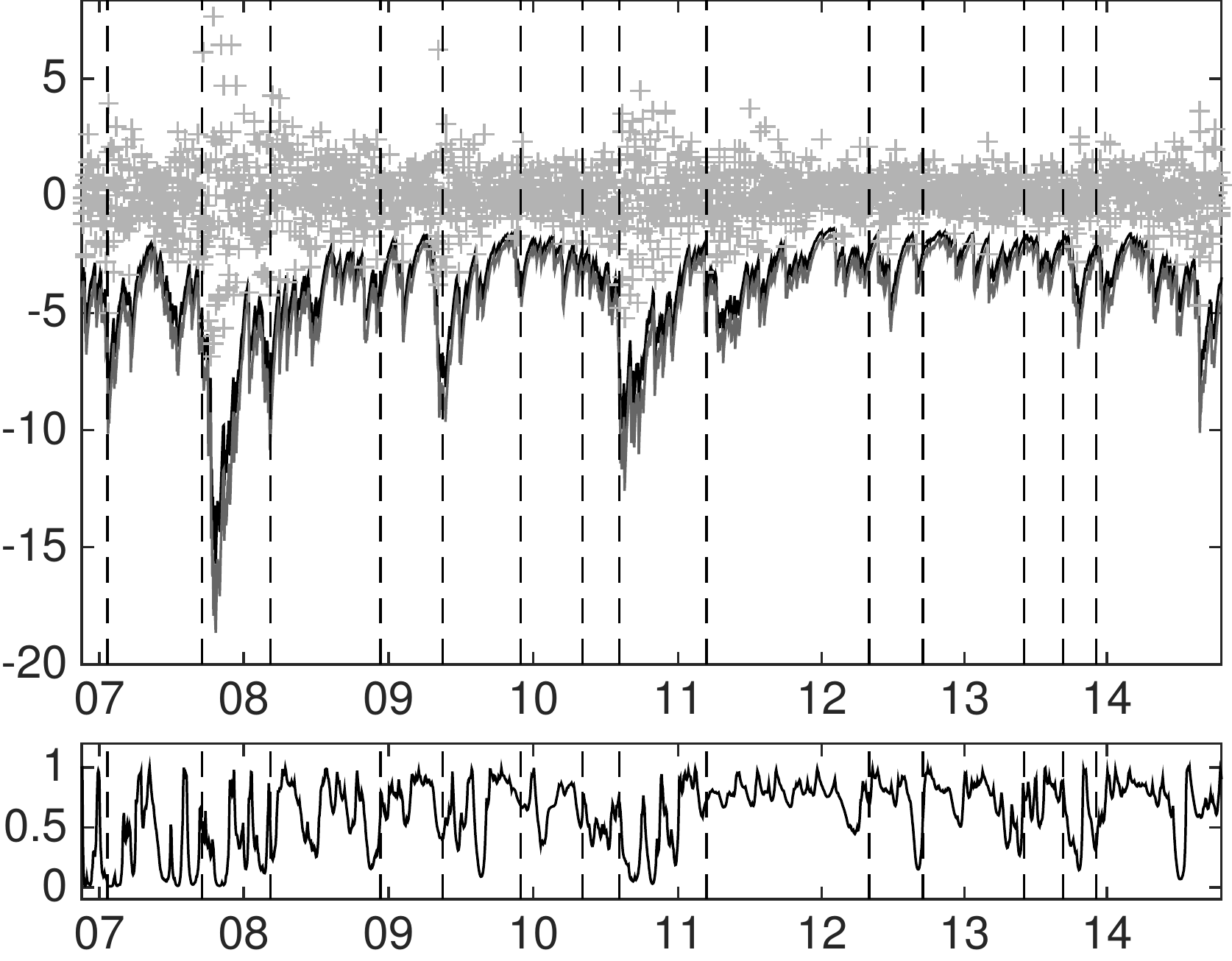}}\quad
\subfloat[Sweden]{\label{fig:Sweden_CoVaR}\includegraphics[width=0.22\textwidth]{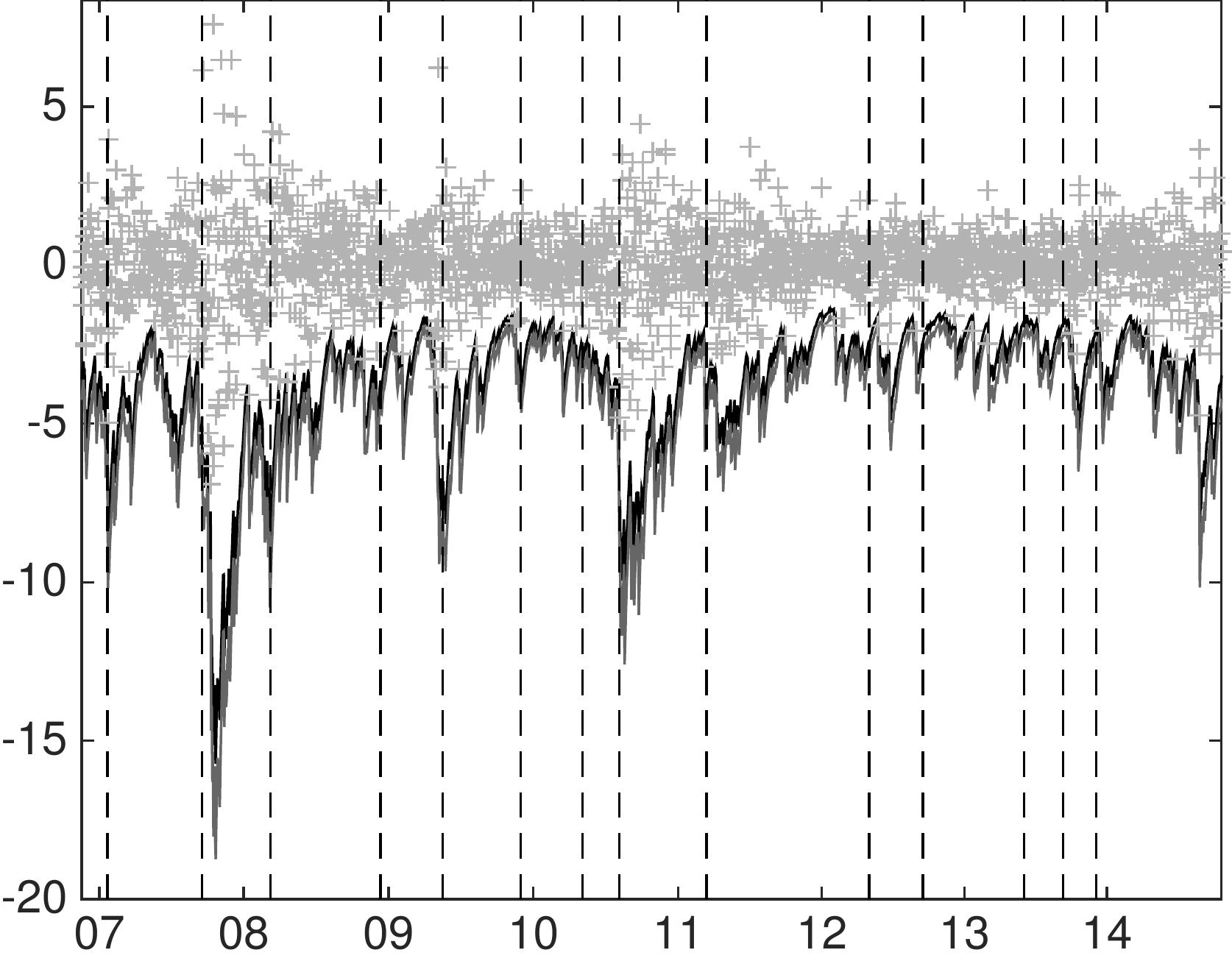}}\quad
\subfloat[United Kingdom]{\label{fig:United Kingdom_CoVaR}\includegraphics[width=0.22\textwidth]{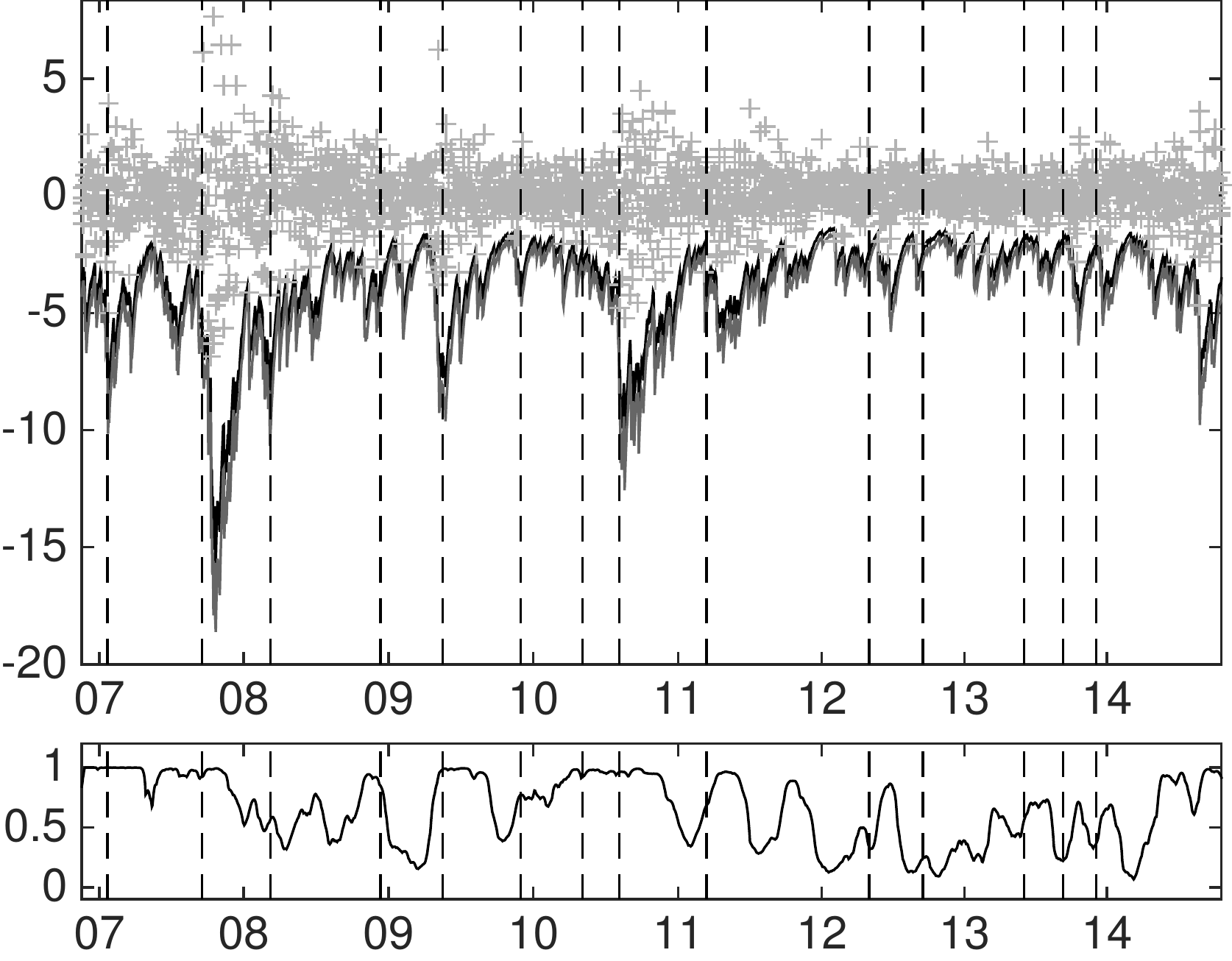}}
%
\caption{\footnotesize{Predicted CoVaR$_{t\vert t-1}^{S\vert j}$ {\it (black line)} and CoES$_{t\vert t-1}^{S\vert j}$ {\it (light grey)}, for $j=1,2,\dots,11$ estimated using the selected models. True observations over the out--of--sample period are depicted in grey. The black line in the bottom figures represents the smoothed estimates probabilities of the high systemic risk state, i.e., $\mathbb{P}\left(S_{t+1}\mid\bY_{1:t}\right)$, for $t=1,2,\dots,T$ for the SGASC specification.
Vertical dashed lines represent major financial downturns: for a detailed description see Table \ref{tab:fin_crisis_timeline}.}}
\label{fig:CoVaR_CoES_ALL}
\end{sidewaysfigure}

%
\noindent In this section, we apply the estimated models and the systemic risk measures introduced in Section \ref{sec:risk_measures} to asses the systemic risk contribution of each country index to the overall European equity market. To this end, one step ahead rolling forecasts are performed using $H=2000$ out--of--sample observations covering the period beginning on November 12, 2007 to the end of the sample. Parameter estimates are updated every 25 observations for a total of 80 refits. The estimated coefficients are quite stable during the forecast period suggesting that the underline structure of the economy is well described by the selected models. The estimated coefficients over the out--of--sample period are not reported to save space and are available upon request to the second author. The selected models are employed to predict the CoVaR and the CoES of the overall European system $\mathrm{M}$, measured by the Market index (MKT), conditional to distress events affecting each European region, measured by the corresponding country indexes, i.e., CoVaR$_{\mathrm{M}\vert j}^{\tau_1\vert\tau_2}$ and CoES$_{\mathrm{M}\vert j}^{\tau_1\vert\tau_2}$, for $j\in\{\mathrm{AU, BEL, DEN, FR, GER, HUN, IT, NET, SPA, SWE, UK}\}$. The confidence levels $\left(\tau_1,\tau_2\right)$ are fixed at 5\% which means that we use the situation where the country specific index is below its 5\% marginal Value--at--Risk level as conditional distress event.
%
%
\noindent Figure \ref{fig:CoVaR_CoES_ALL} reports, for the forecast period, the predicted CoVaR$_{\mathrm{M}\vert j}^{\tau_1\vert\tau_2}$ and CoES$_{\mathrm{M}\vert j}^{\tau_1\vert\tau_2}$. Vertical dashed lines represent the major financial downturns experienced by the European economic system during the period 2007--2014. A timeline of the major European financial crisis is provided in Table \ref{tab:fin_crisis_timeline}. In the bottom panel of each subfigure, for the SGASC specifications, we report the estimated out--of--sample smoothed probabilities, $\mathbb{P}\left(S_t=1\mid\bY_{1:T+H}\right)$. Figure \ref{fig:CoVaR_CoES_ALL} gives insights about the dynamic evolution of the systemic risk contributions during the different economic and financial phases the European system experienced since the end of 2007. As documented by Figure \ref{fig:CoVaR_CoES_ALL}, the CoVaR and the CoES systemic risk measures suddenly adapt to the relevant changes affecting the underlying financial system's wealth, such as those experienced during the turbulent phases before the bails out of Portugal and Greece in 2011.
Concerning instead the interpretation of the smoothed probabilities of the hidden Markov chain in Figure \ref{fig:CoVaR_CoES_ALL}, it is worth stressing that the dynamic non--linear evolution of the copula dependence parameter as well as that of the degrees--of--freedom prevents a clear and straightforward identification of the latent states with given characteristics of the conditional distribution. To further explain the nature of the identification problem, let consider the possibility that for some assets we could observe, for example, that high levels of linear dependence are associated with high values of the degrees--of--freedom parameter. In those cases, the interpretation of low and high probability levels in the bottom panels of Figure \ref{fig:CoVaR_CoES_ALL} becomes cumbersome. Furthermore, another relevant point to be addressed is that the interpretation of the states is not homogeneous across countries. As an example, Denmark in Figure \ref{fig:Denmark_CoVaR} suddenly experienced an abrupt decrease of the smoothed probability from a state of low correlation/high tail dependence to a state of high correlation/moderate tail dependence. This is not the case, for example, of Germany where, during the same period, the system switched from a state of high correlation/high tail dependence to a state of low correlation and low tail dependence. To gain further insights about the identification issues, we report in Table \ref{tab:MS_state_identification} the long run dependence measures as well as two measures of persistence in terms of expected duration, conditional to the Markovian state. Concerning the conditional expected durations, we report the half life (HL) index as a measure of speed reversion of the dependence parameter dynamics and the implied duration of each regime (D) of the Markov chain. Moreover, we also consider the long run correlation $\bar{\rho}_l$ and tail dependence $\bar{\varpi}_l$, for $l=1,2$. Except for Austria, which is characterised by moderate long run tail dependence in each regime, all the remaining SGASC specifications clearly identify two regimes of high and low long run tail dependence and correlation. European countries seams to be also quite heterogeneous with respect to the persistence of the conditional dependence parameter and the expected duration of the Markov chian. Indeed, Spain and Germany display high persistence in the conditional dependence dynamic but low persistence in the Markov chain. On the contrary, Austria reports low persistence in the conditional correlation but very high persistence in the Markov chain. These findings suggest that different European countries react differently to the new information coming from the market. Moreover, we note that SGASC models usually display higher persistence in terms of HL of the conditional process compared to the simpler GAScop specifications. Indeed, the additional flexibility gained from the switching behaviour of SGASC models, permits to disentangle the information that impacts only on the memory of the dependence process and that impacting only on some characteristic of the dependence structure such as the levels of correlation and tail dependence.
%
\subsection{Backtesting the CoVaR systemic risk measure}
\label{sec:backtesting_covar}
%
\noindent Now we move to the evaluation of the CoVaR forecasts. Since the CoVaR risk measure is essentially a modified version of the Value--at--Risk, we can employ the usual VaR backtesting procedures such as the unconditional (UC) and conditional coverage (CC) tests of  \cite{kupiec.1995} and \cite{christoffersen.1998}, the Actual over Expected (AE) ratio and the mean and maximum absolute deviation (ADmean, ADMax) considered by \cite{mcaleer_daveiga.2008}. The only difference between our context and the usual VaR backtesting procedure, concerns the evaluation of the \qmo hitting sequence\qmcsp of returns exceeding the CoVaR levels. In line with our definition of CoVaR in equation \eqref{eq:covar_def}, we follow the approach of \cite{girardi_ergun.2013} who consider the series of returns jointly exceeding both the CoVaR and VaR levels as the proper hitting sequences. In Table \ref{tab:UC_sgasc} we report, for each country, the AE, ADmean, ADmax, as well as the p--values of the UC and CC tests. We observe that, except for BEL, the null hypothesis of correct coverage of the conditional lower tail of the joint distribution at the 5\% confidence level is never rejected. ADmean suggests that, on average, CoVaR violations are of the order of magnitude of about 0.3\%, while the ADmax suggests that, during period of distress, CoVaR absolute deviations could exceed the 3.5\%.
%
\subsection{Ranking European countries}
\label{sec:euro_ranking}
\begin{sidewaysfigure}[t]
\centering
\subfloat[Austria]{\label{fig:Austria_DCoVaR}\includegraphics[width=0.22\textwidth]{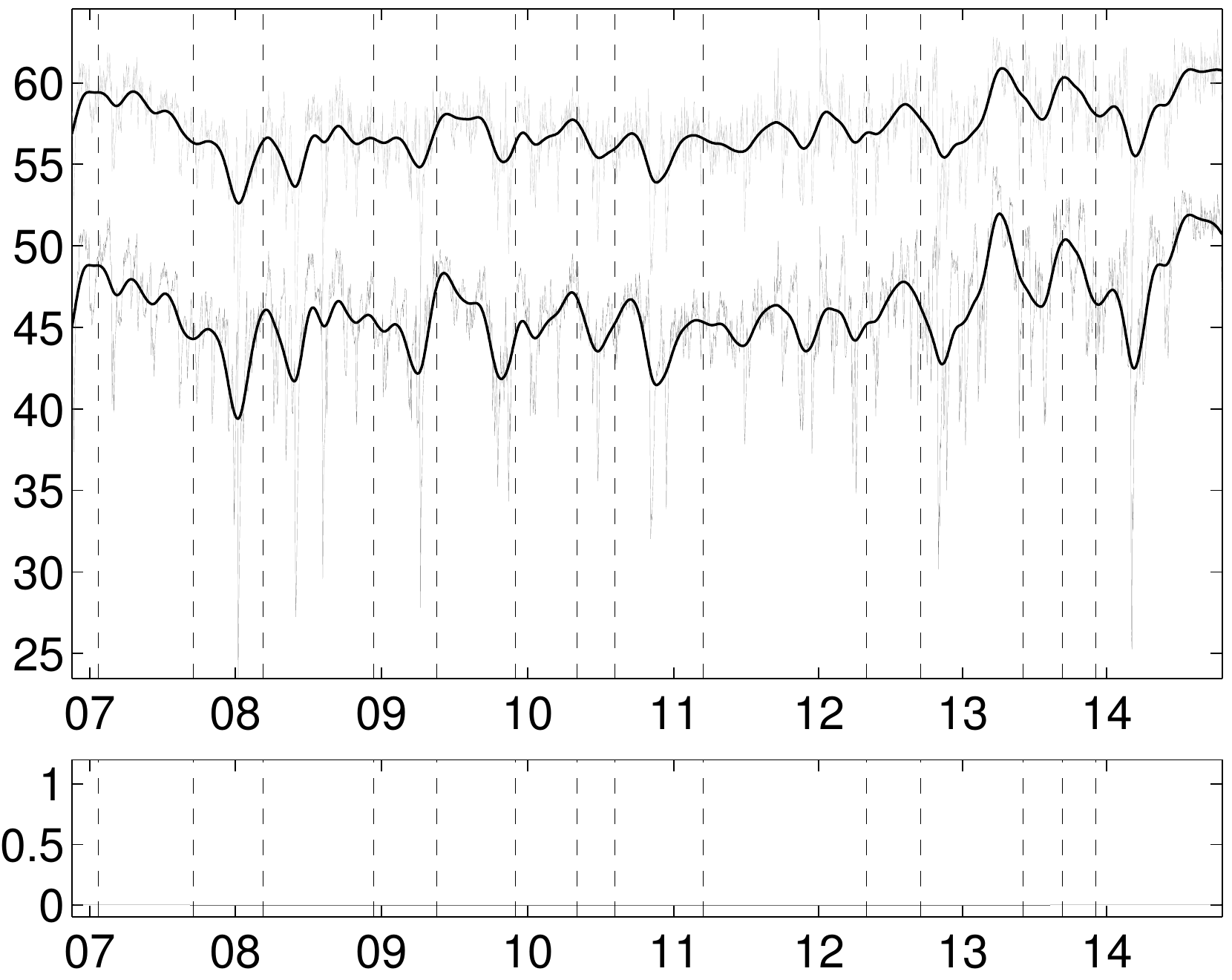}}\quad
\subfloat[Belgium]{\label{fig:Belgium_DCoVaR}\includegraphics[width=0.22\textwidth]{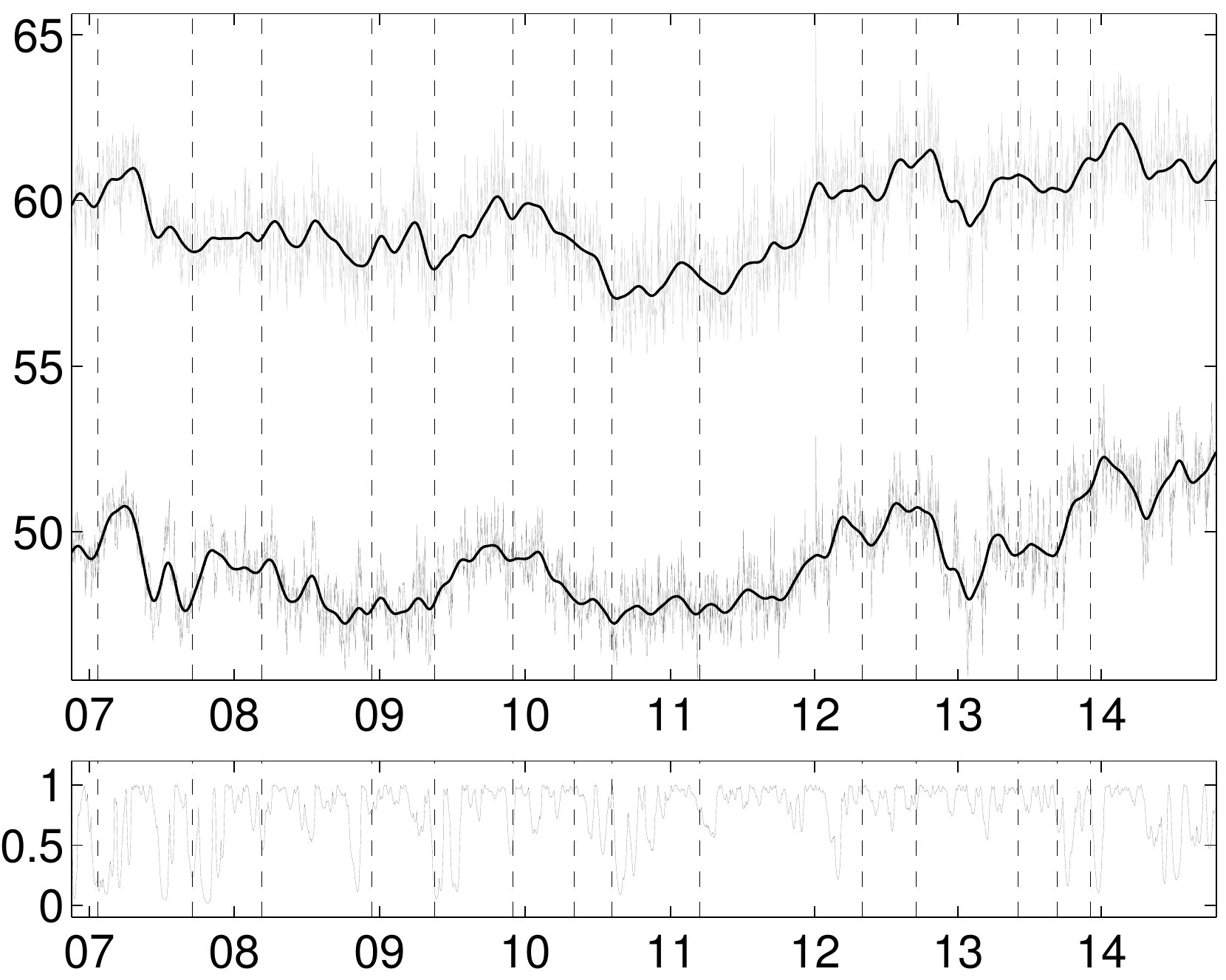}}\quad
\subfloat[Denmark]{\label{fig:Denmark_DCoVaR}\includegraphics[width=0.22\textwidth]{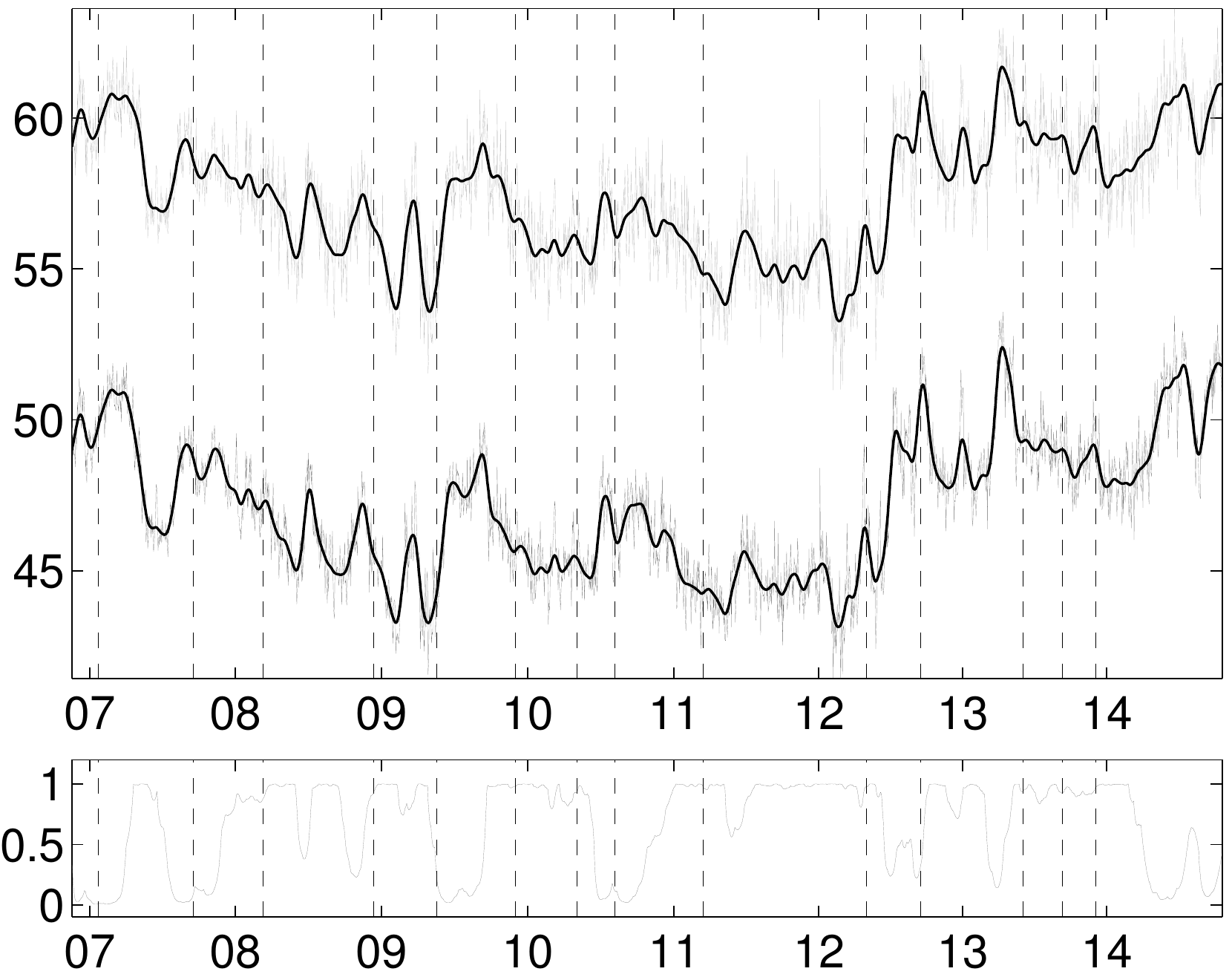}}\quad
\subfloat[France]{\label{fig:France_DCoVaR}\includegraphics[width=0.22\textwidth]{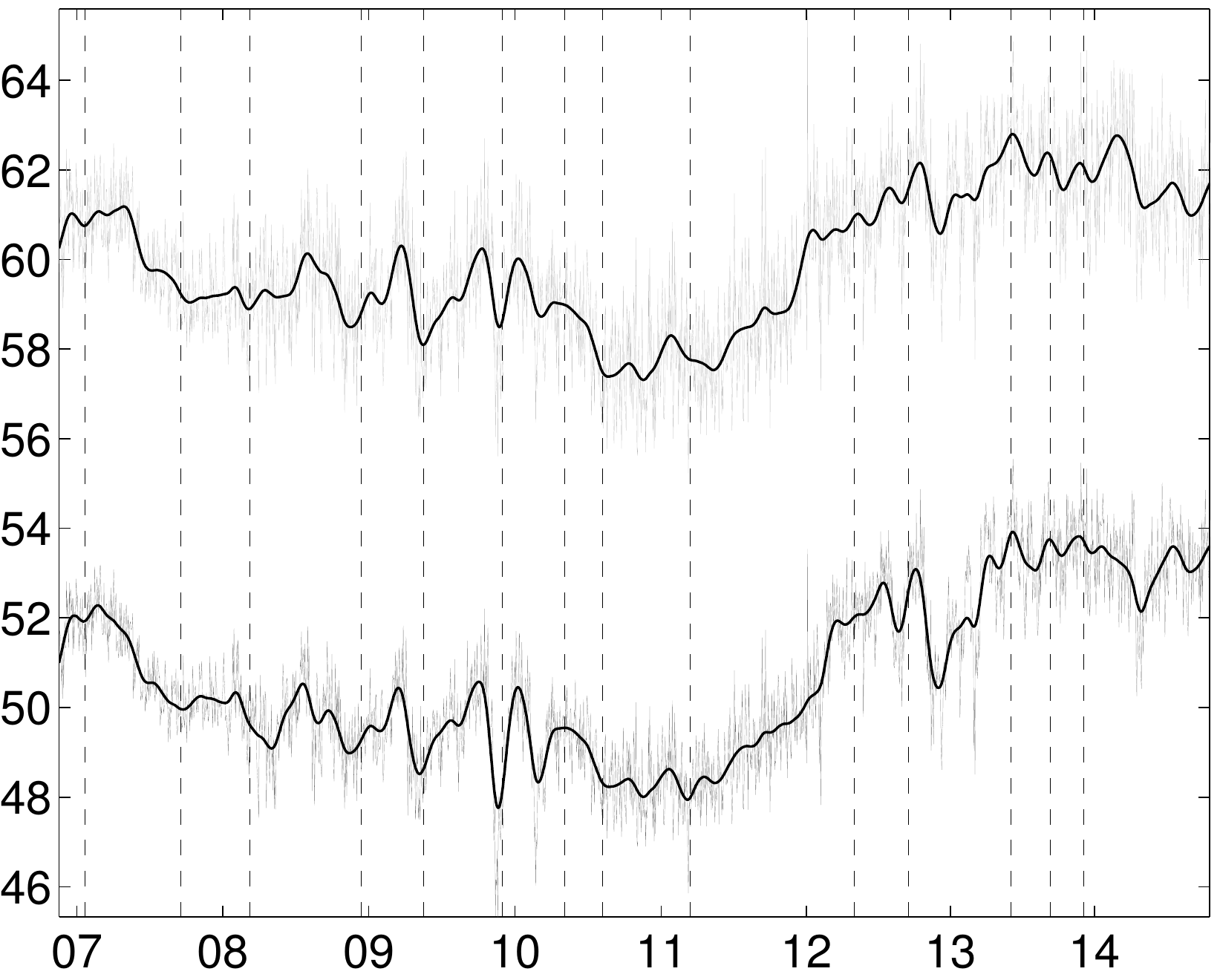}}\\
\subfloat[Germany]{\label{fig:Germany_DCoVaR}\includegraphics[width=0.22\textwidth]{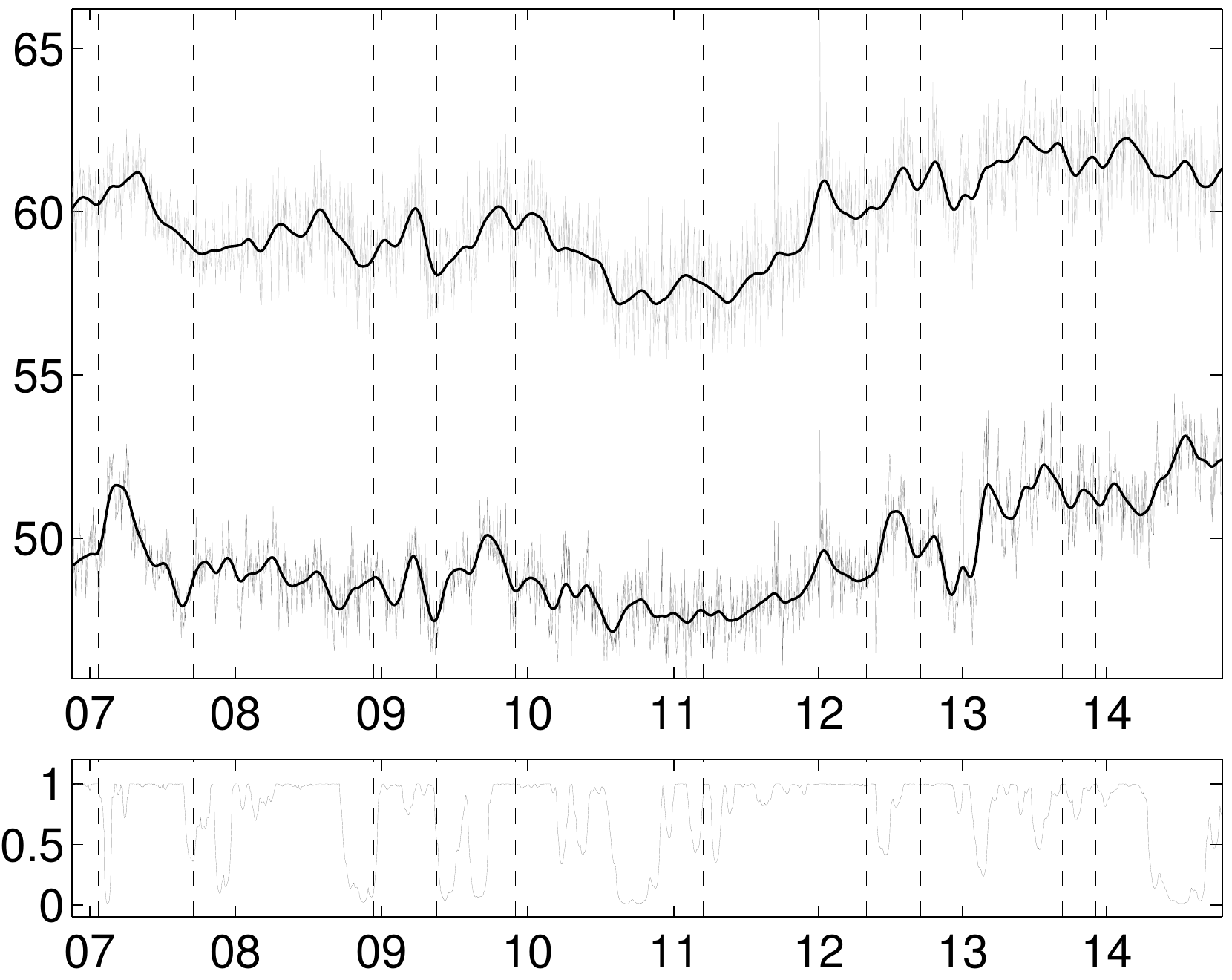}}\quad
\subfloat[Hungary]{\label{fig:Hungary_DCoVaR}\includegraphics[width=0.22\textwidth]{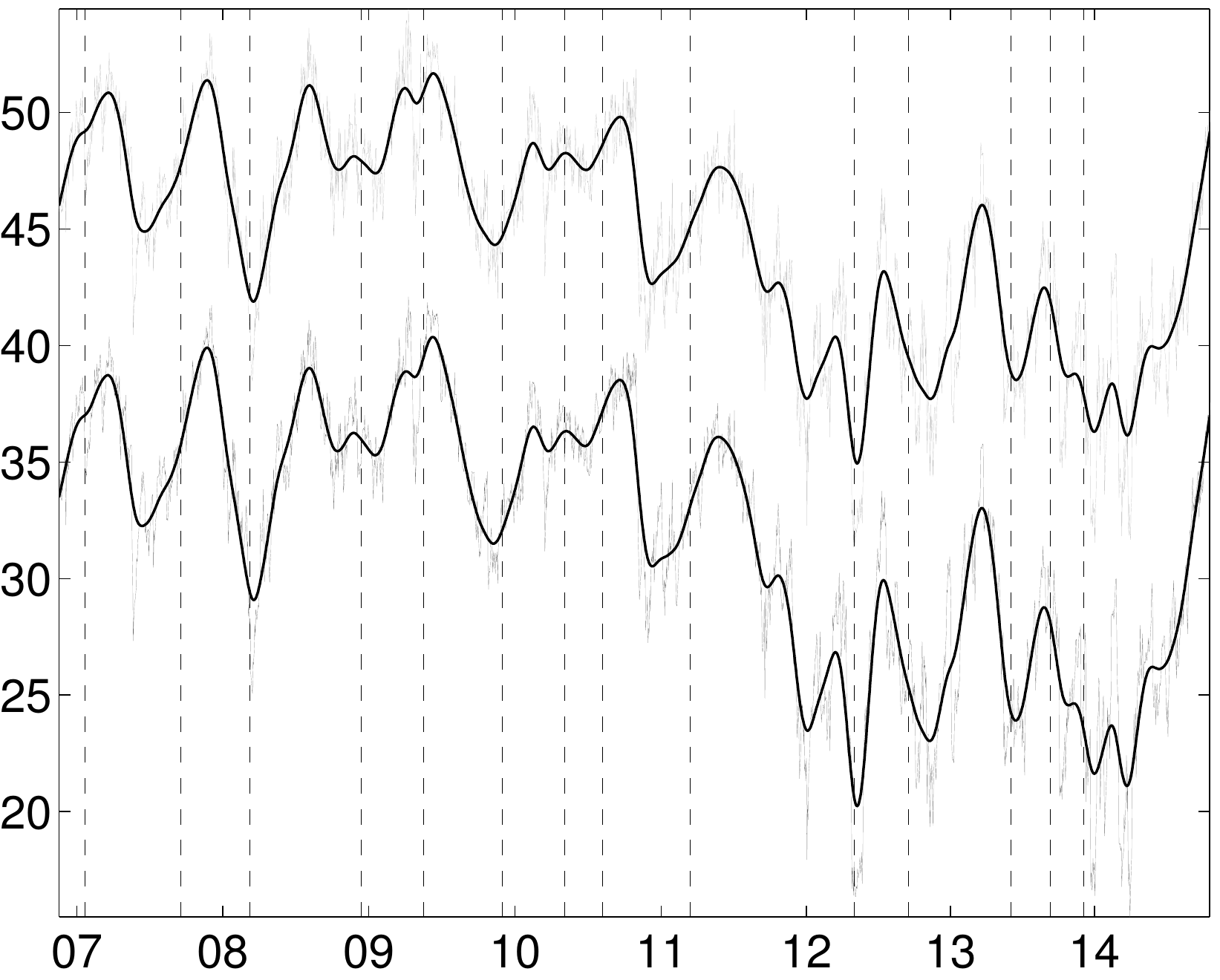}}\quad
\subfloat[Italy]{\label{fig:Italy_DCoVaR}\includegraphics[width=0.22\textwidth]{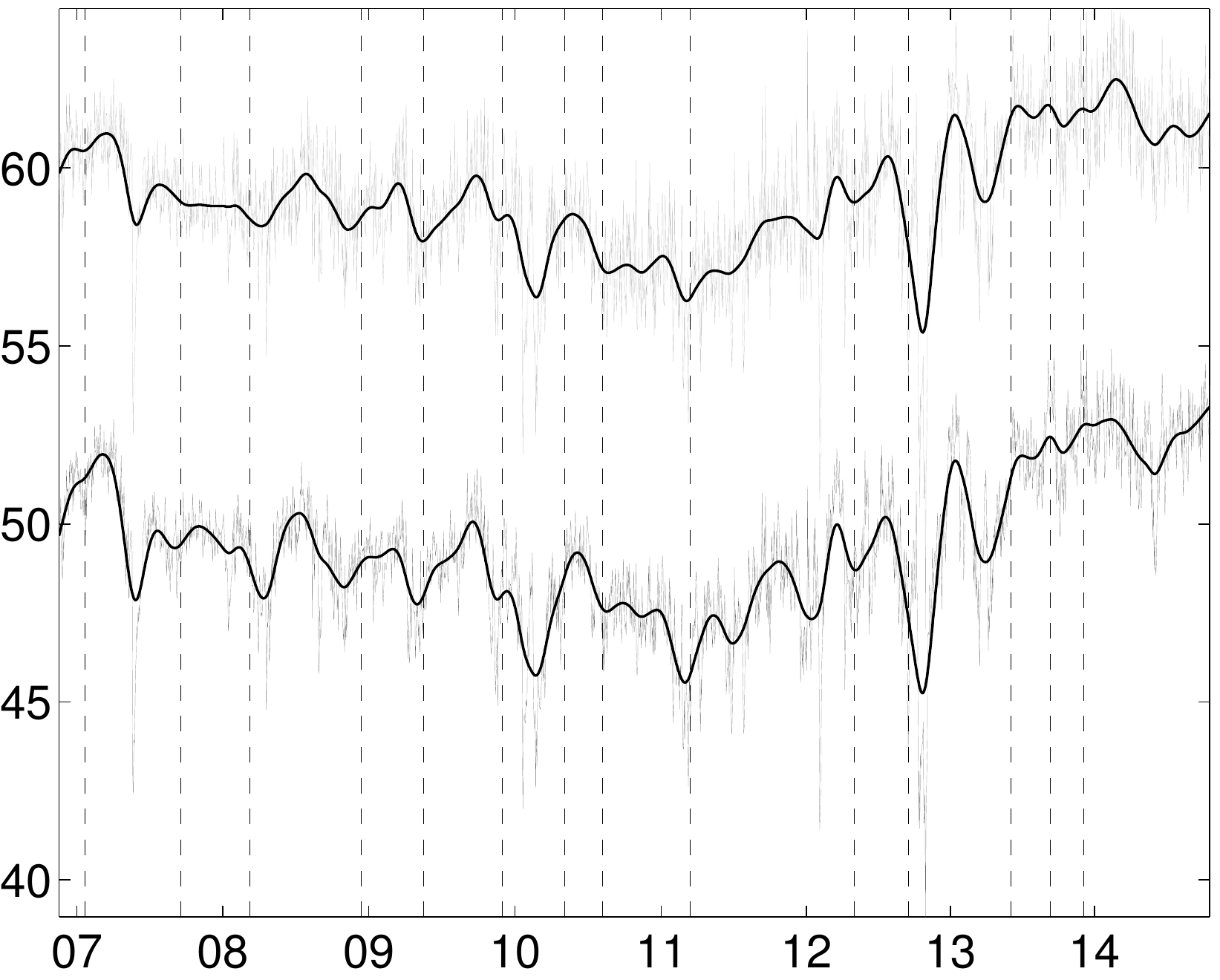}}\quad
\subfloat[Netherlands]{\label{fig:Netherlands_DCoVaR}\includegraphics[width=0.22\textwidth]{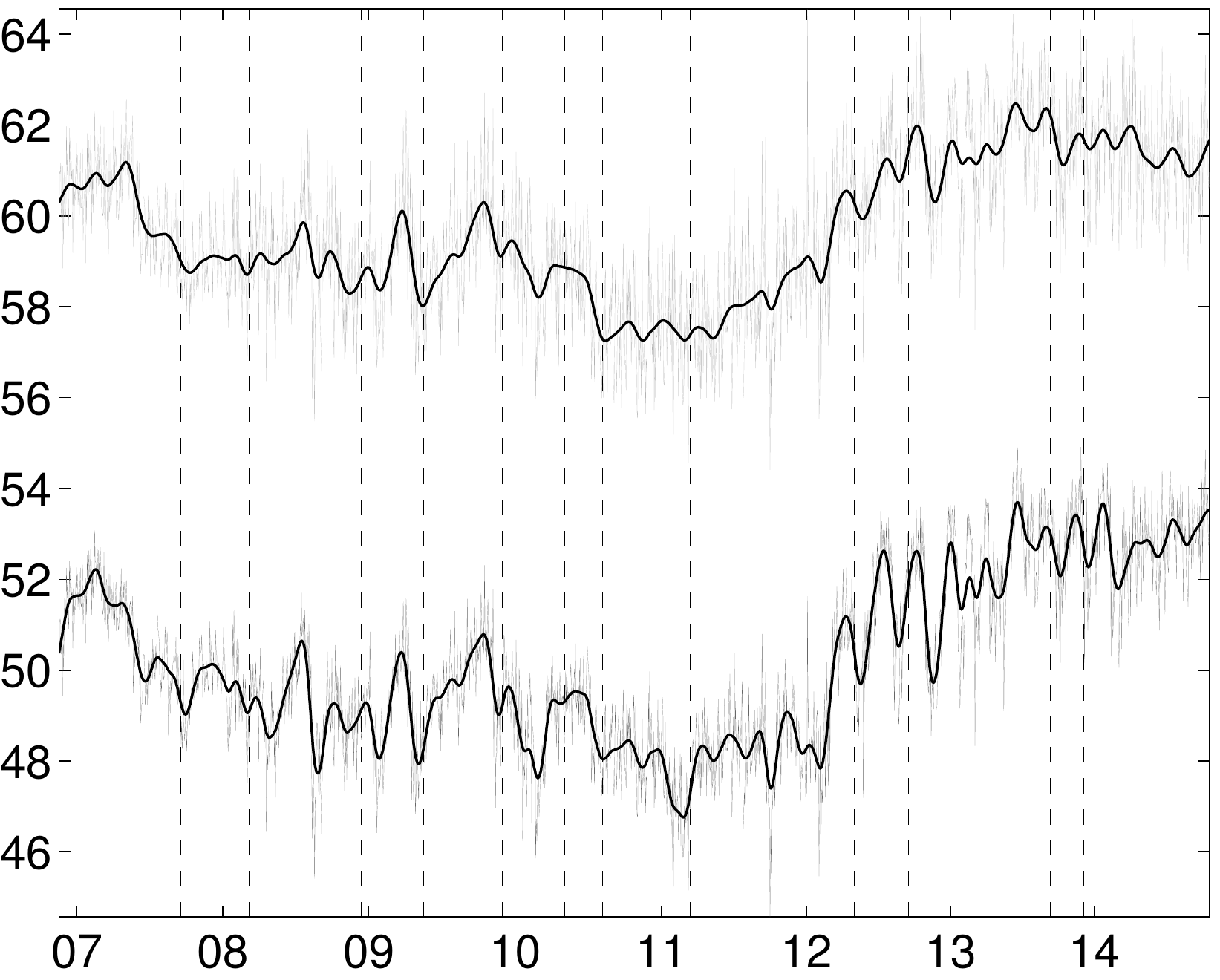}}\\
\subfloat[Spain]{\label{fig:Spain_DCoVaR}\includegraphics[width=0.22\textwidth]{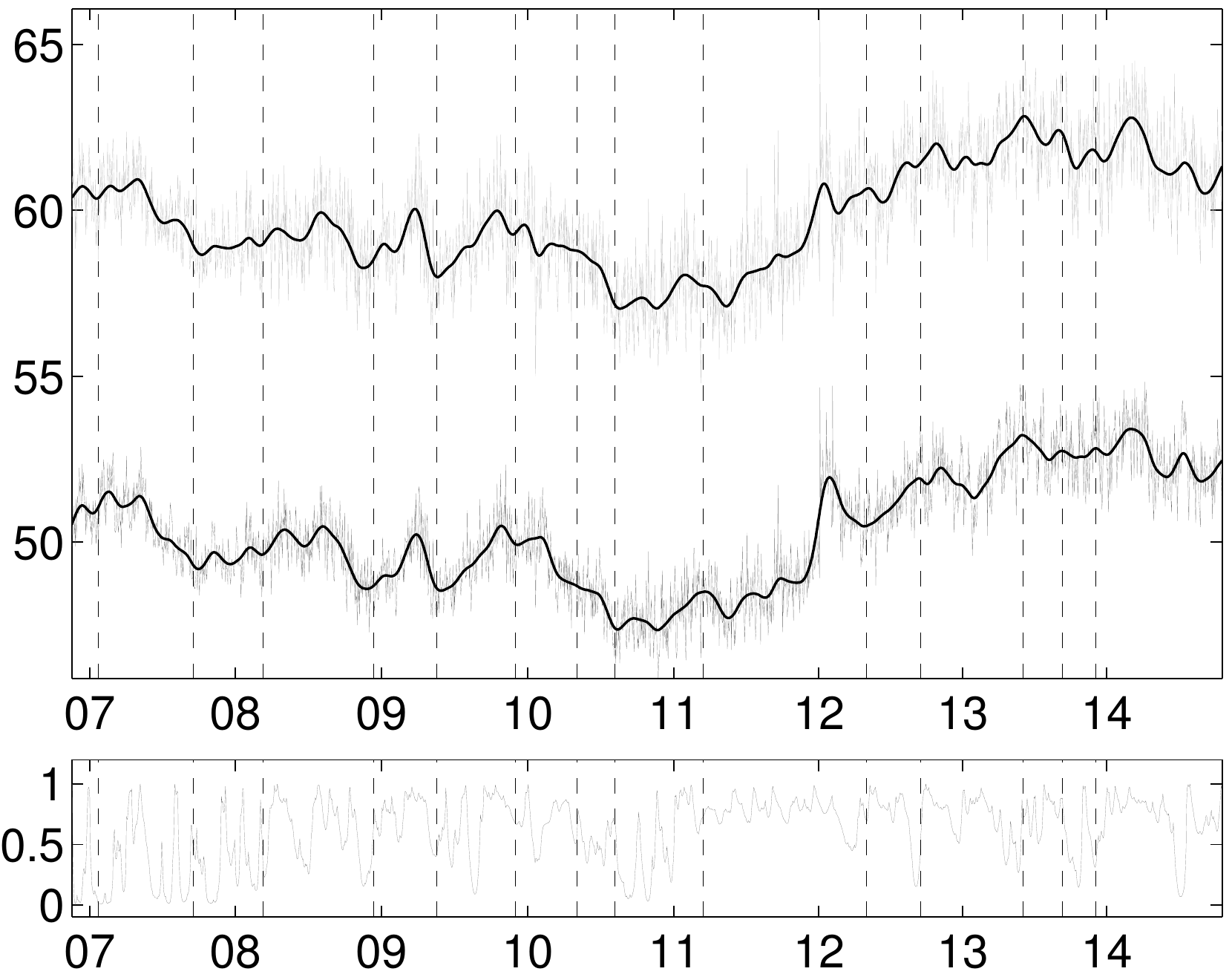}}\quad
\subfloat[Sweden]{\label{fig:Sweden_DCoVaR}\includegraphics[width=0.22\textwidth]{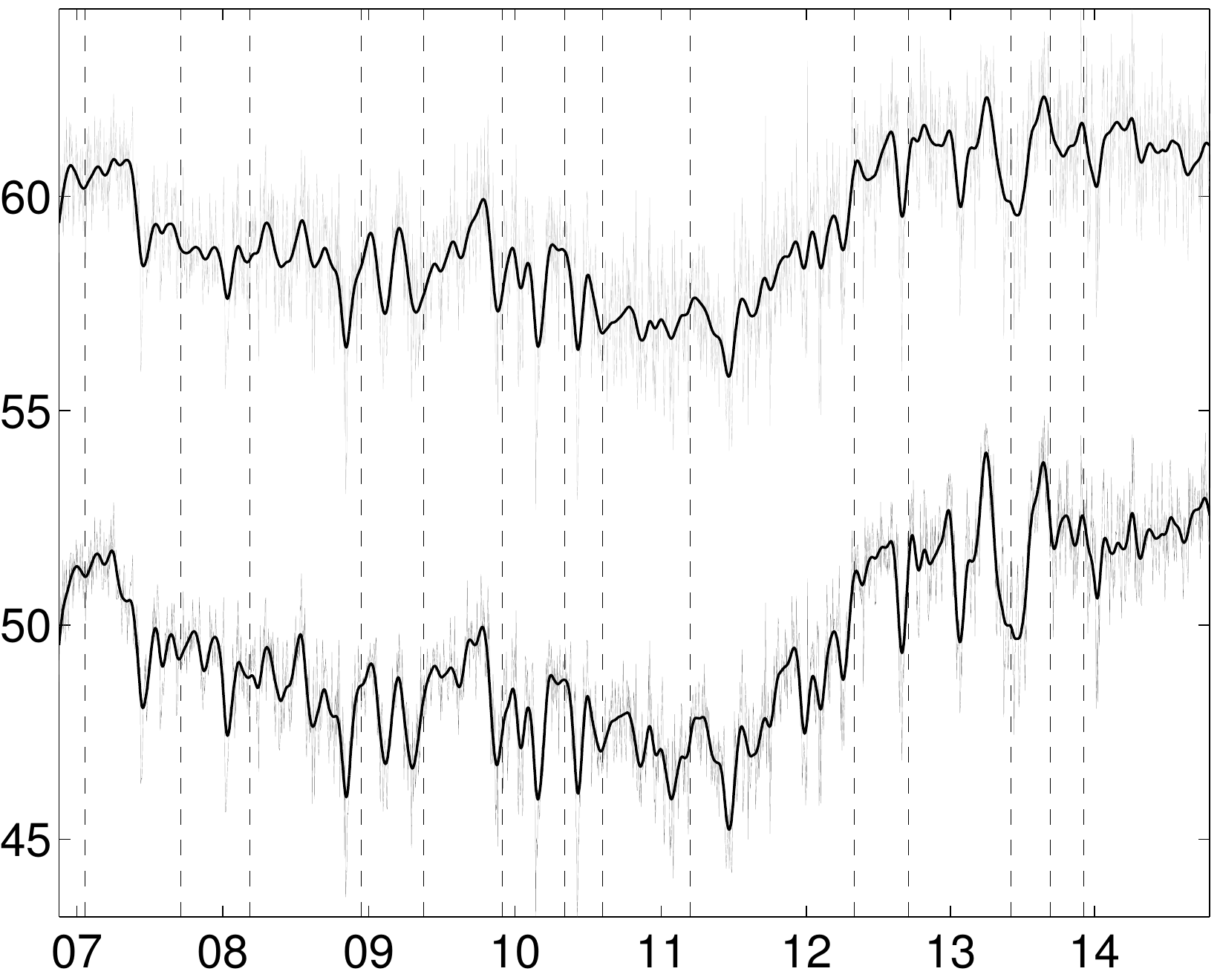}}\quad
\subfloat[United Kingdom]{\label{fig:United Kingdom_DCoVaR}\includegraphics[width=0.22\textwidth]{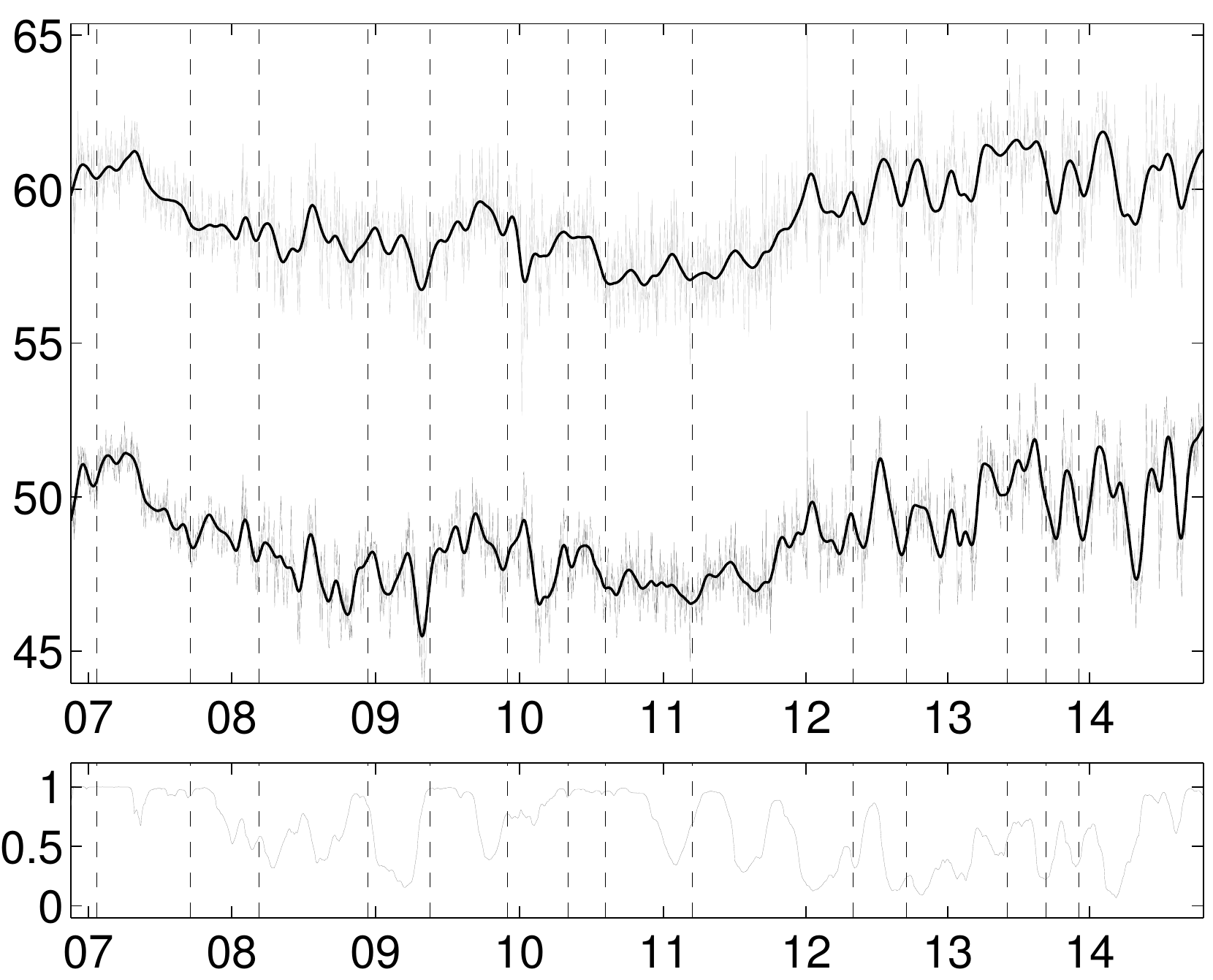}}
%
\caption{\footnotesize{Predicted $\Delta$CoVaR$_{t\vert t-1}^{S\vert j}$ {\it (light grey line)} and $\Delta$CoES$_{t\vert t-1}^{S\vert j}$ {\it (grey line)}, for $j=1,2,\dots,11$ estimated using the selected models. The black lines represented smoothed levels. The black line in the bottom figures represents the smoothed estimates probabilities of the high systemic risk state, i.e. $\mathbb{P}\left(S_{t+1}\mid\bY_{1:t}\right)$, for $t=1,2,\dots,T,$ for the SGASC specification.
Vertical dashed lines represent major financial downturns: for a detailed description see Table \ref{tab:fin_crisis_timeline}.}}
\label{fig:DCoVaR_DCoES_ALL}
\end{sidewaysfigure}

\begin{figure}[p]
\begin{center}
\begin{tabular}{c}
\includegraphics[width=0.5\linewidth]{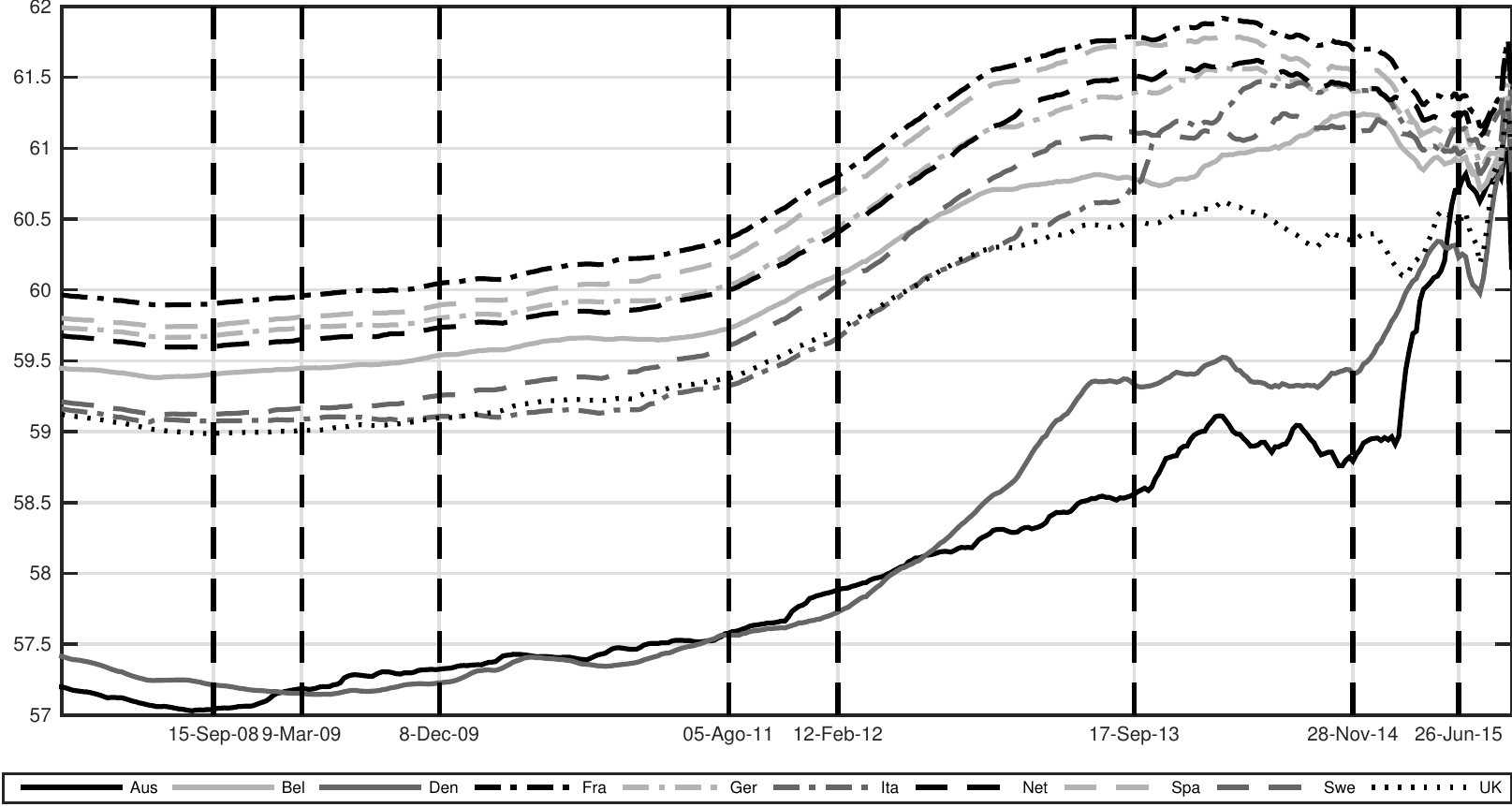}\\
\includegraphics[width=0.5\linewidth]{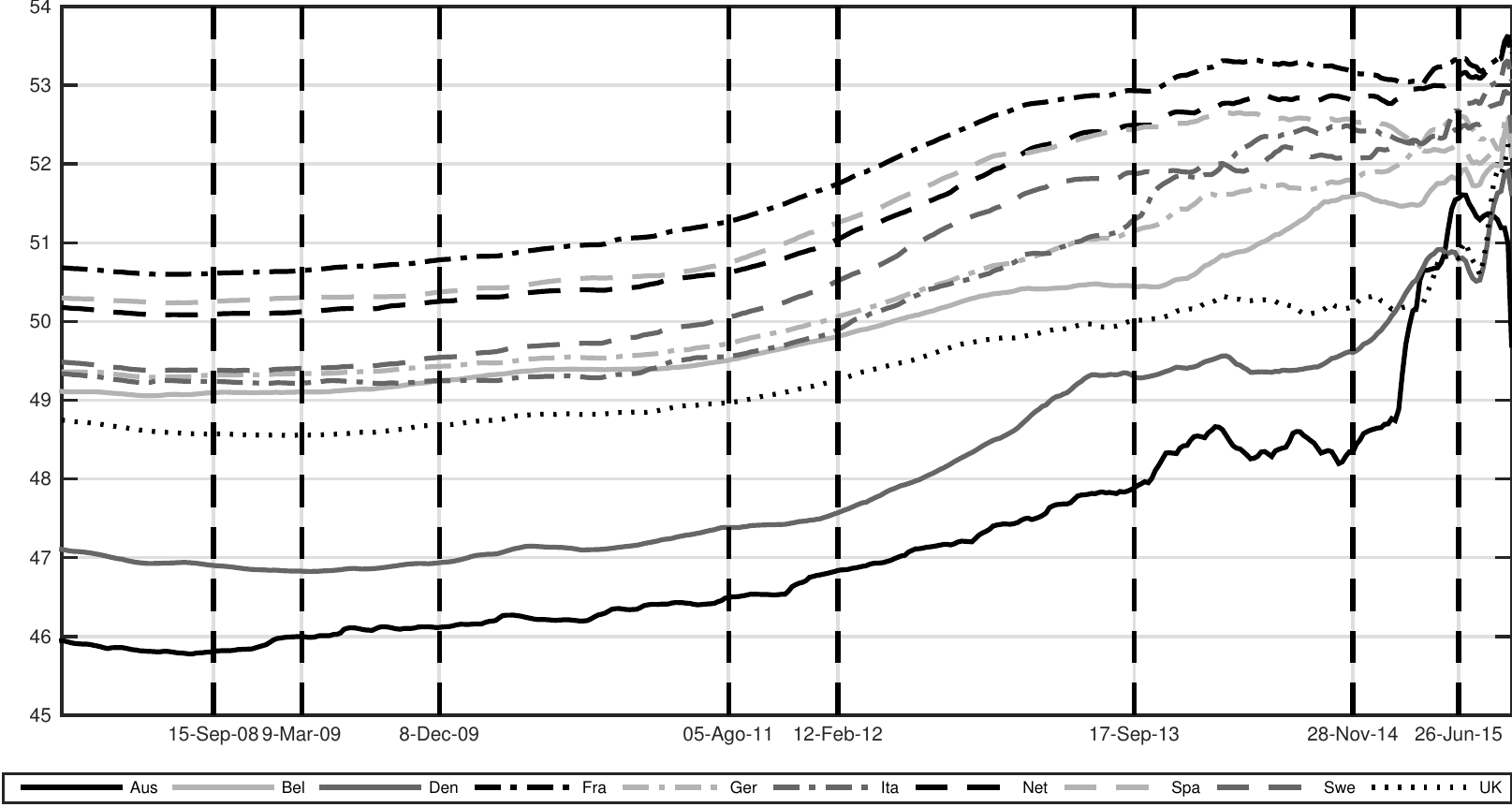}
\end{tabular}
\caption{\footnotesize{Out--of--sample evolution of the $\Delta$CoVaR {\it (top graph)} and $\Delta$CoES {\it (bottom graph)} weekly averages. Vertical dashed lines denote the LehmanÕs failure (September 15, 2008), the peak of the onset of the recent global financial crisis (March 9, 2009), the downgrading of the US sovereign by S\&P (August 5, 2011), the protraction of the Greek austerity package (February 12, 2012), the drop of car sales to the lowest recorded level (September 17, 2013), the Italian unemployment rate reaches the record high since the 1977 (November 28, 2014), the Greek government unilaterally broke off negotiations with the Eurogroup (June 26, 2015). A complete timeline of the crisis events can be found in Table \ref{tab:fin_crisis_timeline}.}}
\label{fig:DCoVaR_DCoES_SS}
\end{center}
\end{figure}

%
\noindent The $\Delta\mathrm{CoVaR}$ and $\Delta\mathrm{CoES}$ risk measures are particularly useful in order to rank the countries in terms of their systemic risk contributions. Note that, the resulting ranking does not tell anything about the intrinsic riskiness profile of the particular country, but instead, it is informative about the role each country plays in the overall European system. This point is particularly relevant when systemic risk analyses are carried on. Indeed, the countries on the top of the rank are going to be those who play a central role into the overall European equity market. The daily evolution of the $\Delta$CoVaR and $\Delta$CoES, calculated as in equation \eqref{eq:deltaCoVaR} and \eqref{eq:deltaCoES}, are plotted in Figure \ref{fig:DCoVaR_DCoES_ALL} along with the superimposed smoothed estimates. The visual inspection of the figure reveals three relevant phenomena. First, as expected, both risk measures provide the same underlying systemic risk signal, with the $\Delta$CoES being always below the $\Delta$CoES with almost the same dynamic pattern. Second, $\Delta$CoVaR and $\Delta$CoES seem to display a quite heterogeneous behaviour across countries. Third, the systemic risk importance of the considered countries increases during the second part of the sample. The only two exceptions are Austria, which has constant systemic risk contributions over the whole sample, and Hungary, which instead is less systemically important over the period 2011--2015. Figure \ref{fig:DCoVaR_DCoES_SS} reports the dynamic evolution of the out--of--sample monthly averages of the $\Delta\mathrm{CoVaR}$ and the $\Delta\mathrm{CoES}$ risk measures, and provides insights on the relative importance of each country to the overall European equity market. Monthly $\Delta$CoVaR and $\Delta$CoES are calculated by averaging the daily levels plotted in Figure \ref{fig:DCoVaR_DCoES_ALL}.
Concerning HUN, the $\Delta\mathrm{CoVaR}$ and the $\Delta\mathrm{CoES}$ risk measures are substantially lower than the estimated levels of others countries and are not reported in Figure \ref{fig:DCoVaR_DCoES_SS}. According to the ranking, over the last decade, France is the country with the highest systemic risk contribution, followed by Spain, Germany and Netherlands. For those countries the ranking remains unchanged over the whole out--of--sample period, while for other countries such as, Italy and UK we observe a huge increase and decrease of the $\Delta$CoVaR just after the onset of the European crisis, respectively. During the last part of the sample we observe an increase of the systemic risk even for AU and DEN, which were at the bottom of the ranking prior to the European crisis. Our results are strictly in line with the findings recently reported by \cite{engle_etal.2015}. To summarise, we found that the systemic risk contribution of each European country grows during the ESDC of 2010--2011 reaching its highest level in mid--2014. After a short decreasing period in the last part of 2014, the magnitude as well as the relative importance each country has in terms of systemic risk contribution, sharply changed. In the very last part of our sample, just after the decision of the Greek government to unilaterally broke off negotiations with the Eurogroup on June 26, 2015, the $\Delta$CoVaR and $\Delta$CoES measures suddenly rise up to the highest levels experienced since the mid--2014, for all the European countries. It is worth mentioning that the main difference between the $\Delta$CoVaR and the $\Delta$CoES concerns the end of the out--of--sample period. Indeed, during this subperiod the systemic signal provided by the $\Delta$CoES risk measure is clearly better then that of the $\Delta$CoVaR which displays the tendency to shrink toward the same level.
%
\subsection{Dependence measures forecast and contagion effects}
\label{sec:contagion_dependence}
\begin{sidewaysfigure}[!t]
\centering
\subfloat[Austria]{\label{fig:Austria_DepMeas}\includegraphics[width=0.22\textwidth]{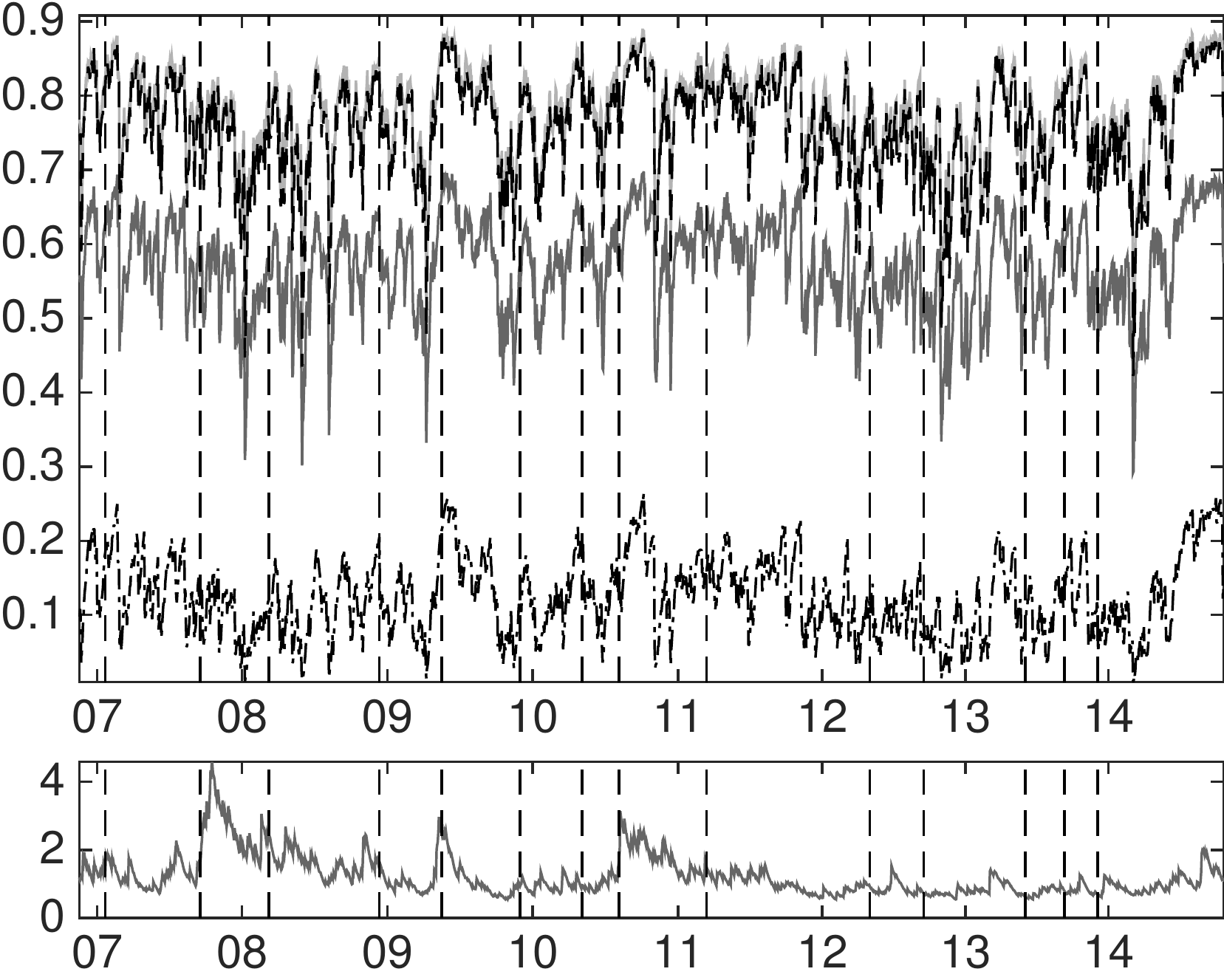}}\quad
\subfloat[Belgium]{\label{fig:Belgium_DepMeas}\includegraphics[width=0.22\textwidth]{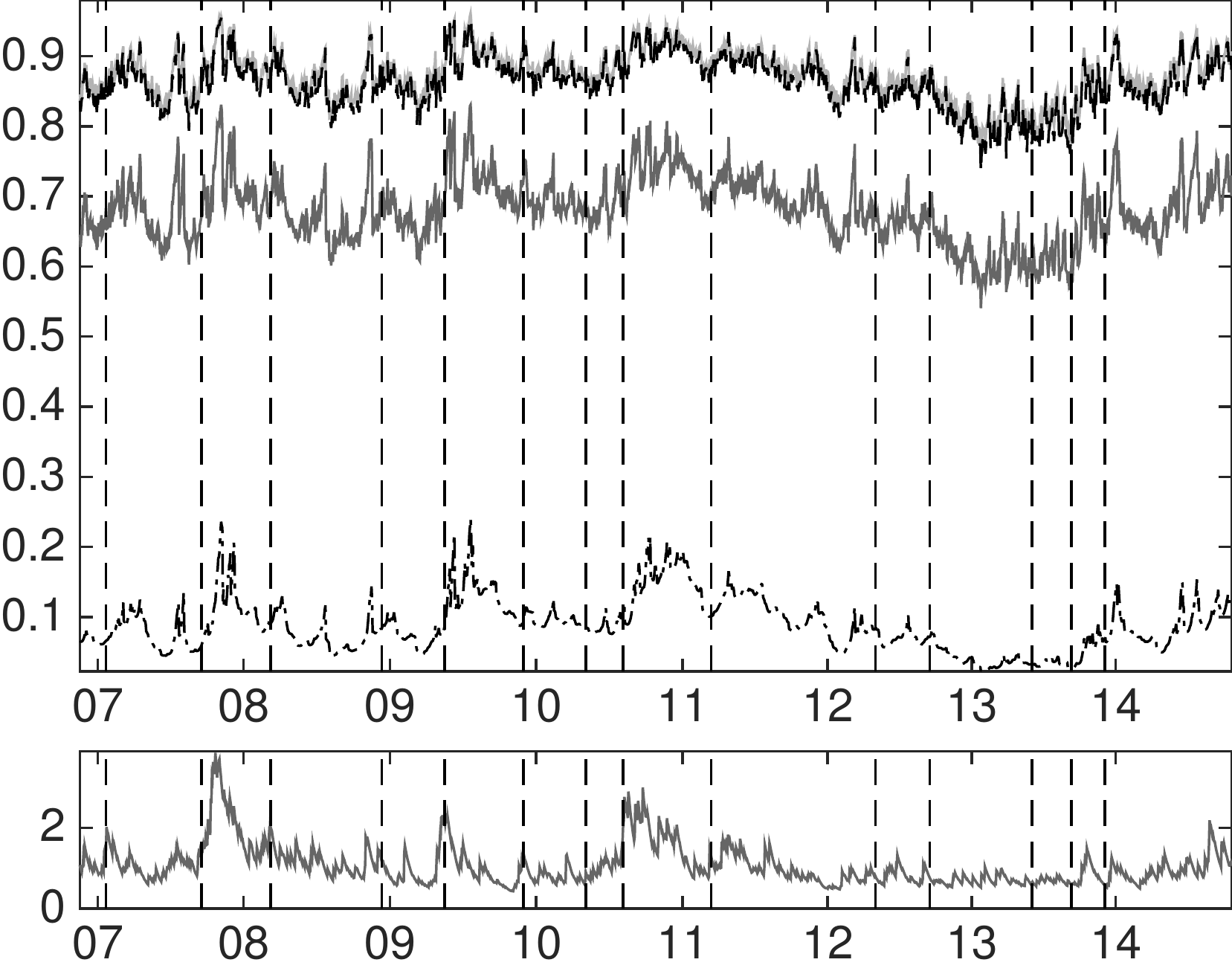}}\quad
\subfloat[Denmark]{\label{fig:Denmark_DepMeas}\includegraphics[width=0.22\textwidth]{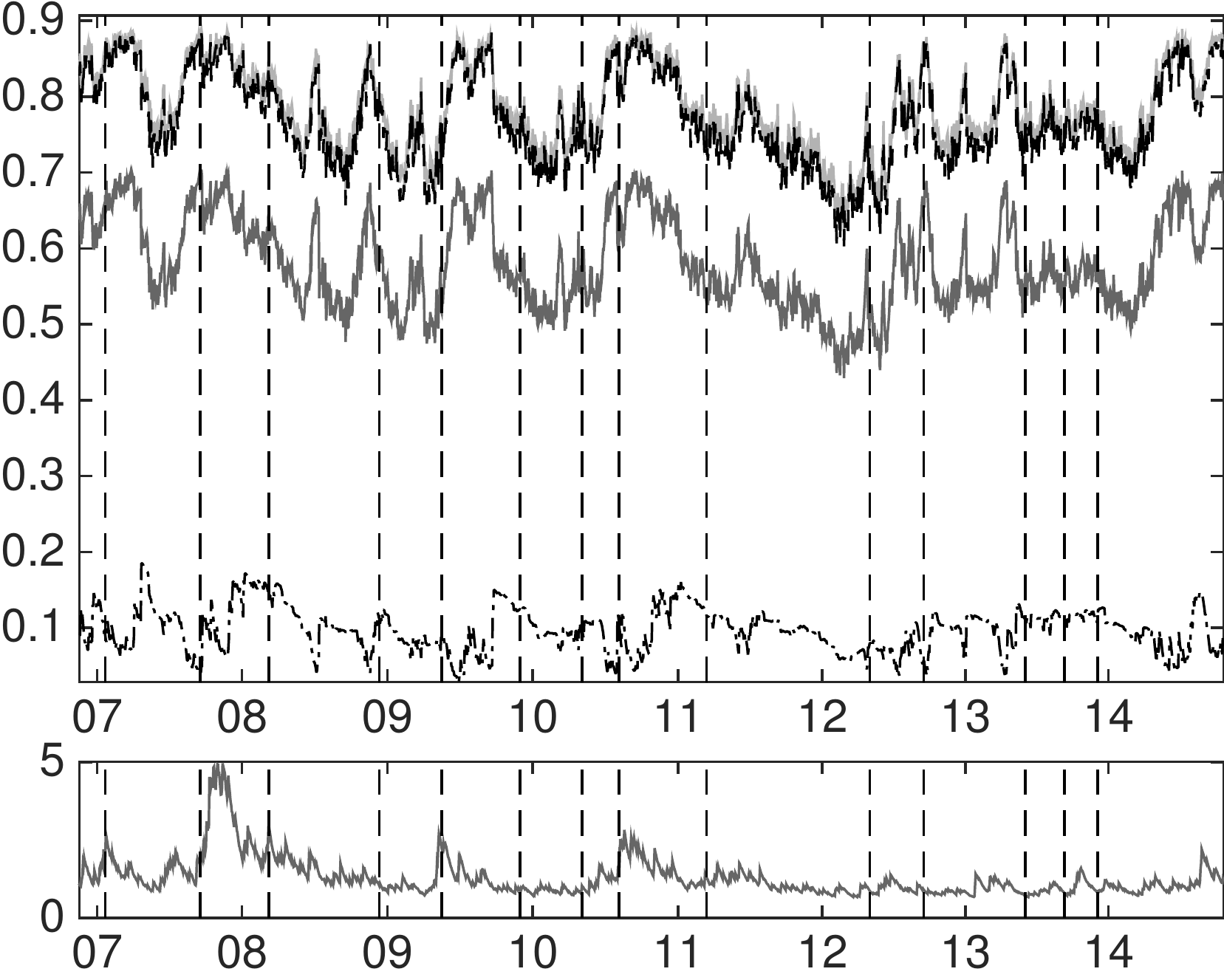}}\quad
\subfloat[France]{\label{fig:France_DepMeas}\includegraphics[width=0.22\textwidth]{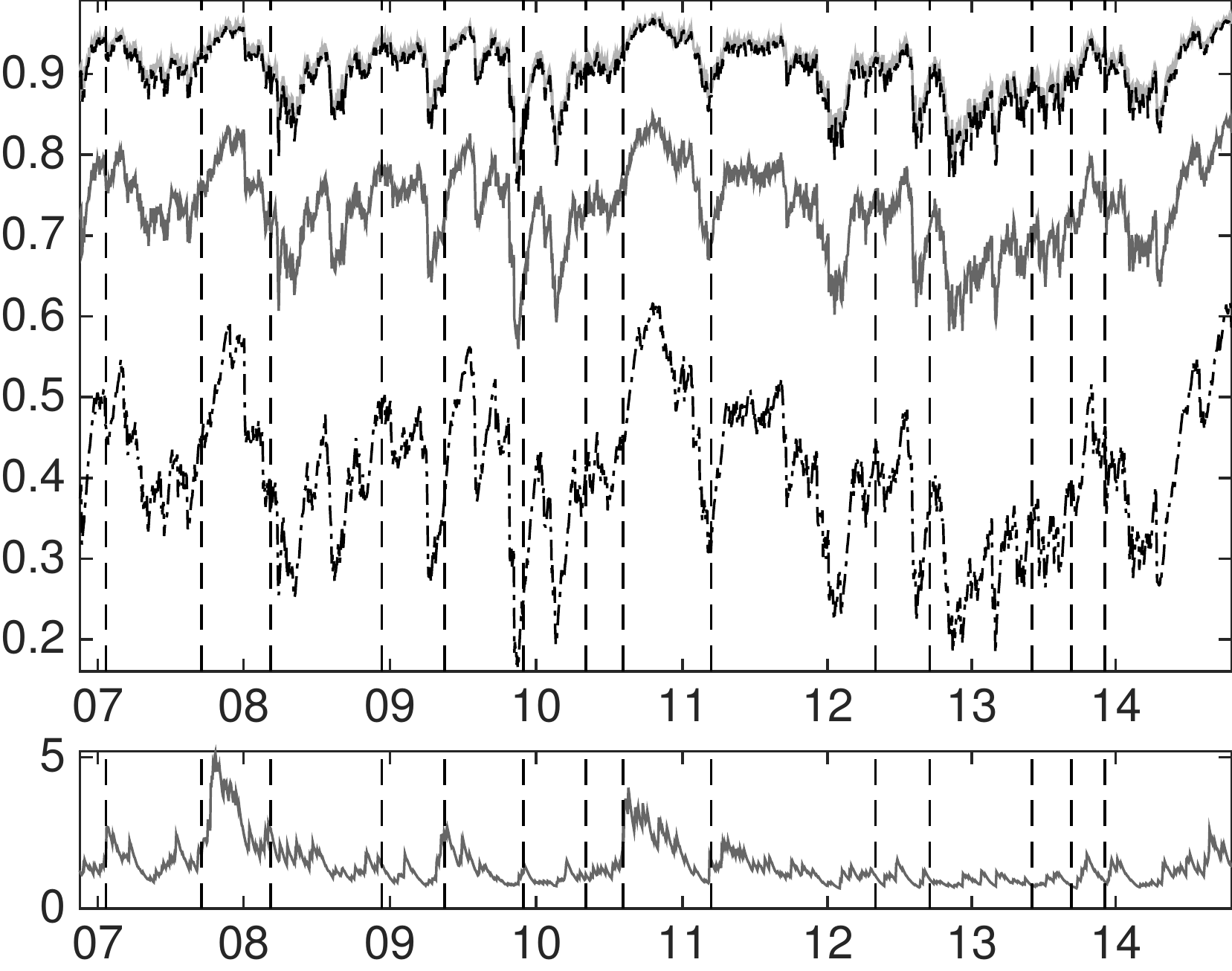}}\\
\subfloat[Germany]{\label{fig:Germany_DepMeas}\includegraphics[width=0.22\textwidth]{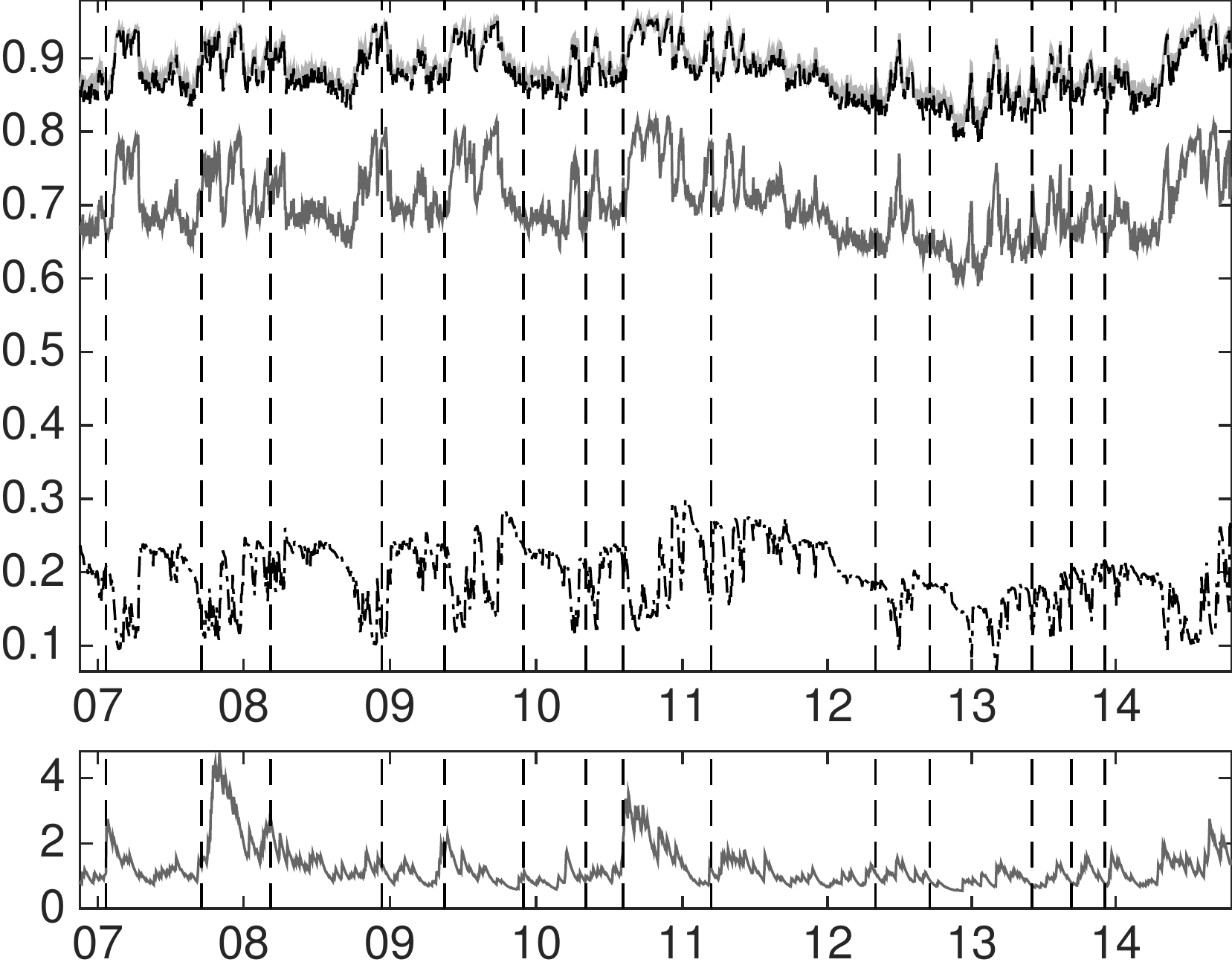}}\quad
\subfloat[Hungary]{\label{fig:Hungary_DepMeas}\includegraphics[width=0.22\textwidth]{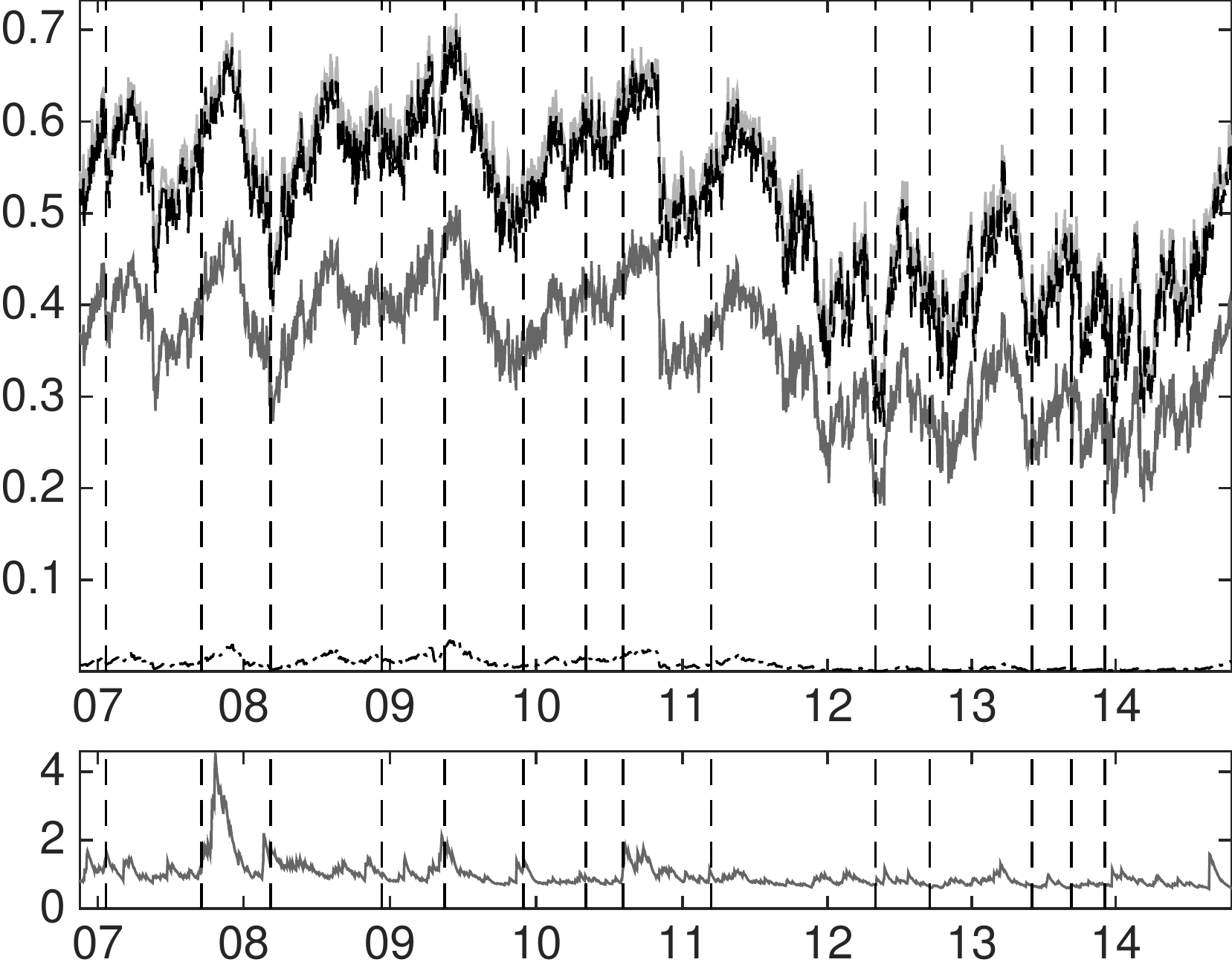}}\quad
\subfloat[Italy]{\label{fig:Italy_DepMeas}\includegraphics[width=0.22\textwidth]{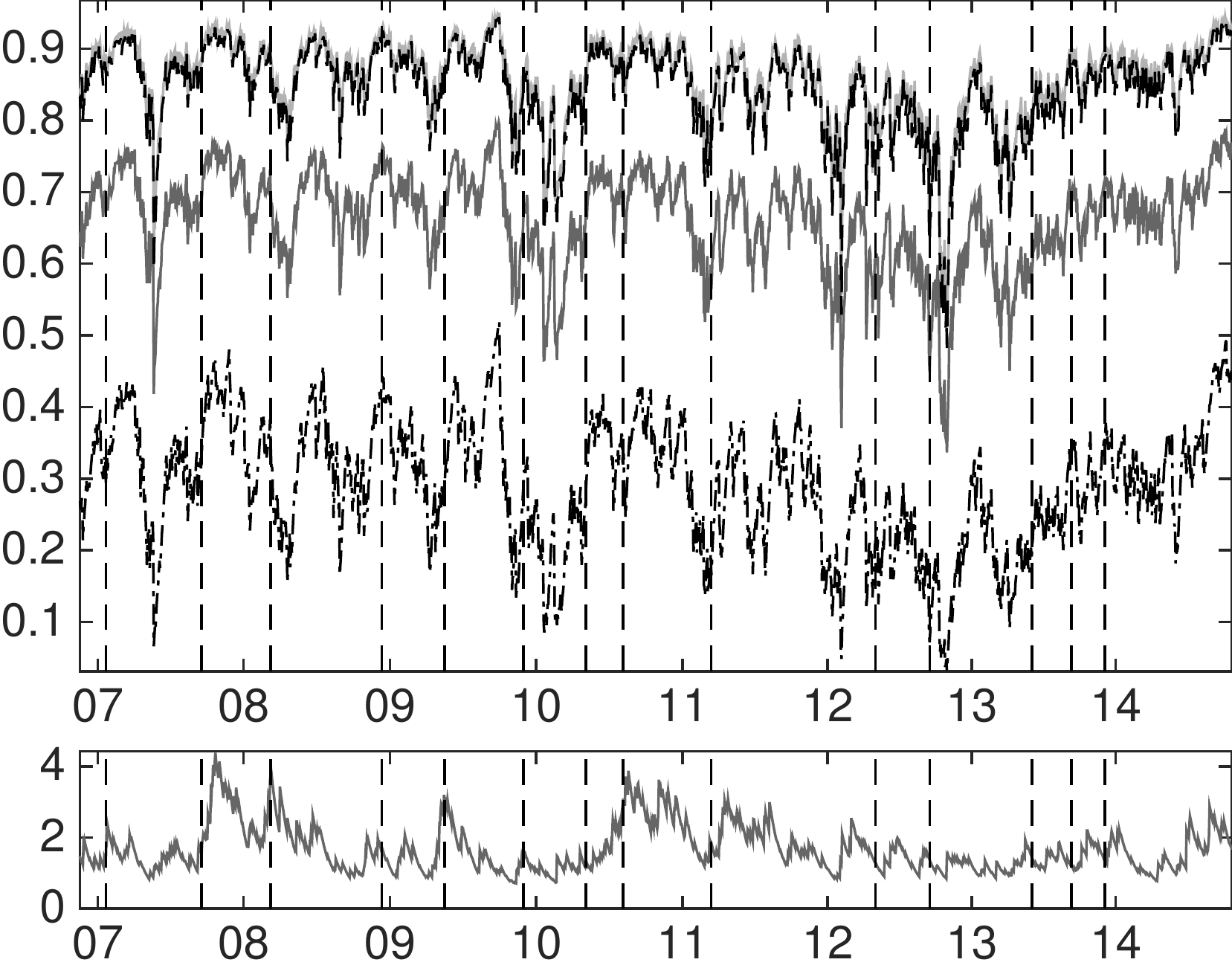}}\quad
\subfloat[Netherlands]{\label{fig:Netherlands_DepMeas}\includegraphics[width=0.22\textwidth]{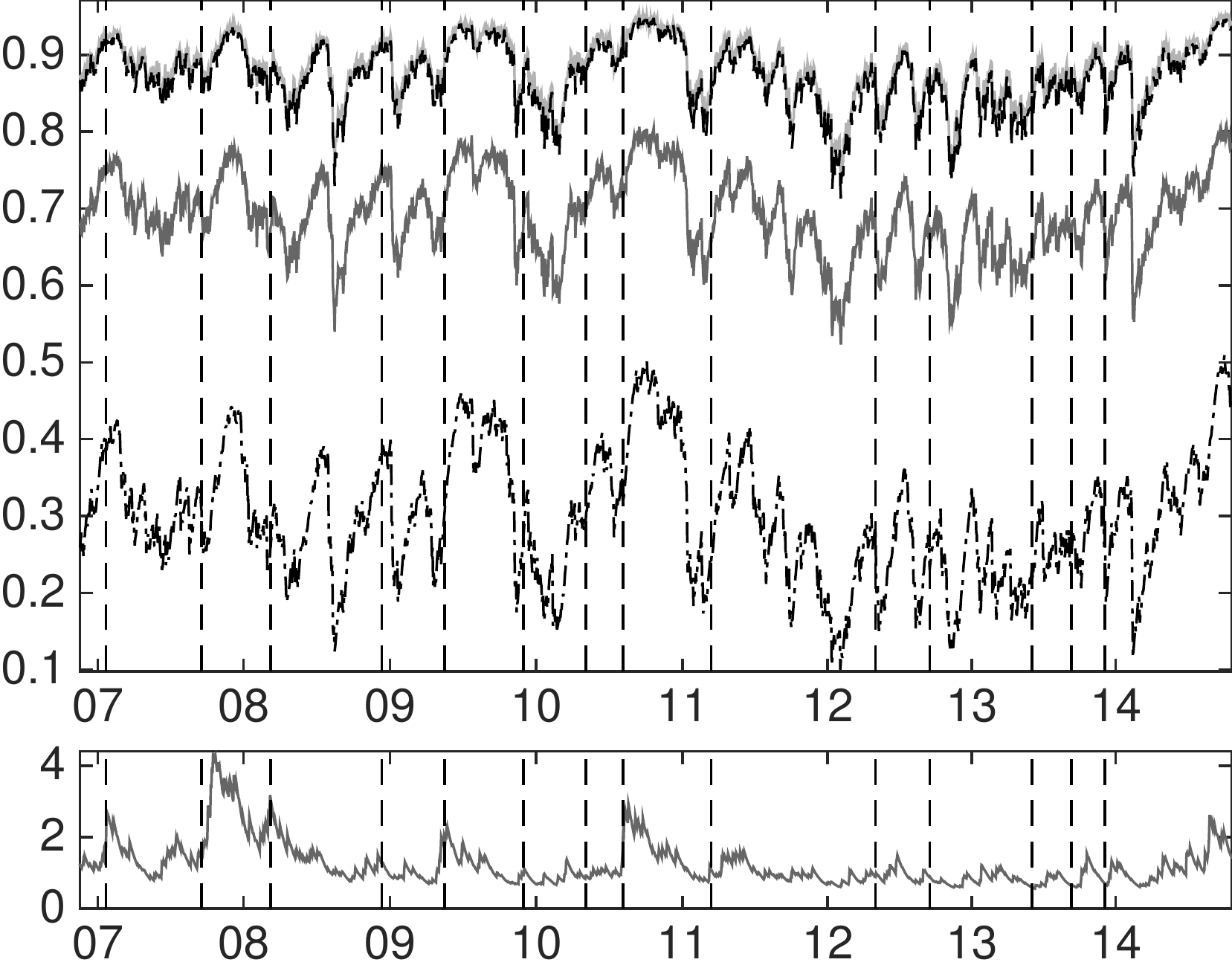}}\\
\subfloat[Spain]{\label{fig:Spain_DepMeas}\includegraphics[width=0.22\textwidth]{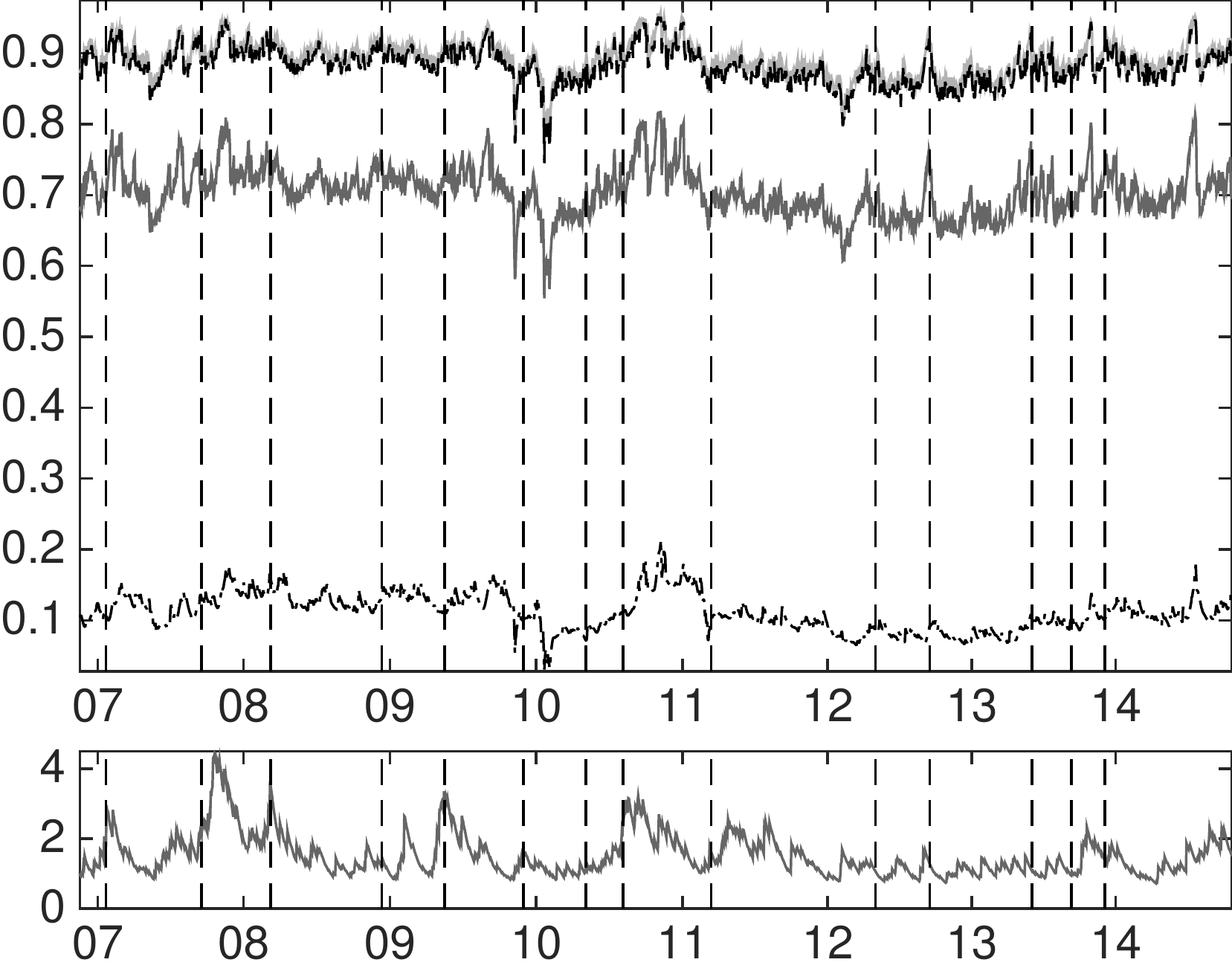}}\quad
\subfloat[Sweden]{\label{fig:Sweden_DepMeas}\includegraphics[width=0.22\textwidth]{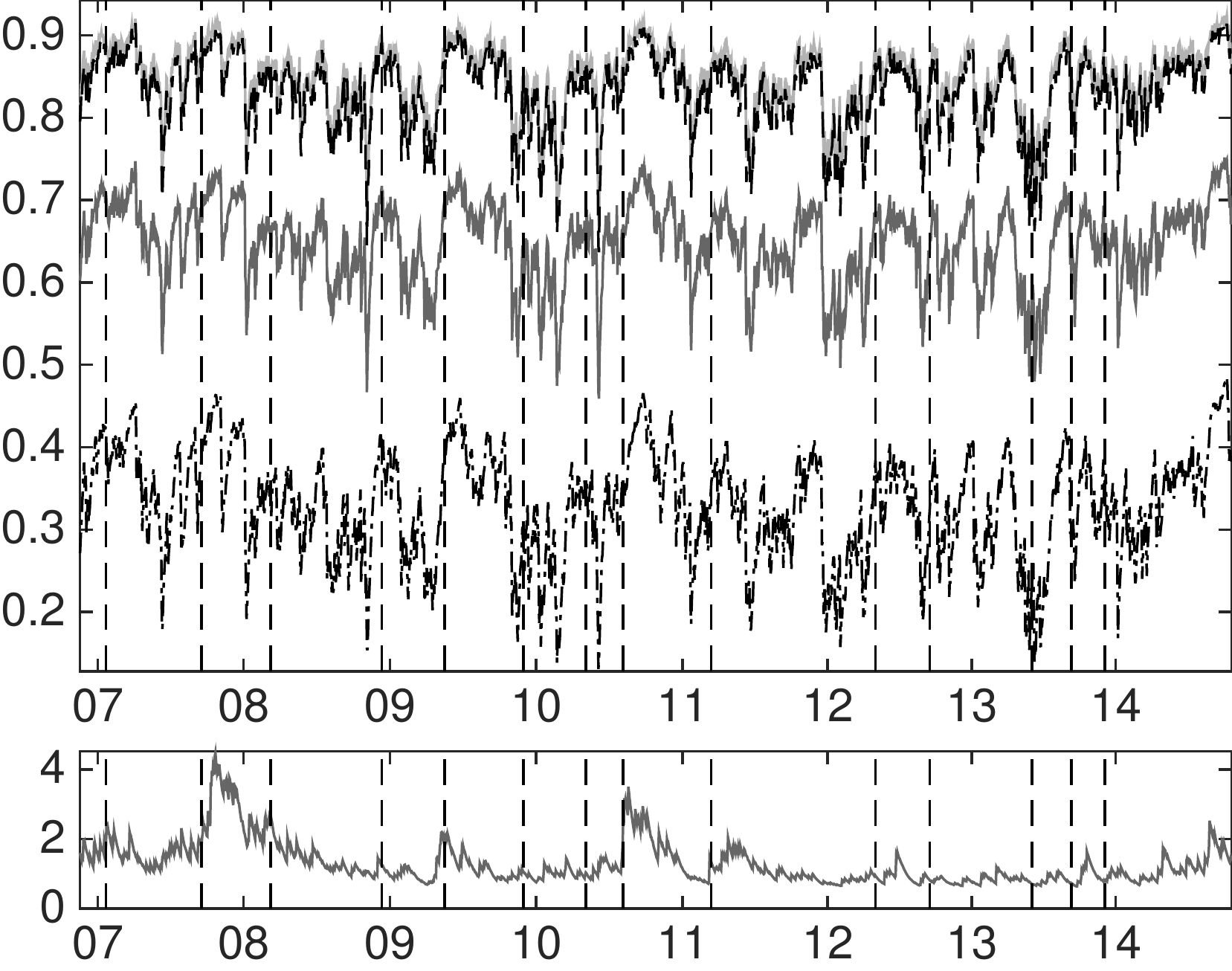}}\quad
\subfloat[United Kingdom]{\label{fig:United Kingdom_DepMeas}\includegraphics[width=0.22\textwidth]{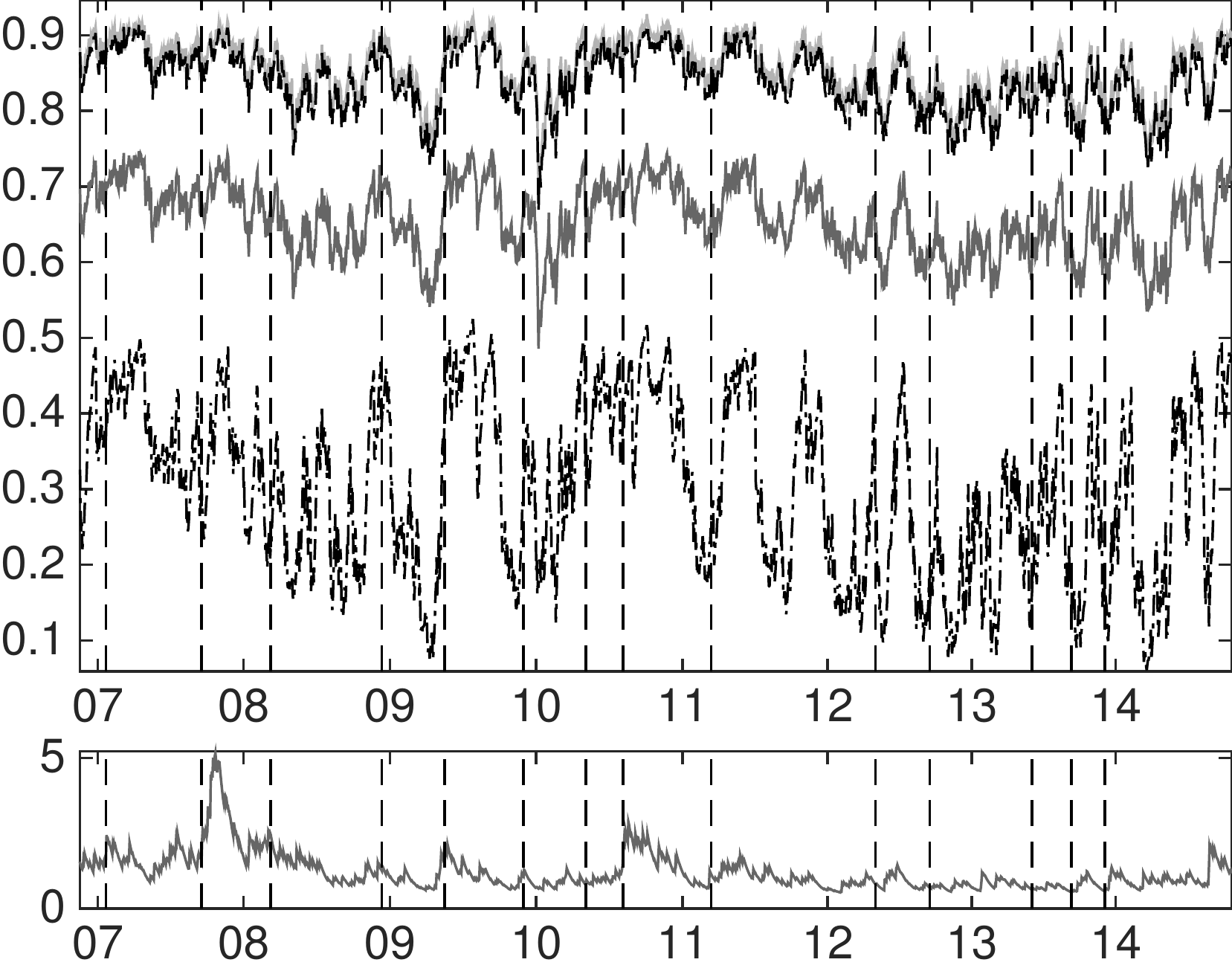}}
\caption{\footnotesize{Predicted dependence measures: linear correlation {\it (light grey line)}, Kendall's tau {\it (grey line)}, Spearman's rho {\it (black line)}, negative tail dependence {\it (dotted black line)}. The black line in the bottom figures represents the predicted marginal volatilities, i.e. $\hat{\sigma}_{j,t+1\vert t}$, for $t=1,2,\dots,T$ and $j=1,2,\dots,11$.
Vertical dashed lines represent major financial downturns: for a detailed description see Table \ref{tab:fin_crisis_timeline}.}}
\label{fig:CorrelMeas_ALL}
\end{sidewaysfigure}

%
\noindent The SGASC model can be effectively used to forecast several measures of association between the random variables such as the linear correlation, the concordance measures Spearman's rho and Kendall's tau and the coefficients of tail dependence. The Markovian nature of the SGASC model, and its flexibility to accommodate the marginals' specification, entails that the evaluation of such dependence measures can become cumbersome. The only exception concerns the evaluation of upper and lower tail dependence. In fact, since these measures only depend on the copula specification, the tail dependence coefficients of the predictive distribution for SGASC models are going to be equal to the convex linear combination of the components specific ones, conditional to the Markovian states. The linear correlation coefficient, the Spearman's rho and the Kendall's tau, instead, are not analytically available, and the only achievable solution is to rely on simulation methods. Following the approach of \cite{chollete_etal.2009}, we evaluate the linear correlation, the Spearman's rho and the Kendall's tau coefficients empirically based on 10000 simulated draws from the joint predictive distribution defined in equation \eqref{eq:hmm_predictive_general} and we repeated the procedure for the whole out--of--sample period. Figure \ref{fig:CorrelMeas_ALL} reports the predicted dependence measures, along with the corresponding filtered marginal volatilities. Figure \ref{fig:CorrelMeas_ALL} provides clear evidence of changing dependence structures during periods of financial turmoil, revealing also that the SGASC model adequately predicts the upward shifts in the dependence measures at economically relevant dates represented by vertical dashed lines. In this respect, particularly evident is the onset of the European sovereign debt crisis of 2010 which coincides with a suddenly increase of the dependence measures for all the investigated countries. Furthermore, by comparing Figure \ref{fig:CorrelMeas_ALL} and \ref{fig:DCoVaR_DCoES_SS} it emerges that the ordering induced by the lower tail dependence measures almost corresponds to that introduced by the systemic risk contributions. The only exception concerns UK which reports high tail dependence and moderate levels of both $\Delta$CoVaR and $\Delta$CoES. A possible explanation for this result should be ascribed to the fact that tail dependence at penultimate levels may be significantly stronger than in the limit, see \cite{manner_segers.2011}. Consistently with previous results, Hungary reports a close to zero tail dependence index over the whole out--of--sample period.
%
\section{Conclusion}
\label{sec:conclusion}
%
\noindent In this paper, we propose a new MS dynamic copula model (SGASC) that explicitly accounts for the presence of different Markovian regimes as well as a smooth within--regime dynamic evolution of the dependence parameters. Specifically, exploiting the recent advantages for score driven processes of \cite{creal_etal.2013} and \cite{harvey.2013}, we allow the state dependent copula parameters to be updated using the scaled score of the conditional copula distribution. This choice allows us to achieve a greater level of flexibility in the context of dynamic copula models, and it also introduces a stochastic behaviour for the class of GAS models in a natural and effective way. The SGASC model nests several alternative copula specifications with increasing levels of complexity such as the static copula, the Markov switching copula of \cite{jondeau_rockinger.2006} and \cite{chollete_etal.2009} as well as the GAS copula specification of \cite{creal_etal.2013} and \cite{harvey.2013}. Model parameters are consistently estimated by employing the Inference Function for Margins (IFM) two step estimator, where the second step is performed by tailoring the EM algorithm of \cite{dempster_etal.1977} to this class of models. The proposed estimation methodology is consistently equivalent to a single step estimation as long as the marginal conditional distributions do not depend on the latent Markovian states. \newline
\indent Although the SGASC model can be effectively used to understand the dynamic evolution of the non linear dependence among financial assets, here we focus our attention on the systemic risk measurement. Indeed, another relevant contribution of the paper is to introduce and estimate the Conditional Value--at--Risk (CoVaR) and the Conditional Expected Shortfall (CoES) risk measures of \cite{adrian_brunnermeier.2011,adrian_brunnermeier.2014} and \cite{girardi_ergun.2013}, for the class of SGASC models.\newline
\indent In the empirical part of the paper, we provide a comprehensive study to analyse the evolution of the systemic risk in Europe over the past decade. The considered period includes the recent Global Financial Crisis 2007--2008 and the European Sovereign Debt Crisis of 2010--2011. As regards our results, we found empirical evidence in support of the SGASC specification for more than half of the selected European countries, while the remaining countries are found to be well represented by the simpler GAS copula specification of \cite{creal_etal.2013} and \cite{harvey.2013}. The country individual systemic risk contribution as well as the overall systemic risk level in Europe is evaluated by means of the CoVaR and CoES risk measures and their $\Delta$ counterparts. Our empirical results confirm that, the proposed SGASC model is able to explain and predict the systemic risk evolution in an effective way. According to \cite{engle_etal.2015}, France is found to be the most systemically important country followed by Spain, Netherlands and Sweden. We find that the systemic risk contributions grow during the ESDC of 2010--2011 reaching its highest level in mid--2014. Then, after a short decreasing period in the last part of 2014, the magnitude as well as the relative importance of each country's systemic risk contribution, sharply changed. In the very last part of our sample, just after the decision of the Greek government to unilaterally broke off negotiations with the Eurogroup on June 26, 2015, the $\Delta$CoVaR and $\Delta$CoES measures suddenly rise up to the highest levels experienced since the mid--2014, for all the European countries. Moreover, we find evidence that, during this final period, the $\Delta$CoES risk measure provides a more clear systemic ranking. To conclude our empirical analysis, we also investigate the evolution of several dependence measures. Our main finding reveals that the ordering of systemic importance induced by the $\Delta$CoVaR and $\Delta$CoES risk measures is almost preserved by the lower tail dependence coefficient.
%
\section*{Acknowledgments}
%
\noindent This research is supported by the Italian Ministry of Research PRIN 2013--2015, ``Multivariate Statistical Methods for Risk Assessment'' (MISURA), and by the ``Carlo Giannini Research Fellowship'', the ``Centro Interuniversitario di Econometria'' (CIdE) and ``UniCredit Foundation''. We would especially like to thank Prof. Fabrizio Durante and Prof. Tommaso Proietti for their helpful comments and discussions on a previous version of the paper. We would like to express our sincere thanks to all the participants to the session ES23, \qmo Dependence models and copulas: Applications\qmc, of the \qmo 7th International Conference of the ERCIM WG on Computational and Methodological Statistics\qmcsp (ERCIM, 2014), for their constructive comments that greatly contributed to improving the final version of the paper. The revision of the paper has greatly benefited from a discussion with Prof. Andrew Harvey at the \qmo NBP Workshop on Forecasting\qmcsp held in Warsaw on September 28--29, 2015.
%
\clearpage
\newpage
\appendix
\section{Tables and figures}
\label{sec:appendix_A}
%
\begin{table}[!th]
\centering
\resizebox{0.9\columnwidth}{!}{%
\begin{tabular}{ll}
\toprule
Date  & Event  \\
\hline
January 21, 2008 & the global stock markets suffer their largest fall since September 2001.\\
March 16, 2008 & the Bear Stearns acquisition by JP Morgan Chase.\\
September 15, 2008 & the Lehman's failure.\\
March 9, 2009 & the peak of the onset of the recent GFC.\\
December 8, 2009 & the downgrading of Greece's credit rating from A- to BBB+ by Fitch ratings agency.\\
May 18, 2010 & the Greece achievement of 18bn USD bailout from EFSF, IMF and bilateral loans.\\
November 29, 2010 & the Ireland achievement of 113bn USD bailout from EU, IMF and EFSF.\\
May 05, 2011 & the Portugal bailout from ECM.\\
August 05, 2011 & the S\&P downgrading of US sovereign debt.\\
March 16, 2013 & the Cyprus achievement of 13bn USD bailout from ECM.\\
April 07, 2013 & the conference of the Portuguese Prime Minister regarding the high court's block of austerity plans.\\
April 30, 2013 & the approval of the Cyprus bailout by the Euro Parliament.\\
September 17, 2013 & the drop of car sales to the lowest recorded level in the Euro area.\\
June 03, 2014 & the drop of Eurozone inflation, and the consequent increasing pressure on the Central Bank.\\
September 09, 2014 & the mediterranean countries prepare for further unrest.\\
November 28, 2014 & the Italian unemployment rate reaches the record high since the 1977.\\
June 26, 2015 & the Greek government unilaterally broke off negotiations with the Eurogroup.\\
\bottomrule
\end{tabular}}
\caption{\footnotesize{Financial crisis timeline.}}
\label{tab:fin_crisis_timeline}
\end{table}
%
\begin{sidewaystable}[!th]
\centering
\resizebox{1.0\columnwidth}{!}{%
\begin{tabular}{lllllllllll}
\toprule
Austria & Belgium & Denmark & France & Germany & Hungary & Italy & Netherlands & Spain & Sweden & United Kingdom\\
\hline
Andritz Ag & Ackermans V.Ha & Carlsberg & Airbus Group & Basf Se & Any Biztonsagi & B P Milano     & Ahold & Abengoa & Abb & Admiral Group \\
Buwog Ag & Befimmo-Sicafi & Chr Hansen Hldg & Alcatel Lucent & Continental & Appeninn Vagyonk & Banco Popolare & Akzo Nbel & Abertis & Assa Abloy & Aviva Gb \\
Ca Immo & Bekaert    & Danske Bank & Alstom & Deutsche Telekom & Cig Pannonia & Bmps           & Altice & Acerinox & Astrazeneca & Berkeley Group \\
Conwert & Cofinimmo      & Dsv & Credit Agricole & E On & Elmu & Enel Green Pw  & Arcelormitta & Banco Santander & Boliden & Bg Group \\
Flughafen Wien & Colruyt        & Flsmidth & E.D.F. & Fresenius & Emasz & Enel           & Asml Holding & Bbva & Cellulosa Sca & Bt Group \\
Immofinanz & Delhaize Group & Gn Store Nord & Engie & Heidelbgcement & Fhb Jelzalogbank & Eni            & Gemalt & Enagas & Getinge & G4S \\
Lenzing & Dieteren   & Jyske Bank & Kering & Henkel Kgaa & Graphisoft Park & Exor           & Koninklijke Dsm & Gamesa & Investr & Imperial Tobacco \\
Osterreichi Post & Elia System Op & Maersk & Lvmh & K\&S & Magyar Telekom & Fiat Chrysler  & Nn Group & Iag & L M Ericsson & Inmarsat \\
Raiff Bnk Int & Gbl            & Maersk & Pernod Ricard & Lufthansa & Mol Magyar Olaj & Generali Ass   & Philips & Inditex & Lundinpetroleum & National Grid \\
Rhi & Kbc Groep      & Novo Nordisk & Publicis Groupe & Merck & Otp Bk & Mediobanca     & Randstad Holding & Mapfre & Securitas & Prudential \\
Telekom Austria & Solvay         & Novozymes & Snof & Munich Re Group & Pannergy & Moncler        & Rlx & Mediaset Espana & Skanska Ab & Royal Bank Scot \\
Uniqa & Telenet Grp Hl & Pandora & Solvay & Rwe & Raba Jarmuipari & Stmicroelec.N. & Shell & Ohl & Swedbank & Rsa Ins Grp \\
Vig & Thrombogenics  & Tryg & Technip & Sap Se & Richter Gedeon & Tenaris        & Unibail Rodamco & Red Electrica Co & Swedish Match & Shire \\
Voestalpine & Ucb            & Vestas Wind & Valeo & Siemens & Zwack Unicum & Terna          & Unilever Nv & Sacyr Es & Tele2 & Stanchart \\
Zumtobel & Umicore        & William Demant & Veolia Environ & Vonovia & -- & Unicredit      & Wolters Kluwer & Tecnicas Reunida & Volvo & Wpp \\
\bottomrule
\end{tabular}}
\caption{\footnotesize{Composition of the regional Indexes and country affiliation.}}
\label{tab:IndexList_list}
\end{sidewaystable}
%
%
\begin{table}[!t]
\captionsetup{font={small}, labelfont=sc}
\begin{center}
\begin{small}
\resizebox{0.8\columnwidth}{!}{%
\smallskip
\begin{tabular}{lcccccccccc}\\
\toprule
Name & Min & Max & Mean & Std. Dev. & Skewness & Kurtosis & 1\% Str. Lev. & JB & Kendall's $\tau$ & $\rho(1)$\\
\hline
\multicolumn{9}{l}{\textit{In--sample, from 08/07/1997 to 09/11/2007}}\\
Austria & -6.19 & 3.47 & 0.05 & 0.69 & -0.91 & 10.37 & -1.88 & 5101.1 & 0.35 & 0.02 \\
Belgium & -3.98 & 6.33 & 0.02 & 0.88 & 0.1 & 7.17 & -2.53 & 1542.75 & 0.51 & 0.11 \\
Denmark & -6.71 & 4.86 & 0.06 & 1.08 & -0.3 & 5.32 & -2.8 & 508.47 & 0.5 & 0.02 \\
France & -7.63 & 7.12 & 0.01 & 1.27 & -0.17 & 6.14 & -3.66 & 882.71 & 0.67 & 0.05 \\
Germany & -5.34 & 6.19 & 0.02 & 1.22 & -0.11 & 5.47 & -3.36 & 542.83 & 0.66 & -0.01 \\
Hungary & -4.77 & 4.08 & 0.05 & 1 & -0.14 & 4.27 & -2.62 & 149.62 & 0.34 & 0.03 \\
Italy & -7.75 & 7.48 & 0.01 & 1.18 & -0.2 & 7.09 & -3.53 & 1499.85 & 0.65 & 0.03 \\
Netherlands & -5.31 & 5.93 & 0.01 & 1.24 & -0.07 & 4.93 & -3.36 & 331.75 & 0.67 & 0.02 \\
Spain & -7.2 & 6.09 & 0.05 & 1.03 & -0.35 & 5.92 & -2.79 & 797.23 & 0.59 & -0.01 \\
Sweden & -7.01 & 6.21 & 0.04 & 1.24 & -0.15 & 5.86 & -3.35 & 730.84 & 0.64 & 0.01 \\
United Kingdom & -6.57 & 5.85 & 0.02 & 1.18 & -0.22 & 5.95 & -3.58 & 786.92 & 0.62 & 0 \\
Market & -5.13 & 4.61 & 0.03 & 0.86 & -0.38 & 6.12 & -2.57 & 910.75 & -- & 0.05 \\
\hline
Name & Min & Max & Mean & Std. Dev. & Skewness & Kurtosis & 1\% Str. Lev. & JB & Kendall's $\tau$ & $\rho(1)$\\
\hline
\multicolumn{9}{l}{\textit{Out--of--sample, from 09/11/2007 to 16/10/2015}}\\
Austria & -7.15 & 6.77 & -0.01 & 1.42 & -0.37 & 6.03 & -4.33 & 812.77 & 0.65 & 0.12 \\
Belgium & -5.4 & 6.14 & 0.01 & 1.2 & -0.11 & 5.69 & -3.55 & 610.77 & 0.73 & 0.09 \\
Denmark & -9.49 & 8.44 & 0.02 & 1.47 & -0.22 & 7.56 & -4.41 & 1756.1 & 0.64 & 0.07 \\
France & -9.32 & 10.1 & 0 & 1.63 & -0.17 & 6.88 & -4.99 & 1266.06 & 0.79 & 0.05 \\
Germany & -7.62 & 9.96 & 0 & 1.42 & -0.03 & 8.2 & -4.48 & 2259.45 & 0.73 & 0.02 \\
Hungary & -8.83 & 6.6 & -0.03 & 1.06 & -0.63 & 11.3 & -3.22 & 5886.51 & 0.39 & 0.06 \\
Italy & -7.48 & 8.81 & -0.03 & 1.83 & -0.17 & 4.5 & -5.09 & 198.35 & 0.71 & 0.01 \\
Netherlands & -7.04 & 7.3 & 0.02 & 1.37 & -0.29 & 7.01 & -4.09 & 1372.06 & 0.74 & 0.02 \\
Spain & -7.82 & 11.03 & -0.02 & 1.71 & -0.07 & 5.91 & -4.74 & 709.44 & 0.71 & 0.06 \\
Sweden & -7.76 & 8.9 & 0.02 & 1.46 & -0.03 & 7.49 & -4.48 & 1687.4 & 0.7 & -0.01 \\
United Kingdom & -8.83 & 6.93 & 0.01 & 1.36 & -0.22 & 6.73 & -3.84 & 1178.19 & 0.69 & 0 \\
Market & -6.9 & 7.6 & 0 & 1.29 & -0.24 & 6.93 & -3.89 & 1311.09 & -- & 0.06 \\
\bottomrule
\end{tabular}}
\caption{\footnotesize{Summary statistics of the panel of country specific indexes along with the total Market, for the period beginning on July 8, 1999 and ending on October 16, 2015. The seventh column, denoted by ``1\% Str. Lev.'' is the 1\% empirical quantile of the returns distribution, while the eight column, denoted by ``JB'' is the value of the Jarque-Ber\'a test-statistics. The last two columns report the estimated Kendall $\tau$ with respect to the total Market and the first order empirical autocorrelation of returns, respectively.}}
\label{tab:Index_data_summary_stat}
\end{small}
\end{center}
%
%
\end{table}

\begin{table}[!th]
\centering
\resizebox{0.8\columnwidth}{!}{%
\begin{tabular}{lcccccccccccc}
\toprule
& Austria & Belgium & Denmark & France & Germany & Hungary & Italy & Netherlands & Spain & Sweden & United Kingdom & Market\\
\hline
$\phi_{0,i}$ & $0.063^a$ & $0.045^a$ & $0.08^a$ & $0.028$ & $0.034^c$ & $0.063^b$ & $0.022$ & $0.024$ & $0.069^a$ & $0.043^c$ & $0.032^c$ & $0.05^a$ \\
$\phi_{1,i}$ & $0.031$ & $0.083^a$ & $0.034$ & $0.011$ & $-0.011$ & $0.008$ & $0.002$ & $0.006$ & $-0.003$ & $0.002$ & $0$ & $0.028$ \\
$\varpi_i$ & $0.019^a$ & $0.015^a$ & $0.035^a$ & $0.024^a$ & $0.024^a$ & $0.036^b$ & $0.019^a$ & $0.018^a$ & $0.03^a$ & $0.041^a$ & $0.019^a$ & $0.019^a$ \\
$\vartheta_{1,i}$ & $0.046^a$ & $0.015$ & $0.039^a$ & $0.017$ & $0$ & $0.039^a$ & $0.004$ & $0.008$ & $0.028$ & $0.017$ & $0$ & $0$ \\
$\vartheta_{2,i}$ & $0.079^a$ & $0.149^a$ & $0.098^a$ & $0.103^a$ & $0.138^a$ & $0.05^b$ & $0.131^a$ & $0.108^a$ & $0.103^a$ & $0.13^a$ & $0.144^a$ & $0.169^a$ \\
$\vartheta_{3,i}$ & $0.868^a$ & $0.886^a$ & $0.88^a$ & $0.912^a$ & $0.909^a$ & $0.898^a$ & $0.913^a$ & $0.924^a$ & $0.888^a$ & $0.888^a$ & $0.91^a$ & $0.88^a$ \\
$\eta_i$ & $0.967$ & $0.912^a$ & $0.938^b$ & $0.879^a$ & $0.876^a$ & $0.979$ & $0.875^a$ & $0.938^b$ & $0.891^a$ & $0.895^a$ & $0.925^a$ & $0.824^a$ \\
$\upsilon_i$ & $6.989^a$ & $10.835^a$ & $10.917^a$ & $12.093^a$ & $21.464^b$ & $8.981^a$ & $14.539^a$ & $22.519^b$ & $12.029^a$ & $9.602^a$ & $15.456^a$ & $13.832^a$ \\
\bottomrule
\end{tabular}}
\caption{\footnotesize{In sample parameters estimate of the marginal Skew--Student--t AR(1)--GJR--GARCH(1,1) model, defined in equation \eqref{eq:marginal_model}. The apexes \qmo a\qmc, \qmo b\qmcsp and \qmo c\qmc, denote the rejection of the null hypothesis of not significance of the corresponding parameter, at different confidence levels $1\%$, $5\%$ and $10\%$.}}
\label{tab:Marginal_estimates}
\end{table}
%
\begin{table}[!th]
\centering
\resizebox{0.8\columnwidth}{!}{%
\begin{tabular}{lcccccccccccc}
\hline
& Austria & Belgium & Denmark & France & Germany & Hungary & Italy & Netherlands & Spain & Sweden & United Kingdom & Market\\
\hline
DGT--AR$^{\left(1\right)}$ & $11.37$ & $17.35$ & $15.5$ & $21.18$ & $11.68$ & $21.94$ & $18.16$ & $23.5$ & $13.71$ & $17.66$ & $23.68$ & $20.15$ \\
DGT--AR$^{\left(2\right)}$ & $24.24$ & $21.93$ & $26.15$ & $12.73$ & $11.83$ & $8.88$ & $11.69$ & $31.33^c$ & $10.78$ & $17.24$ & $17.25$ & $13.81$ \\
DGT--AR$^{\left(3\right)}$ & $16.97$ & $19.88$ & $22.97$ & $19.58$ & $21.01$ & $22.02$ & $16.03$ & $20.27$ & $18.32$ & $14.02$ & $26.52$ & $16.93$ \\
DGT--AR$^{\left(4\right)}$ & $23.29$ & $22.37$ & $29.03^c$ & $13.18$ & $11.36$ & $8.6$ & $14.99$ & $31.17^c$ & $14.12$ & $18.73$ & $19.58$ & $18.09$ \\
DGT--H$\left(20\right)$ & $22.42$ & $16.43$ & $38.57^a$ & $12.85$ & $23.83$ & $58.7^a$ & $18.75$ & $18.87$ & $16.55$ & $36.57^a$ & $36.91^a$ & $18.3$ \\
\hline
\end{tabular}}
\caption{\footnotesize{ In sample Goodness--of--Fit test of \cite{tay_etal.1998}. Significance is denoted by superscripts at the $1\%(^a)$, $5\% (^b)$, and $10\% (^c)$ levels. See also \cite{vlaar_palm.1993} and \cite{jondeau_rockinger.2006}.}}
\label{tab:Uniform_test}
\end{table}
%
\begin{table}[!th]
\centering
\resizebox{0.8\columnwidth}{!}{%
\begin{tabular}{lccccccccccc}
\toprule
& Austria & Belgium & Denmark & France & Germany & Hungary & Italy & Netherlands & Spain & Sweden & United Kingdom\\
\hline
GAScop & -779.875 & -1518.162 & -1415.213 & $\mathbf{-2958.542}$ & -2861.274 & $\mathbf{-634.795}$ & $\mathbf{-2616.335}$ & $\mathbf{-2901.243}$ & -2218.804 & $\mathbf{-2510.746}$ & -2343.528 \\
SGASC--L2 & $\mathbf{-781.645}$ & $\mathbf{-1523.257}$ & -1424.369 & -2956.711 & $\mathbf{-2869.538}$ & -622.795 & -2612.144 & -2893.598 & $\mathbf{-2224.875}$ & -2504.244 & $\mathbf{-2347.613}$ \\
SGASC--L3 & -772.105 & -1508.505 & -1408.366 & -2952.458 & -2853.233 & -616.103 & -2605.672 & -2873.243 & -2216.333 & -2491.122 & -2332.282 \\
SGASC--L4 & -755.814 & -1495.789 & -1398.815 & -2937.515 & -2839.278 & -597.071 & -2583.14 & -2870.729 & -2198.038 & -2473.934 & -2314.826 \\
STATcop & -675.052 & -1394.788 & -1371.802 & -2833.301 & -2774.346 & -614.137 & -2496.057 & -2798.156 & -2081.016 & -2469.562 & -2227.762 \\
MScop--L2 & -770.958 & -1498.406 & -1415.98 & -2936.215 & -2845.161 & -633.146 & -2595.88 & -2875.751 & -2175.874 & -2500.511 & -2335.623 \\
MScop--L3 & -774.165 & -1512.3 & $\mathbf{-1426.806}$ & -2952.376 & -2862.936 & -623.068 & -2612.905 & -2895.403 & -2201.949 & -2502.917 & -2341.89 \\
MScop--L4 & -762.478 & -1503.91 & -1414.824 & -2940.936 & -2857.26 & -611.777 & -2598.457 & -2886.811 & -2199.878 & -2489.773 & -2333.726 \\
\bottomrule
\end{tabular}}
\caption{\footnotesize{Akaike Information Criterium for the different specification of the dynamic evolution of the dependence parameter. For each index, the selected specification is denoted in bold.}}
%
\label{tab:copulaAIC}
\end{table}
%
\begin{table}[!th]
\centering
\resizebox{0.80\columnwidth}{!}{%
\begin{tabular}{lccccccccccc}
\toprule
& Austria & Belgium & Denmark & France & Germany & Hungary & Italy & Netherlands & Spain & Sweden & United Kingdom \\
\hline
GAScop & 0.815 & 0.217 & 0.134 & 0.101 & 0.071 & 0.42 & 0.052 & 0.296 & 0.275 & 0.443 & 0.287 \\
SGASC-l2 & 0.975 & 0.038 & 0.13 & 0.096 & 0.046 & 0.419 & 0.047 & 0.027 & 0.122 & 0.358 & 0.756 \\
STATcop & 0.987 & 0.037 & 0.027 & 0.139 & 0.074 & 0.493 & 0.083 & 0.196 & 0.065 & 0.039 & 0.133 \\
MScop-L2 & 0.935 & 0.199 & 0.098 & 0.099 & 0.055 & 0.81 & 0.027 & 0.328 & 0.225 & 0.048 & 0.565 \\
MScop-L3 & 0.923 & 0.236 & 0.127 & 0.013 & 0.048 & 0.729 & 0.309 & 0.214 & 0.154 & 0.027 & 0.862 \\
MScop-L4 & 0.902 & 0.089 & 0.163 & 0.011 & 0.012 & 0.75 & 0.103 & 0.228 & 0.142 & 0.013 & 0.876 \\
\bottomrule
\end{tabular}}
\caption{\footnotesize{Test on the correct copula specification. The table reports the p--values of the Anderson--Darling goodness of fit test obtained using $1000$ bootstrap resamples, see \cite{manner_reznikova.2012} and \cite{patton.2012}.}}
\label{tab:ADTest}
\end{table}
\begin{table}[!th]
\centering
\resizebox{0.80\columnwidth}{!}{%
\begin{tabular}{lccccccccccc}
\toprule
& Austria & Belgium & Denmark & France & Germany & Hungary & Italy & Netherlands & Spain & Sweden & United Kingdom\\
\hline
$\omega$ & 0.081 & 0.009 & 0.33 & 1 & 0.001 & 1 & 0.179 & 1 & 0.034 & 1 & 0.005 \\
$\alpha$ & 0.147 & 0.377 & 0.712 & 0.117 & 0.109 & 1 & 0.657 & 0.037 & 0.118 & 1 & 0 \\
$\beta$ & 0.909 & 0.009 & 0.307 & 1 & 0.005 & 1 & 0.344 & 1 & 0.001 & 0.995 & 0.005 \\
$\nu$ & 0.826 & 0.721 & 0.001 & 0.057 & 0.303 & 1 & 0.445 & 1 & 0.83 & 0.681 & 0.094 \\
\bottomrule
\end{tabular}}
\caption{\footnotesize{Likelihood Ratio p--values for the hypothesis of regime independence of each parameter combination in the set $\{\omega, \beta,\alpha,\nu\}$. Numbers in boxes represent those specifications with the lowest AIC across that with a p--value below the significance level of 5\% per each country.}}
\label{tab:LR}
\end{table}
\begin{table}[!th]
\centering
\resizebox{0.80\columnwidth}{!}{%
\begin{tabular}{lccccccccccc}
\toprule
& Austria & Belgium & Denmark & France & Germany & Hungary & Italy & Netherlands & Spain & Sweden & United Kingdom\\
\hline
$\omega_1$ & $0.08459^a$ & $0.00133^a$ & $0.00895^a$ & $0.03873^a$ & $0.00462^a$ & $0.0186^a$ & $0.09411^a$ & $0.04758^a$ & $3e-05^a$ & $0.17078^a$ & $0.00059^a$ \\
$\beta_1$ & $0.90574^a$ & $0.99864^a$ & $0.99385^a$ & $0.98516^a$ & $0.99954^a$ & $0.98316^a$ & $0.96179^a$ & $0.98173^a$ & $0.99998^a$ & $0.92881^a$ & $0.99869^a$ \\
$\alpha_1$ & $0.02438^a$ & $0.01183^a$ & $0.01185^a$ & $0.04425^a$ & $0.0017^a$ & $0.01939^a$ & $0.0665^a$ & $0.04423^a$ & $0.0031^a$ & $0.0595^a$ & $0.00975^a$ \\
$\omega_2$ & $0.1471^a$ & $0.08396^a$ & $0.22526^a$ & -- & $0.00425^a$ & -- & -- & -- & $0.05423^a$ & -- & $0.21305^a$ \\
$\beta_2$ & $0.9233^a$ & $0.98971^a$ & $0.92492^a$ & -- & $0.99802^a$ & -- & -- & -- & $0.99197^a$ & -- & $0.92514^a$ \\
$\alpha_2$ & $0.07714^a$ & $0.03378^a$ & $0^a$ & -- & $0.00569^a$ & -- & -- & -- & $0.04082^a$ & -- & $0.04297^a$ \\
$\nu_1$ & $16.55112^a$ & $40.19022^a$ & $17.41423^a$ & $15.584^a$ & $150^a$ & $27.13906^a$ & $15.73629^a$ & $18.51802^a$ & $34.39654^a$ & $11.9017^a$ & $33.91062^a$ \\
$\nu_2$ & $20.91397^a$ & $149.99458^a$ & $149.96147^a$ & -- & $20.62929^a$ & -- & -- & -- & $137.29938^a$ & -- & $8.76538^a$ \\
$\gamma_{12}$ & $0.00079^a$ & $0.15478^a$ & $0.04524^a$ & -- & $0.01017^a$ & -- & -- & -- & $0.1096^a$ & -- & $0.01788^a$ \\
$\gamma_{21}$ & $0.00047^a$ & $0.02963^a$ & $0.00639^a$ & -- & $0.04714^a$ & -- & -- & -- & $0.04806^a$ & -- & $0.01838^a$ \\
\bottomrule
\end{tabular}}
\caption{\footnotesize{Parameters estimate of the SGASC--L2 and GAScop models with Student--t copula. The apexes \qmo a\qmc, \qmo b\qmcsp and \qmo c\qmc, denote the rejection of the null hypothesis of not significance of the corresponding parameter, at different confidence levels $1\%$, $5\%$ and $10\%$.}}
\label{tab:copulaCoef}
\end{table}
%
\begin{table}[!th]
\centering
\resizebox{0.80\columnwidth}{!}{%
\begin{tabular}{lccccccccccc}
\toprule
& Austria & Belgium & Denmark & France & Germany & Hungary & Italy & Netherlands & Spain & Sweden & United Kingdom\\
\hline
$\bar\rho_1$ & 0.42 & 0.454 & 0.62 & 0.861 & 0.998 & 0.501 & 0.841 & 0.861 & 0.581 & 0.832 & 0.222 \\
$\bar\rho_2$ & 0.742 & 0.997 & 0.903 & -- & 0.789 & -- & -- & -- & 0.996 & -- & 0.888 \\
$\bar\varpi_1$ & 0.016 & 0 & 0.052 & 0.282 & 0.692 & 0.005 & 0.246 & 0.241 & 0.004 & 0.296 & 0 \\
$\bar\varpi_2$ & 0.086 & 0.66 & 0.006 & -- & 0.125 & -- & -- & -- & 0.584 & -- & 0.465 \\
${\rm HL}_1$ & 7.518 & 555.2 & 132.628 & 69.633 & 4783.995 & 45.248 & 25.924 & 56.414 & 33868.316 & 13.495 & 542.058 \\
${\rm HL}_2$ & 11.281 & 194.099 & 14.419 & -- & 477.173 & -- & -- & -- & 225.685 & -- & 14.034 \\
${\rm D}_1$ & 2114.657 & 32.75 & 155.415 & -- & 20.211 & -- & -- & -- & 19.808 & -- & 53.395 \\
${\rm D}_2$ & 1271.144 & 5.461 & 21.105 & -- & 97.355 & -- & -- & -- & 8.124 & -- & 54.935 \\
\bottomrule
\end{tabular}}
\caption{\footnotesize{State identification. For each state $l=1,2,\dots,L$, $\bar{\rho}_l,\bar\varpi_l$ denote the long run value of the dependence parameter and that of the tail dependence coefficient of the student--t copula, respectively.
$\bar{\rho}_l=\xp\left(\rho_{l,t}\mid S_t=l\right)=\lambda\left(\frac{\omega_l}{1-\beta_l}\right)$,
where $\lambda\left(\cdot\right)$ is defined in equation \eqref{eq:mapping}.
${\rm HL}_l$ is the half life of the GAS dynamics, which can be calculated as the solution of the equation $1-\exp\left(-\beta_l^{{\rm HL}_l\lambda^{-1}\left(\bar{\rho}_l\right)}\right)=\frac{\bar\rho_l\left(1+\exp\left(-\beta_l^{{\rm HL}_l\lambda^{-1}\left(\bar{\rho}_l\right)}\right)\right)}{2}$, ${\rm D}_l=\frac{\gamma_{ll}}{1-\gamma_{ll}}$ represents the expected duration (in days) of state $l$ and $\bar{\varpi}_l=\xp\left(\lim_{u\to 0}\mathbb{P}\left(u_S\leq u\mid u_j\leq u\right)\mid S_t=l\right)$.}}
\label{tab:MS_state_identification}
\end{table}
%
%
\begin{table}[!th]
\centering
\resizebox{0.8\columnwidth}{!}{%
\begin{tabular}{lccccccccccc}
\toprule
& Austria & Belgium & Denmark & France & Germany & Hungary & Italy & Netherlands & Spain & Sweden & United Kingdom\\
\hline
AE & 1.58 & 1.91 & 1.65 & 1.68 & 1.55 & 1.2 & 1.6 & 1.68 & 1.29 & 1.89 & 1.85 \\
UC & 0.14 & 0.04 & 0.13 & 0.14 & 0.18 & 0.66 & 0.16 & 0.12 & 0.46 & 0.06 & 0.07 \\
CC & 0.34 & 0.04 & 0.13 & 0.32 & 0.4 & 0.61 & 0.15 & 0.12 & 0.66 & 0.17 & 0.07 \\
ADMax & 3.45 & 3.51 & 3.5 & 3.52 & 3.52 & 3.13 & 3.51 & 3.52 & 3.53 & 3.49 & 3.52 \\
ADMean & 0.27 & 0.34 & 0.29 & 0.3 & 0.28 & 0.19 & 0.28 & 0.3 & 0.23 & 0.33 & 0.33 \\
\bottomrule
\end{tabular}}
\caption{\footnotesize{Actual over expected (A/E) ratios, p--values of the Unconditional Coverage (UC) and Conditional Coverage (CC) tests of \cite{kupiec.1995} and \cite{christoffersen.1998}, Maximum absolute deviation (ADmax) and Mean absolute deviation (ADmean) for the CoVaR$_t^{S\vert j}$, $j=1,2,\dots,11$. The AE indicator is calculated as the ratio between the realised and expected CoVaR$_t^{S\vert j}$, $j=1,2,\dots,11$ exceedances, while the hitting values of the UC and CC tests are calculated using the procedure suggested by \cite{girardi_ergun.2013}.}}
\label{tab:UC_sgasc}
\end{table}
\clearpage
\newpage
%
%
%
%
%
%
%
%
%
%

%
%
\clearpage
\newpage
%
%

%

\end{document}